\documentclass[%
 aip,pop,
 amsmath,amssymb,
 reprint,
]{revtex4-1}

\usepackage{graphicx, xcolor}
\usepackage{float} 
\usepackage[caption = false]{subfig} 
\graphicspath{ {./images/}}
\usepackage{dcolumn}
\usepackage{bm}
\usepackage[utf8]{inputenc}
\usepackage[T1]{fontenc}
\usepackage{mathptmx}
\usepackage{afterpage}

\usepackage{dsfont} 

\newcommand{\rd}[1]{\mathrm{#1}} 

\draft 

\begin{document}

\preprint{AIP/123-QED}

\title{Linear Dispersion Theory of Parallel Electromagnetic Modes 
       for Regularized Kappa-Distributions}

\author{Edin Husidic}
\email[]{eh@tp4.rub.de}
\affiliation{Institut für Theoretische Physik, Lehrstuhl IV:
Plasma-Astroteilchenphysik, Ruhr-Universität Bochum, D-44780 Bochum, Germany}
\affiliation{Centre for Mathematical Plasma Astrophysics, 3001 Leuven Belgium}

\author{Marian Lazar}
\affiliation{Institut für Theoretische Physik, Lehrstuhl IV:
Plasma-Astroteilchenphysik, Ruhr-Universität Bochum, D-44780 Bochum, Germany}
\affiliation{Centre for Mathematical Plasma Astrophysics, 3001 Leuven Belgium}

\author{Horst Fichtner}
\affiliation{Institut für Theoretische Physik, Lehrstuhl IV:
Plasma-Astroteilchenphysik, Ruhr-Universität Bochum, D-44780 Bochum, Germany}
\affiliation{Research Department, Plasmas with Complex Interactions,
Ruhr-Universität Bochum, D-44780 Bochum, Germany}

\author{Klaus Scherer}
\affiliation{Institut für Theoretische Physik, Lehrstuhl IV:
Plasma-Astroteilchenphysik, Ruhr-Universität Bochum, D-44780 Bochum, Germany}
\affiliation{Research Department, Plasmas with Complex Interactions,
Ruhr-Universität Bochum, D-44780 Bochum, Germany}

\author{Patrick Astfalk}
\affiliation{Max-Planck-Institut f\"ur Plasmaphysik, D-85748 Garching, Germany}
 \affiliation{2-36-6 Imado, Taito-ku, Tokyo-to, 111-0024, Japan}

\date{\today}

\begin{abstract}     

The velocity particle distributions measured in-situ 
in space plasmas deviate from Maxwellian (thermal) 
equilibrium, showing enhanced suprathermal tails which 
are well described by the standard Kappa-distribution (SKD). 
Despite its successful application, the SKD is frequently 
disputed due to a series of unphysical implications like 
diverging velocity moments, preventing a macroscopic 
description of the plasma. The regularized Kappa-distribution
(RKD) has been introduced to overcome these limitations, but 
the dispersion properties of RKD-plasmas are not explored yet. 
In the present paper we compute the wavenumber dispersion of 
the frequency and damping or growth rates for the electromagnetic 
modes in  plasmas characterized by the RKD. This task is
accomplished by using the grid-based kinetic dispersion solver 
\emph{LEOPARD} developed for arbitrary gyrotropic distributions
[Ref.~\onlinecite{Astfalk-and-Jenko-2017}]. By reproducing previous 
results obtained for the SKD and Maxwellian, we validate the 
functionality of the code. Furthermore, we apply the isotropic as 
well as the anisotropic RKDs to investigate stable electromagnetic
electron-cyclotron (EMEC) and ion-cyclotron (EMIC) modes as well as 
temperature-anisotropy-driven instabilities, both for the case 
$T_\perp / T_\parallel > 1$ (EMEC and EMIC instabilities) and for the
case $T_\perp / T_\parallel < 1$ (proton and electron firehose 
instabilities), where $\parallel$ and $\perp$ denote directions 
parallel and perpendicular to the local time-averaged magnetic field. 
Provided that the cutoff parameter $\alpha$ is small enough, the 
results show that the RKDs reproduce the dispersion curves of the SKD 
plasmas at both qualitative and quantitative levels. For higher values,
however, physically significant deviation occurs.
\end{abstract}


\maketitle 




\textbf{Credits and permissions}: This article may be downloaded for 
personal use only. Any other use requires prior permission of the author 
and AIP Publishing. This article appeared in 
\textit{Phys. Plasmas} 27, 042110 (2020) 
and may be found at \url{https://doi.org/10.1063/1.5145181}. 

\section{Introduction}\label{sec:introduction} 

The existence of nonthermal particle populations in the 
heliospheric plasma is well confirmed by various in-situ 
measurements.\cite{Maksimovic-et-al-2005,Zouganelis-et-al-2005, Marsch-2006}
The velocity distributions of plasma particles deviate 
from thermal (Maxwellian) equilibrium, exhibiting nonthermal 
features like suptrathermal tails and temperature anisotropies, 
which one expects in general to be the case for any dilute and 
low-collisional plasma in space.\cite{Schlickeiser-2002} 
Non-equilibrium particle populations present in space plasmas, 
in particular in the solar wind and terrestrial magnetosphere,
can explain the enhanced fluctuations reported in these environments. 
Whenever a particle velocity distribution deviates from a Maxwellian, 
there is free energy available that can be accessed by linear eigenmodes 
of the system, if a certain threshold is crossed.\cite{Lazar-et-al-2013} 
This can lead to an exponential growth of instabilities, entertaining 
the wave fluctuations that carry the energy throughout the 
plasma.\citep{Baumjohann-and-Treumann-1997} 

A widely used tool to describe the solar wind particle distributions 
is the (isotropic) standard Kappa-distribution 
(SKD)\cite{Pierrard-and-Lazar-2010}
\begin{equation}\label{eq:olbert_kappa}
f_\rd{SKD}(v) =  \frac{n}{\pi^{3/2}\,\Theta^3} 
\frac{\Gamma(\kappa + 1)}{\kappa^{3/2}\,\Gamma(\kappa - 1/2)}
\left(1 + \frac{v^2}{\kappa\,\Theta^2} 
\right)^{-\kappa - 1}
\end{equation}
with $v$ being the particle speed, $\Theta$ the most 
probable speed related to kinetic 
temperature\cite{Vasyliunas-1968, Lazar-et-al-2015, Lazar-et-al-2016}, 
$n$ the particle number density and the $\kappa$-parameter 
associated with the high-energy power-law decrease of the distribution. 
The SKD has originally been introduced by Olbert\cite{Olbert-1968} and 
Vasyli\={u}nas\cite{Vasyliunas-1968} in 1968 to describe electron 
distributions in Earth's magnetosphere, and since then has been employed 
in the interplanetary environment,\cite{Maksimovic-et-al-1997, Maksimovic-et-al-2005, Pierrard-and-Lazar-2010} in the interstellar 
and intergalactic medium,\cite{Davelaar-et-al-2018, de-Avillez-et-al-2018} 
and even applied in experimental 
physics.\cite{Webb-et-al-2012, Elkamash-and-Kourakis-2016} 
This power-law function shows enhanced high-energy tails, 
extending from a Maxwellian core, and turns (approximately) 
into the same Maxwellian for $\kappa \to \infty$. A generalization 
of the SKD is the standard Bi-Kappa-distribution (SBKD)
\begin{equation}\label{eq:skd}
  \begin{split}
    f_\rd{SBKD}(v) = &\frac{n}{\pi^{3/2}\,\Theta_\parallel\,\Theta_\perp^2} 
    \frac{\Gamma(\kappa + 1)}{\kappa^{3/2}\,\Gamma(\kappa - 1/2)}\\ 
    &\times \left(1 + \frac{v_\parallel^2}{\kappa\,\Theta_\parallel^2}
    + \frac{v_\perp^2}{\kappa\,\Theta_\perp^2}\right)^{-\kappa - 1}\,\rd{,}
  \end{split}
\end{equation}
which has been regularly applied to describe plasmas with anisotropic 
distributions,\cite{Lazar-and-Poedts-2008,Lazar-et-al-2017} 
where $\parallel$ and $\perp$ denote, respectively, directions 
parallel and perpendicular to the local background magnetic field.

Despite their current involvement in a wide variety of applications 
in space plasmas, these standard models are still controversial in 
the plasma physics community, mainly due to the diverging velocity 
moments for low values of $\kappa$, which prevents a macroscopic 
description of these plasma systems. More exactly, the existence of 
moments of order $l$ impose the 
restriction\cite{Scherer-et-al-2017,Scherer-et-al-2019a} 
$\kappa > (l + 1)/2$ on the $\kappa$-parameter. Indeed, the SKD is 
defined in terms of $\Theta$, which is directly related to 
the second order moment (i.e., kinetic temperature) that requires 
$\kappa > 3/2$. Further unphysical implications of the SKD are mainly 
concerning the entropy of the physical system and the contribution of 
superluminal particles to macroscopic 
quantities.\cite{Fichtner-et-al-2018, Scherer-et-al-2019b}
Regarding the validity of the SKD in non-relativistic applications, 
as for any non-relativistic theory in order to be valid, the 
contribution from superluminal particles, i.e., particles 
with $v > c$, to macroscopic quantities (e.g., pressure) must 
be negligible. However, Scherer et al.\cite{Scherer-et-al-2019b} 
showed that for values of $\kappa < 2$ superluminal particles may 
contribute significantly to the pressure. The RKD offers an elegant 
and straightforward way to reduce these effects, which otherwise 
would have been solved by adopting a relativistic Kappa model. 
However, a relativistic Kappa does not solve the divergence of higher 
order moments but complicates very much the kinetic approach. 
In the solar wind (kinetic) temperatures of electrons and ions are 
nonrelativistic but comparable, meaning that ions have speeds much 
lower than electrons, such that the regularization applied for 
electrons to reduce the effects of their superluminal populations 
may not be necessary for the ions. Additionally, Gaelzer and 
Ziebell\cite{Gaelzer-and-Ziebell-2014} found that a small-gyroradius 
treatment, which is used in the study of kinetic Alfv\'{e}n waves, 
is seriously compromised when employing the SKD. To fix these 
problematic features a regularized (isotropic) Kappa-distribution 
(RKD) has been introduced in the form\cite{Scherer-et-al-2017}
\begin{equation}\label{eq:rkd}
  f_\rd{RKD}(v) = n\,N_\rd{RKD} 
  \left(1 + \frac{v^2}{\kappa\,\Theta^2}\right)^{-\kappa - 1} 
  \exp \left(\frac{-\alpha^2\,v^2}{\Theta^2}\right)\,.
\end{equation}
The normalization constant is such that 
$\int \rd{d}^3 v\,f_{\rd{RKD}} = n$ is fulfilled, and is given by
\begin{equation}\label{eq:n_rkd}
  N_\rd{RKD} = \frac{1}{(\pi\,\kappa)^{3/2}\,\Theta^3}\, 
  U\left(\frac{3}{2},\frac{3-2\,\kappa}{2},\alpha^2\,\kappa\right)\,\rd{,}
\end{equation}
where $U(a,c,x)$ denotes the Kummer function\cite{Oldham-et-al-2000} 
(also known as Tricomi function). The RKD consists of a power-law term 
and a Maxwellian cutoff, which is independent of $\kappa$, 
but contains a regularization parameter $\alpha$, which needs to be 
positive but small enough in order to keep the main implication of the 
distribution and allow for convergent velocity moments and small 
contribution of superluminal particles. Despite its apparent ad-hoc 
application, there is also observational justification for the development 
of the RKD.\cite{Steenberg-and-Moraal-1999,Fisk-and-Gloeckler-2012}
 
\begin{figure}[t!]
  \includegraphics[width=.4\textwidth]{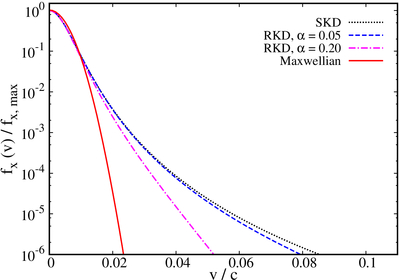}
    \caption{Comparison of two RKDs ($\kappa = 2$, $\alpha = 0.05$, 
             blue dashed line, and $\alpha = 0.2$, pink
             dashed-dotted line), the Maxwellian ($\alpha = 0$, 
             $\kappa \to \infty$, solid red line), and the SKD 
             ($\alpha = 0$, $\kappa = 2$, dotted black line). 
             All distributions (normalized to their maximums) are 
             plotted vs.\ particle speed $v_\perp \equiv v$
             (normalized to the speed of light) at 
             $v_\parallel = 0$.}\label{fig:distributions}
\end{figure}

Figure~\ref{fig:distributions} illustrates the impact of the 
regularization parameter $\alpha$ on the distribution function, 
where two RKDs ($\kappa = 2$; $\alpha = 0.05$ and $\alpha = 0.2$) 
are compared with the corresponding SKD ($\alpha = 0$) and Maxwellian 
($\alpha = 0$ and $\kappa \to \infty$). By applying the RKD, the 
contribution of superluminal particles can become negligible, if 
$\alpha$ is properly chosen. In Figure~\ref{fig:distributions} we can 
see that more particles of higher speed are removed with increasing $\alpha$. 
On the other hand, the $\alpha$-parameter must be small enough, i.e., 
usually $\alpha \ll  1$, in order to keep a reasonable contribution of the 
suprathermals in the high-energy tails of the distribution.

In this paper we consider the (isotropic) RKD to study the 
electromagnetic (EM) stable modes, and for the unstable modes the 
(anisotropic) regularized bi-Kappa-distribution (RBKD), 
given by\cite{Scherer-et-al-2019a}
\begin{equation}\label{eq:rbkd}
  \begin{split}
    f_\rd{RBKD}(v_\parallel,v_\perp) = &n\,N_\rd{RBKD} 
    \left(1 + \frac{v_\parallel^2}{\kappa\,\Theta_\parallel^2} 
    + \frac{v_\perp^2}{\kappa\,\Theta_\perp^2}\right)^{-\kappa - 1} \\
    &\times 
    \exp \left(-\frac{\alpha_\parallel^2\,v_\parallel^2}{\Theta_\parallel^2}
    -\frac{\alpha_\perp^2\,v_\perp^2}{\Theta_\perp^2} \right)
  \end{split}
\end{equation}
and allowing for different regularization parameters $\alpha_\parallel$ 
and $\alpha_\perp$ in directions parallel and perpendicular to the 
magnetic field, respectively. Here, the normalization constant reads\cite{Scherer-et-al-2019a}
\begin{equation}\label{eq:n_rbkd}
  N_\rd{RBKD} = \frac{1}{(\pi\,\kappa)^{3/2}\,
  \Theta_\parallel\,\Theta_\perp^2\,W}\,,
\end{equation}
where the quantity $W$ is the integral
\begin{equation}\label{eq:w_with_kummer_f}
  W = \int\limits_{0}^{1} \rd{d}t\,U \left(\frac{3}{2}, 
  \frac{3 - 2\,\kappa}{2}, \kappa\left[\alpha_\perp^2 +
  \left(\alpha_\parallel^2 - \alpha_\perp^2 \right)t^2 \right]\right) \,,
\end{equation}
which can be solved analytically if $\alpha_\parallel = \alpha_\perp$, 
and numerically otherwise.\cite{Scherer-et-al-2019a}

This paper is structured in the following way. In Section~\ref{sec:theory} 
we introduce the dispersion theory and present 
\emph{LEOPARD}\cite{Astfalk-and-Jenko-2017} (acronym for Linear 
Electromagnetic Oscillations in Plasmas with Arbitrary 
Rotationally-symmetric Distributions), the numerical solver used to solve 
the main dispersion relations. Section~\ref{sec:results} contains the 
results, with detailed discussions of the stable modes, the electromagnetic 
electron-cyclotron and ion-cyclotron waves in Section~\ref{subsec:stable_modes}, 
while Sections~\ref{subsec:emec_emic}, \ref{subsec:pfh_efh} and 
\ref{subsec:apara_neq_aperp} examine the instabilities of EM cyclotron and 
firehose modes. We conclude in Section~\ref{sec:summary} with a summary of our 
present results and a discussion of the strengths of \emph{LEOPARD}, as well as 
the computational issues to be fixed in future works.

\section{General Theory \& Numerical Solver}\label{sec:theory}  

The hot and dilute plasmas from space are propitious to departures from 
thermal equilibrium and implicitly to various types of wave instabilities. 
These instabilities are described by solving the kinetic dispersion relation, 
which sets the (complex) wave frequency $\omega = \omega_\rd{\,r} + i\, \gamma$ 
and the wavenumber $k$ in relationship. From the beginning we need to mention 
that the analytical derivation of the dielectric tensor for a RKD, especially 
the anisotropic versions, is not straightforward. For this reason we proceed to 
test and use \emph{LEOPARD}\cite{Astfalk-and-Jenko-2017}, a numerical solver able 
to resolve the dispersion and stability properties of plasmas with arbitrary 
gyrotropic distribution functions defined numerically. This solver works with a 
general theoretical approach which may appeal any of the distribution models, 
including Maxwellians, SKDs and RKDs, to compute the dispersion curves for the 
wave frequency as well as growth or damping rate of the electromagnetic waves in 
a hot, homogeneous and magnetized plasma. 

In the following we consider the plasma fluctuations 
to be sufficiently weak, so that linear theory can be 
applied to describe them. Typically, to derive the linear 
kinetic dispersion relation $\omega(k)$, one may start from  
the Vlasov-Maxwell equations and perform a Fourier and Laplace 
transformation of the perturbed quantities, which gives temporally 
and spatially variations for the EM fields and particle (flux) 
densities. These are inserted into Maxwell's equations to yield the 
general dispersion relation for nontrivial 
solutions\cite{Baumjohann-and-Treumann-1997, Schlickeiser-2002}
\begin{equation}\label{eq:disp_rel}
  0 = \det \left(\frac{c^2\,k^2}{\omega^2} \left(\frac{\vec{k} \otimes 
  \vec{k}}{k^2} - \mathds{1} \right) + \overleftrightarrow{\epsilon} 
  (\vec{k},\omega) \right)\,\rd{,}
\end{equation}
where $c$ is the speed of light, $k$ is the wavenumber, $\mathds{1}$ is 
the unit tensor, and $\otimes$ is the dyadic product of two vectors. 
The dielectric tensor $\overleftrightarrow{\epsilon} (\vec{k},\omega)$ 
in Eq.~\eqref{eq:disp_rel} is defined as
\begin{equation}\label{eq:dielectric_tensor}
  \overleftrightarrow{\epsilon} (\vec{k},\omega) = 
  \mathds{1} + \frac{i}{\omega\,\epsilon_0} \,
  \overleftrightarrow{\sigma}(\vec{k},\omega)
\end{equation}
with $\epsilon_0$ being the permittivity of free space 
and $\overleftrightarrow{\sigma}(\vec{k},\omega)$ the 
conductivity tensor. By linearizing the collision-free 
Vlasov-Maxwell equations the components of the dielectric 
tensor can be derived via\cite{Brambilla-1998, Schlickeiser-2002}
\begin{equation}\label{eq:diel_tensor_equation}
  \begin{split}
    \epsilon_{\,ij}(\vec{k},\omega) = 
    &\delta_{ij} - 2\,\pi \sum_\rd{s} \frac{\omega^2_{\rd{ps}}}{\omega^2}
    \bigg\lbrace \int\limits_{-\infty}^{\infty}\rd{d}v_\parallel \\
    &\times
    \int\limits_{0}^{\infty}\rd{d}v_\perp\,v_\perp^2 \sum_{n=-\infty}^{\infty} 
    \frac{\omega}{\omega - n\,\Omega_\rd{s} - k_\parallel\,v_\parallel}\,
    Q_{ij}^{\rd{s},n}\bigg\rbrace\,\rd{.}
  \end{split}
\end{equation}
Here, $\delta_{ij}$ denotes the Kronecker delta, $\omega_\rd{ps}$ the 
plasma frequency, $\Omega_\rd{s}$ the cyclotron frequency of species $s$ 
and $n$ the Bessel function index. The elements $Q_{ij}^{\rd{s},n}$ 
contain the derivatives of the particle velocity distribution function 
and are computed as described in 
Refs.~\onlinecite{Brambilla-1998,Astfalk-and-Jenko-2017}. The only requirement 
for a distribution function $f_\rd{s}$ is gyrotropy, i.e., 
$\frac{\partial f_\rd{s}}{\partial \phi} = 0$ with $\phi$ being the gyroangle, 
it is otherwise arbitrary. For the cases considered below we choose the 
isotropic RKD (Eq.~\ref{eq:rkd}) for the stable modes in 
Section~\ref{subsec:stable_modes}, and the anisotropic RBKD 
(Eq.~\ref{eq:rbkd}) for the instabilities in Sections~\ref{subsec:emec_emic}, 
\ref{subsec:pfh_efh} and \ref{subsec:apara_neq_aperp}.

The code \emph{LEOPARD} is a grid-based kinetic
dispersion relation solver, which is written in 
Fortran-90 and is based on its predecessor 
\emph{DSHARK}.\cite{Astfalk-et-al-2015} The code 
can be applied to distributions of arbitrary shapes 
and is provided to public 
access\footnote{see {https://github.com/pastfalk/LEOPARD} 
for source code of LEOPARD and a manual
(last accessed \today)} to be used and modified.
\emph{LEOPARD} solves the general linear dispersion relation 
(Eq.~\ref{eq:disp_rel}) by computing the dielectric tensor 
(Eq.~\ref{eq:dielectric_tensor}) with its components in order 
to find complex solutions $\omega(k)$, i.e., the frequencies 
(i.e, real part $\Re(\omega) \equiv \omega_\rd{\,r}$) as well 
as the growth and damping rates (i.e., imaginary part 
$\Im(\omega) \equiv \gamma$) of electromagnetic waves in 
homogeneous plasmas. The code allows for arbitrary wave propagation, 
i.e., parallel and oblique direction of wave propagation with respect 
to the background magnetic field, and for an arbitrary number of 
particle species. For the application of (bi-)Maxwellian velocity 
distributions a separately implemented routine can be used, which 
reduces the computation time significantly. However, other 
distributions have to be provided to the code as a 2D data grid in 
velocity space ($v_\parallel,v_\perp$). 
In the code the distribution function is interpolated on the velocity 
grid with cubic splines, which approximate the function piecewise in a 
continuous and smooth way. Furthermore, the user specifies the wavenumber 
interval (normalized to the skin depth), its resolution and needs to 
provide the code with initial guesses for the real and imaginary part 
of $\omega$. While the code allows for an arbitrary number of particle 
species, the declaration of the first particle species is important, 
as all normalized quantities and values are given in units of that species.

\begin{figure*}[t!]
  \centering
  \includegraphics[width=.35\textwidth]{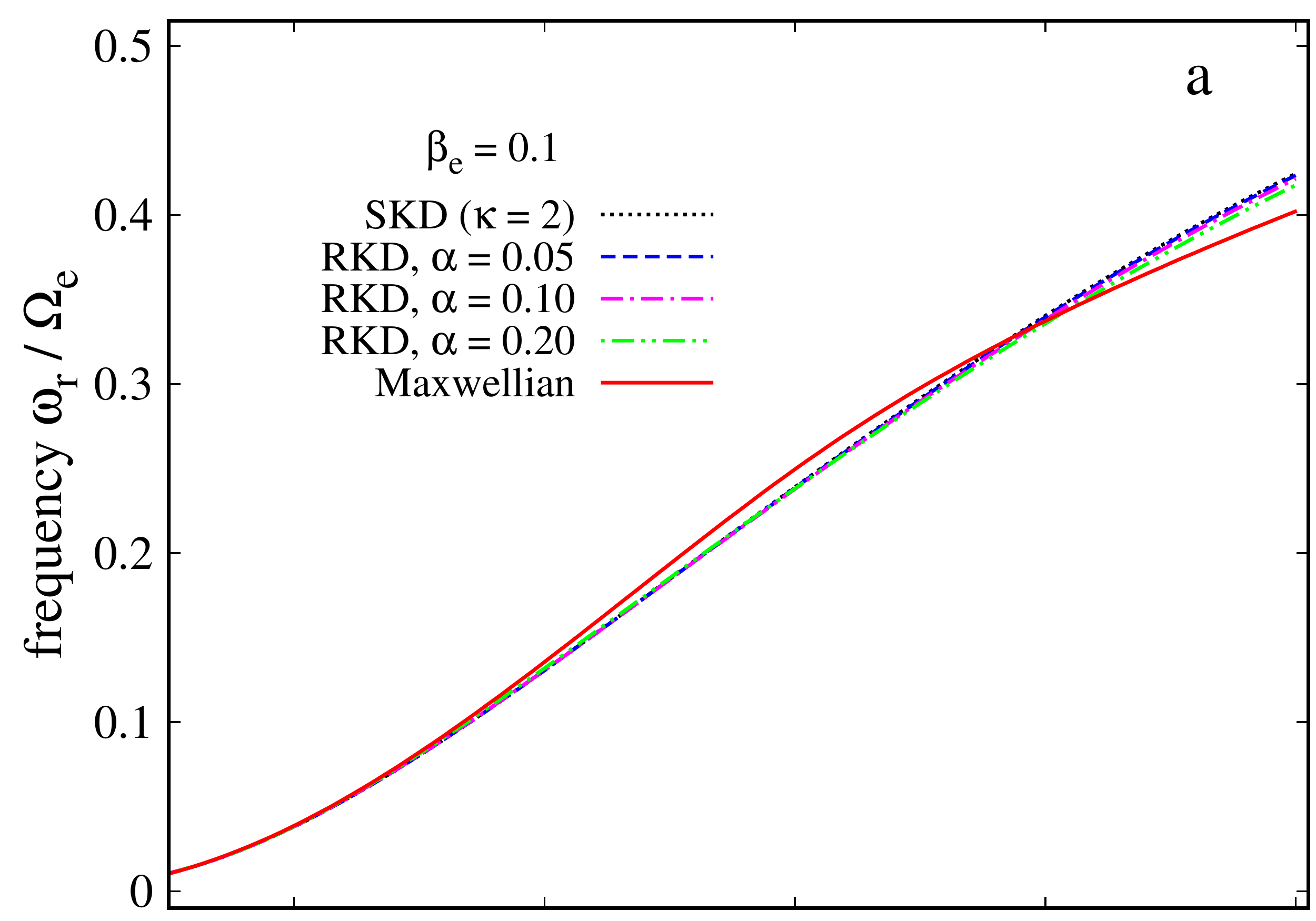}
  \hspace{0.1cm}
  \includegraphics[width=.35\textwidth]{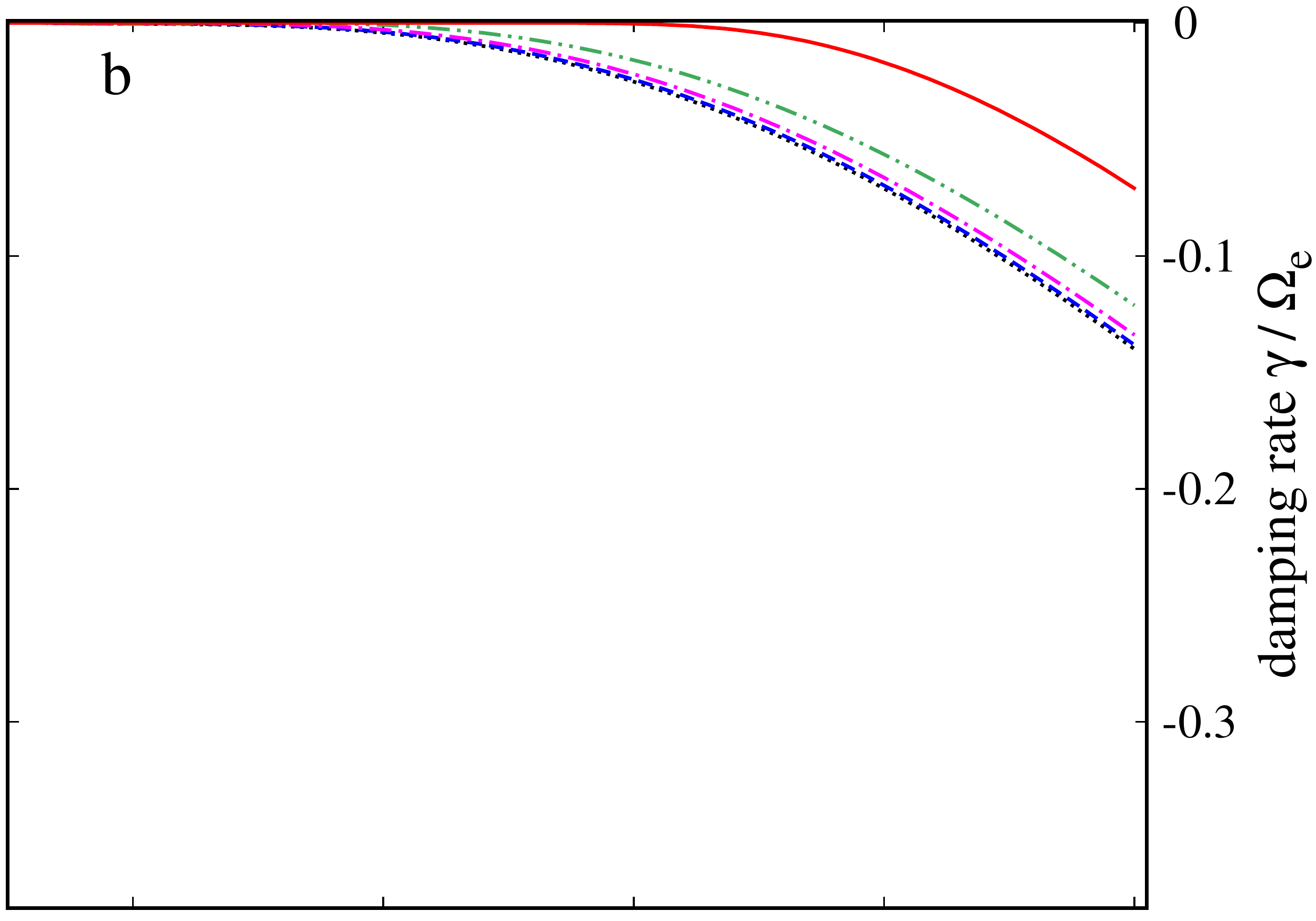}

  \vspace{0.1cm}

  \includegraphics[width=.35\textwidth]{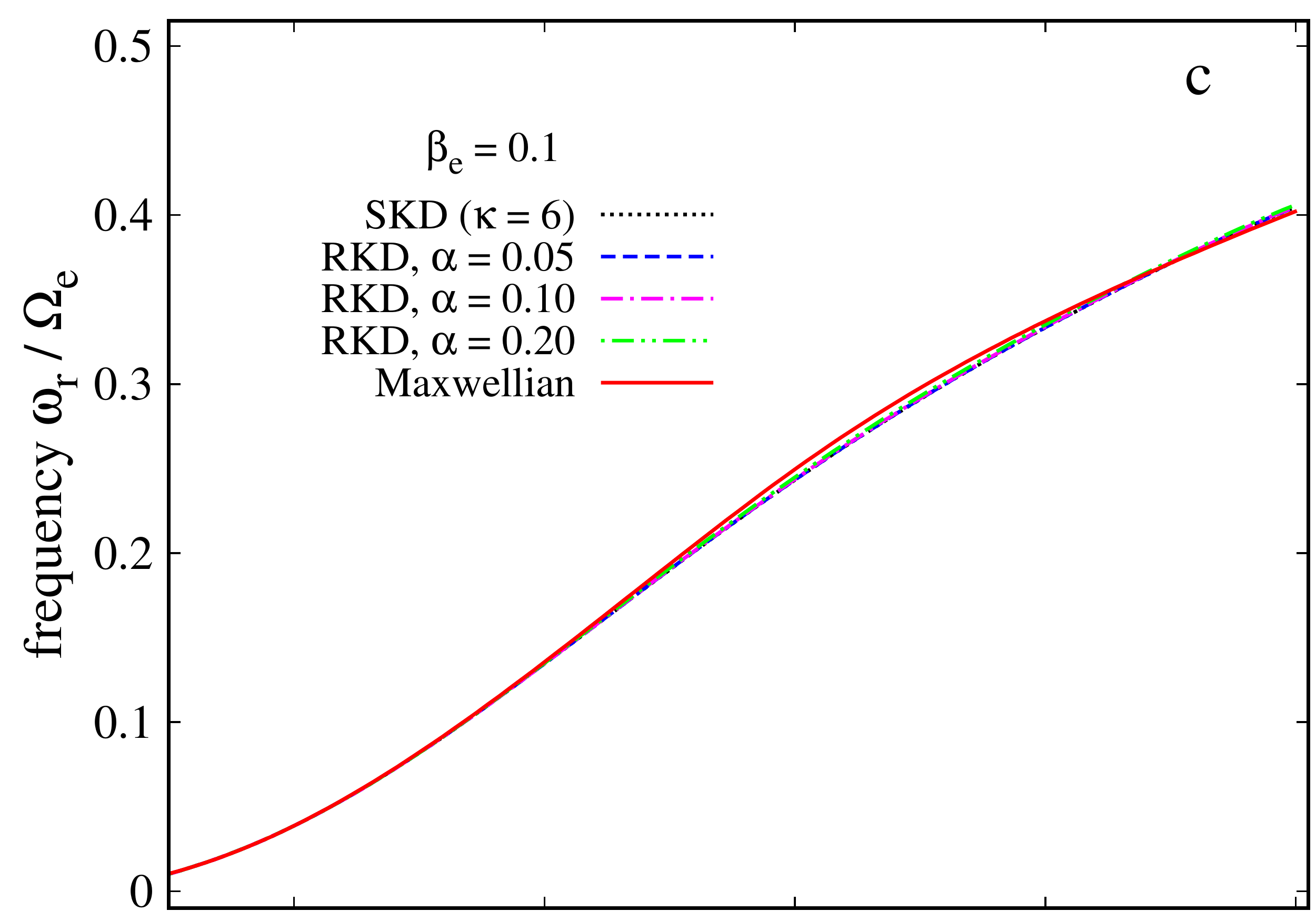}
  \hspace{0.1cm}
  \includegraphics[width=.35\textwidth]{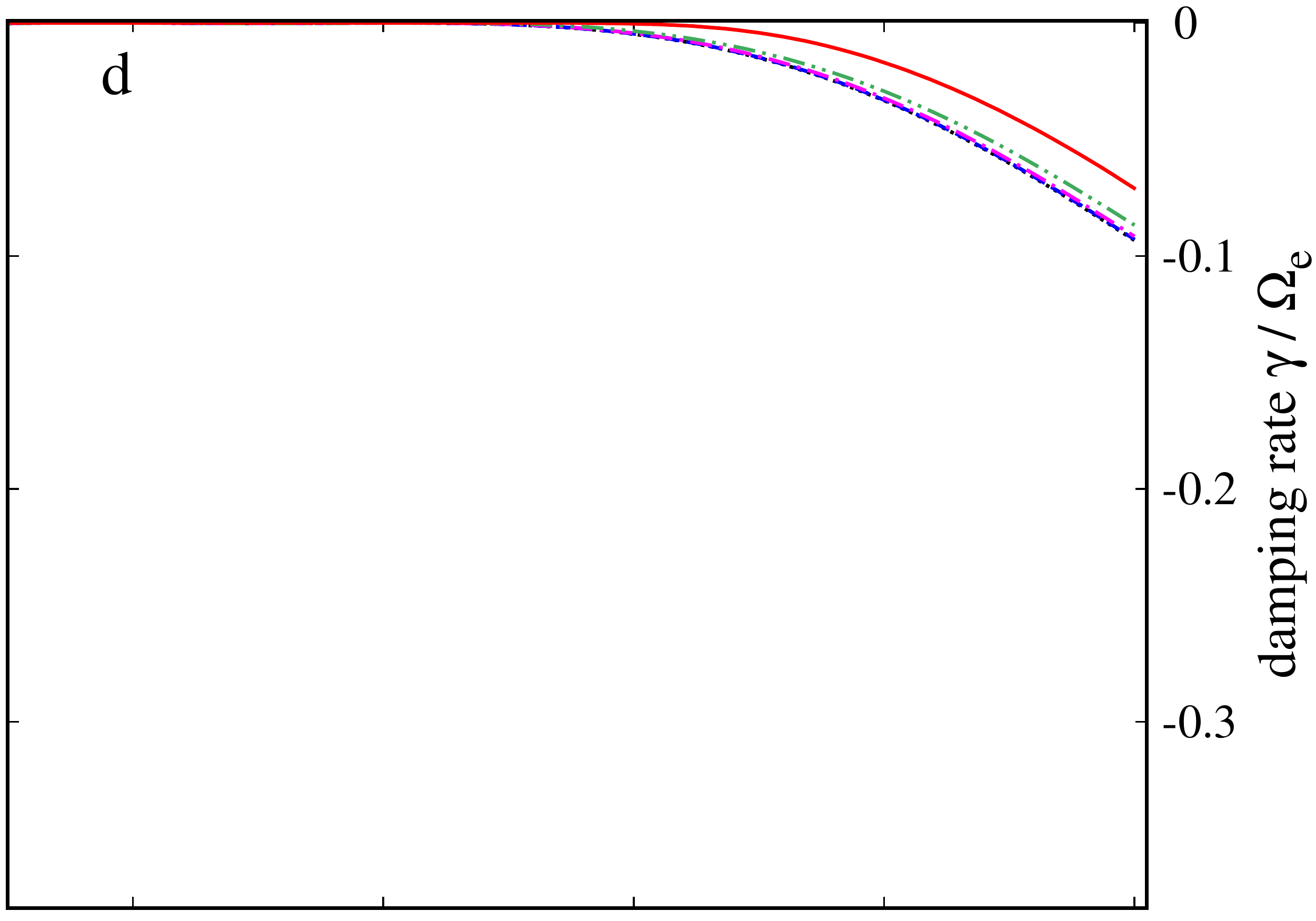}

  \vspace{0.1cm}

  \includegraphics[width=.35\textwidth]{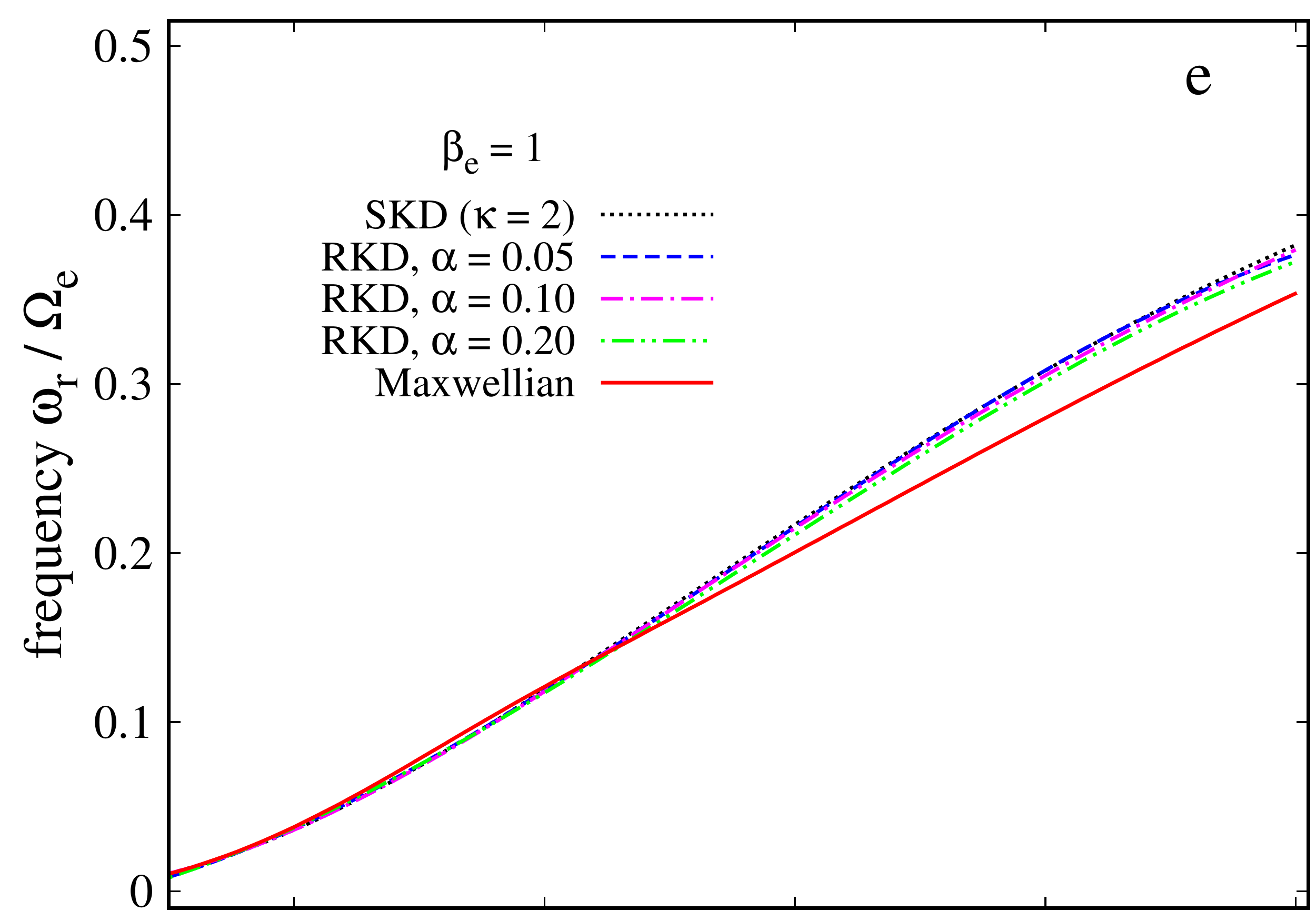}
  \hspace{0.1cm}
  \includegraphics[width=.35\textwidth]{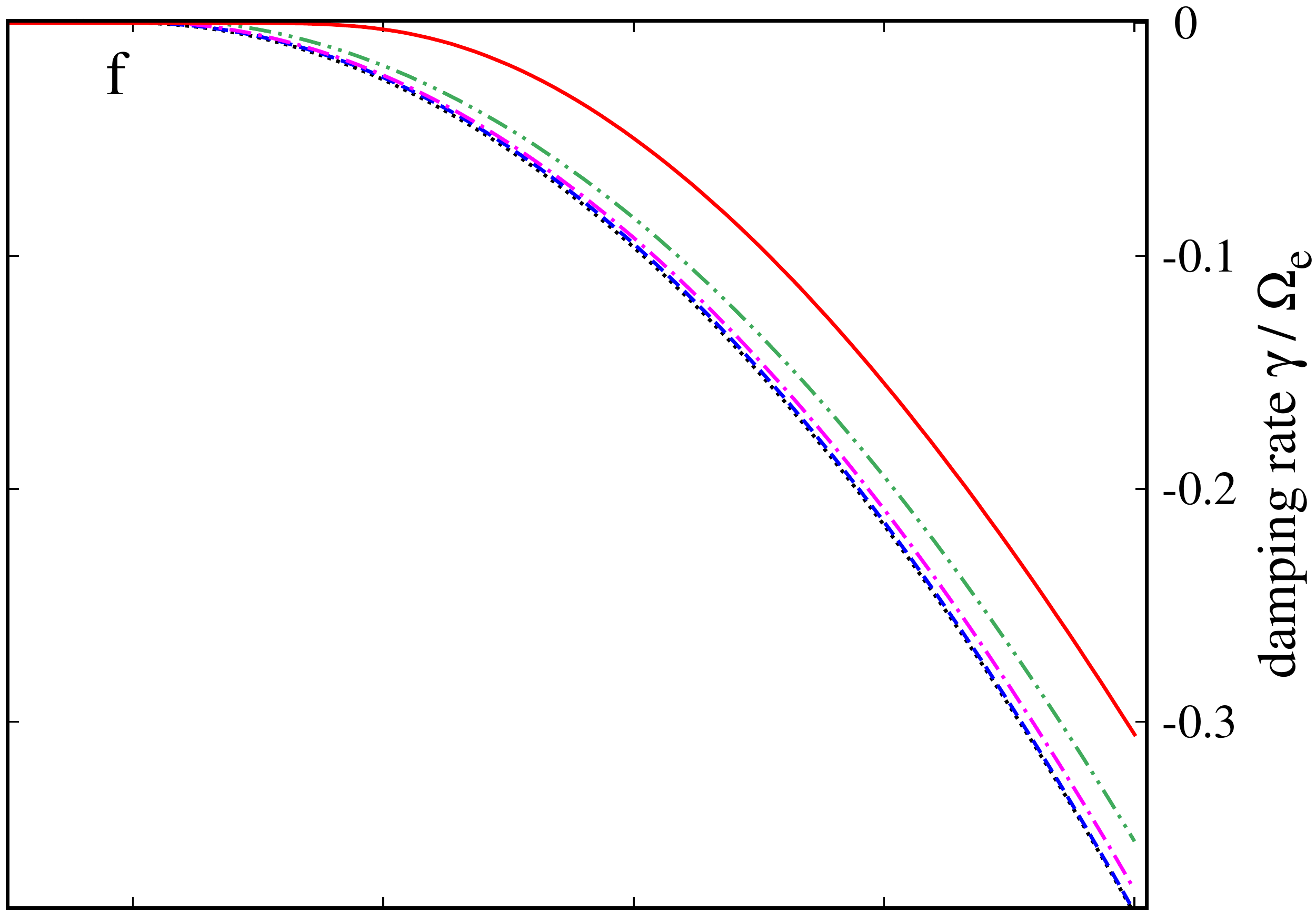}

  \vspace{0.1cm}

  \includegraphics[width=.35\textwidth]{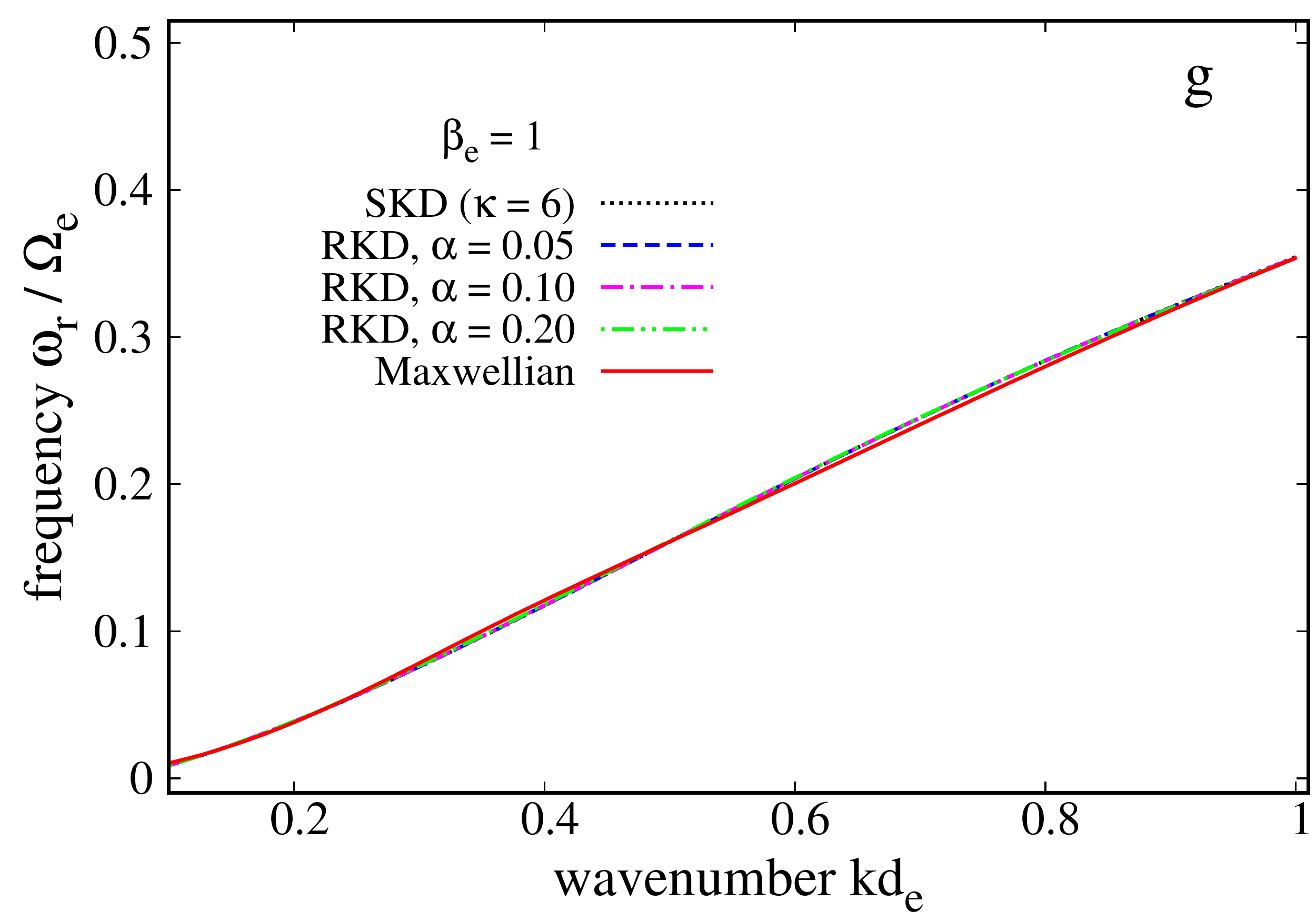}
  \hspace{0.1cm}
  \includegraphics[width=.35\textwidth]{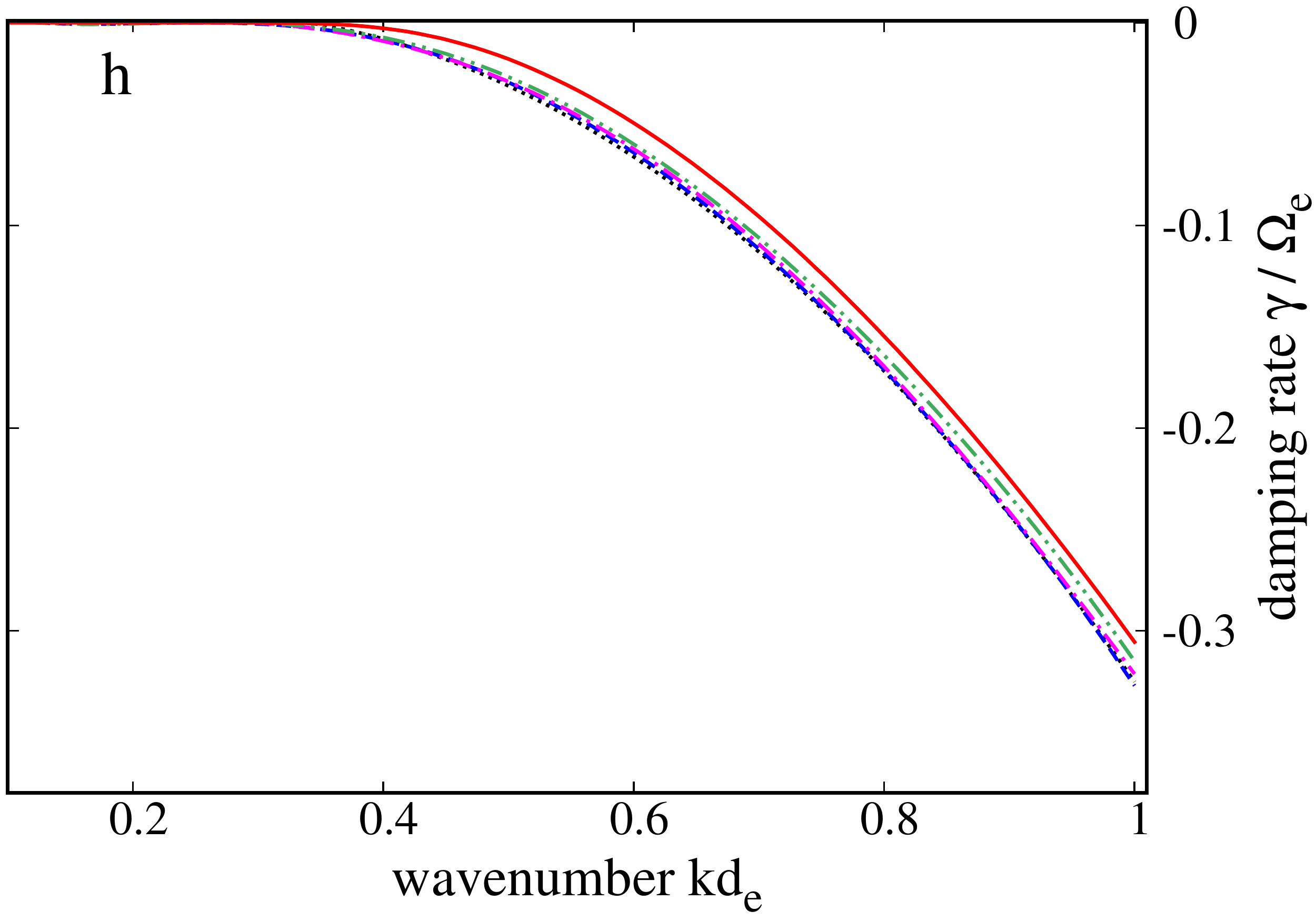}

  \caption{Wavenumber dispersion for the EMEC frequency (left column) and 
           damping rate (right column). Parameters are given in the 
           legends.}\label{fig:emec_rkd}
\end{figure*}

\section{Results}\label{sec:results}            

The solutions of the general dispersion relation (Eq.~\ref{eq:disp_rel}) 
are derived numerically with the numerical solver \emph{LEOPARD}, 
yielding wavenumber dispersion curves for the frequencies as well as the 
growth and damping rates of the considered parallel electromagnetic modes. 
We examine the effects of the regularization parameter $\alpha$ on the 
results and compare them to the ones obtained for the corresponding 
Maxwellian and standard Kappa-distributions, which serve as test cases in 
order to validate the functionality of the numerical code. While damped 
modes in Maxwellian-distributed plasmas are discussed widely, see the textbook 
of Gary,\cite{Gary-1993} studies in general do not cover damped modes in 
Kappa-distributed plasmas. In the following we consider a two-component plasma 
consisting of protons ($\rd{s} = \rd{p}$) and electrons ($\rd{s} = \rd{e}$). 
In Section~\ref{subsec:stable_modes} the isotropic RKD is applied and the 
dispersion curves and damping rates of stable electromagnetic 
electron-cyclotron (EMEC) and ion-cyclotron (EMIC) modes are discussed. 
In Sections.~\ref{subsec:emec_emic} and \ref{subsec:pfh_efh} we analyze the 
instabilities of anisotropic electrons or protons, when described with RBKDs 
with $\alpha_\parallel = \alpha_\perp$. While Section~\ref{subsec:emec_emic} 
presents the EMEC and EMIC instabilities, which are driven by a temperature 
anisotropy $A \equiv T_\perp/T_\parallel > 1$, Section~\ref{subsec:pfh_efh} shows 
the results for the electron firehose (EFH) and proton firehose (PFH) 
instabilities, which occur for opposite anisotropies $A < 1$. 
In Section \ref{subsec:apara_neq_aperp} we revisit the EMEC, EMIC, EFH 
and PFH instabilities, and describe the anisotropic particle species with 
RBKDs with $\alpha_\parallel \neq \alpha_\perp$.

\subsection{Damped Electromagnetic Cyclotron Modes}\label{subsec:stable_modes} 

The electromagnetic electron-cyclotron (EMEC) or whistler 
waves are often identified, directly or indirectly in 
observations of space plasmas, see Refs.~\onlinecite{Stverak-et-al-2008} 
or \onlinecite{Lazar-et-al-2019} and references therein. EMEC waves are 
transverse electromagnetic modes with right-hand circular polarization 
and frequencies between the electron and ion gyrofrequencies, i.e., 
$\Omega_\rd{i} < \omega < \Omega_\rd{e}$, where $\Omega_\rd{e,i}$ are 
absolute values of electron and ion (proton) gyrofrequencies, 
respectively.\cite{Lazar-et-al-2010} For the first time these modes were 
observed as whistlers in Earth's atmosphere,\cite{Hellwell-et-al-1956} 
being enhanced through resonance with energetic electrons. The particles 
in a magnetized plasma are subject to gyromotion and cyclotron resonance 
with these modes, more exactly, with the transverse electric field of the 
parallel EM modes considered here. For particles that gyrate in phase with 
the electric field, the gyrofrequency $\Omega$ matches the Doppler 
(thermally) shifted wave frequency, and the cyclotron resonance condition 
reads\cite{Lutomirski-1970}
\begin{equation}\label{eq:cyclotron_resonance}
  v = \frac{\omega \pm \Omega}{k_\parallel}\,.
\end{equation}
The signs $\pm$ are usually assigned to right-hand and left-hand 
polarized waves, respectively.

\begin{figure*}[t!]
  \centering
  \includegraphics[width=.35\textwidth]{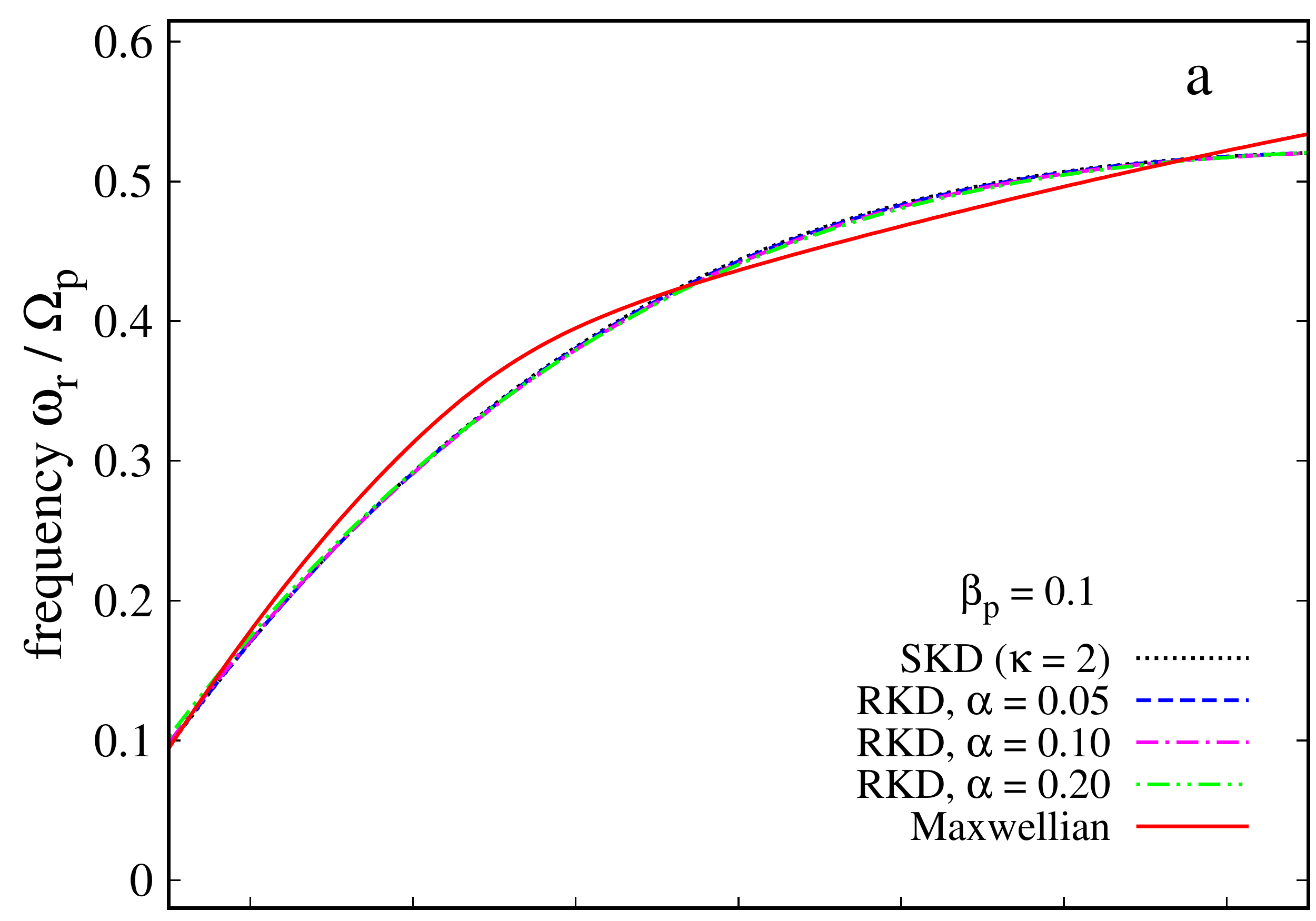}
  \hspace{0.1cm}
  \includegraphics[width=.35\textwidth]{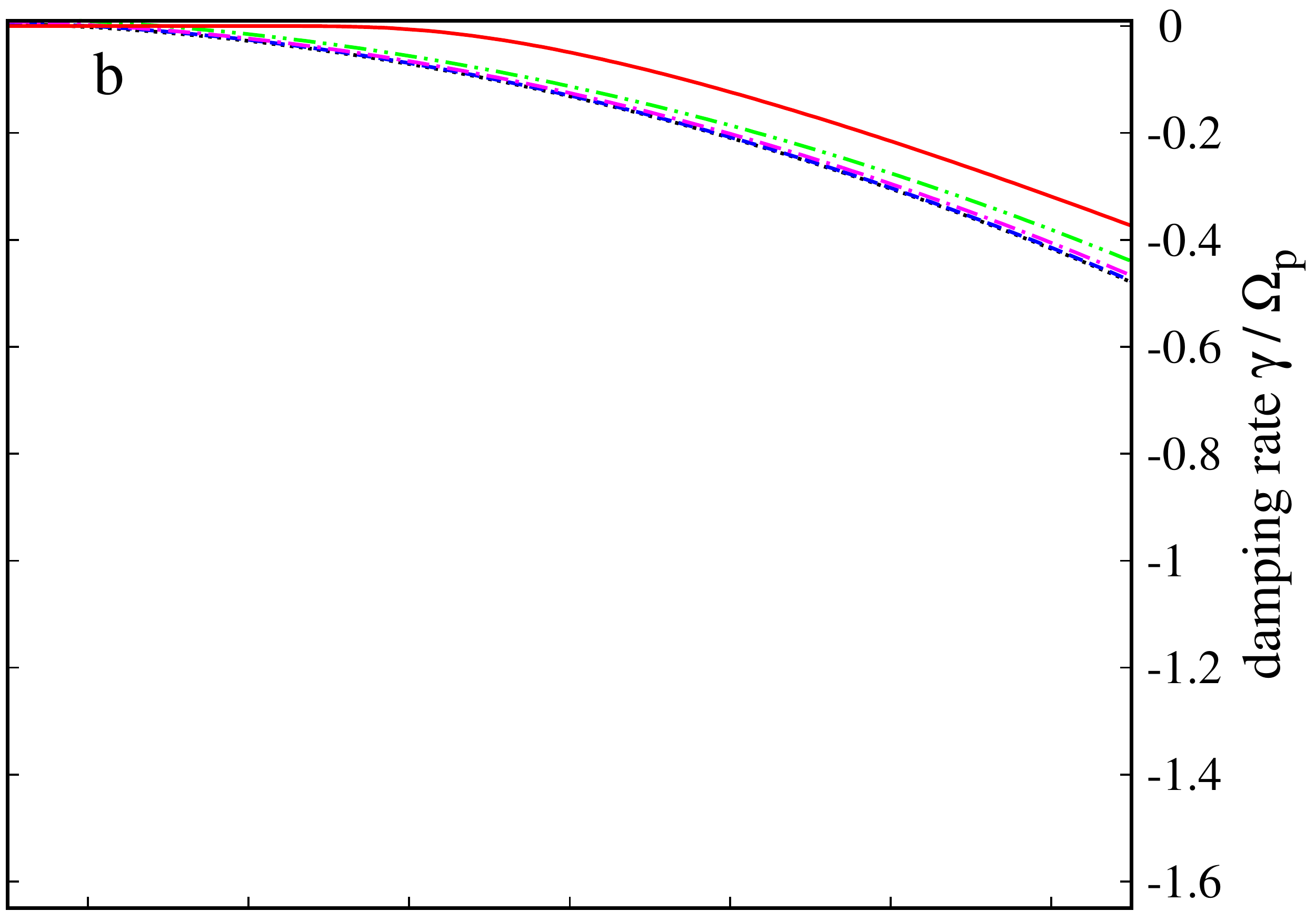}

  \vspace{0.1cm}

  \includegraphics[width=.35\textwidth]{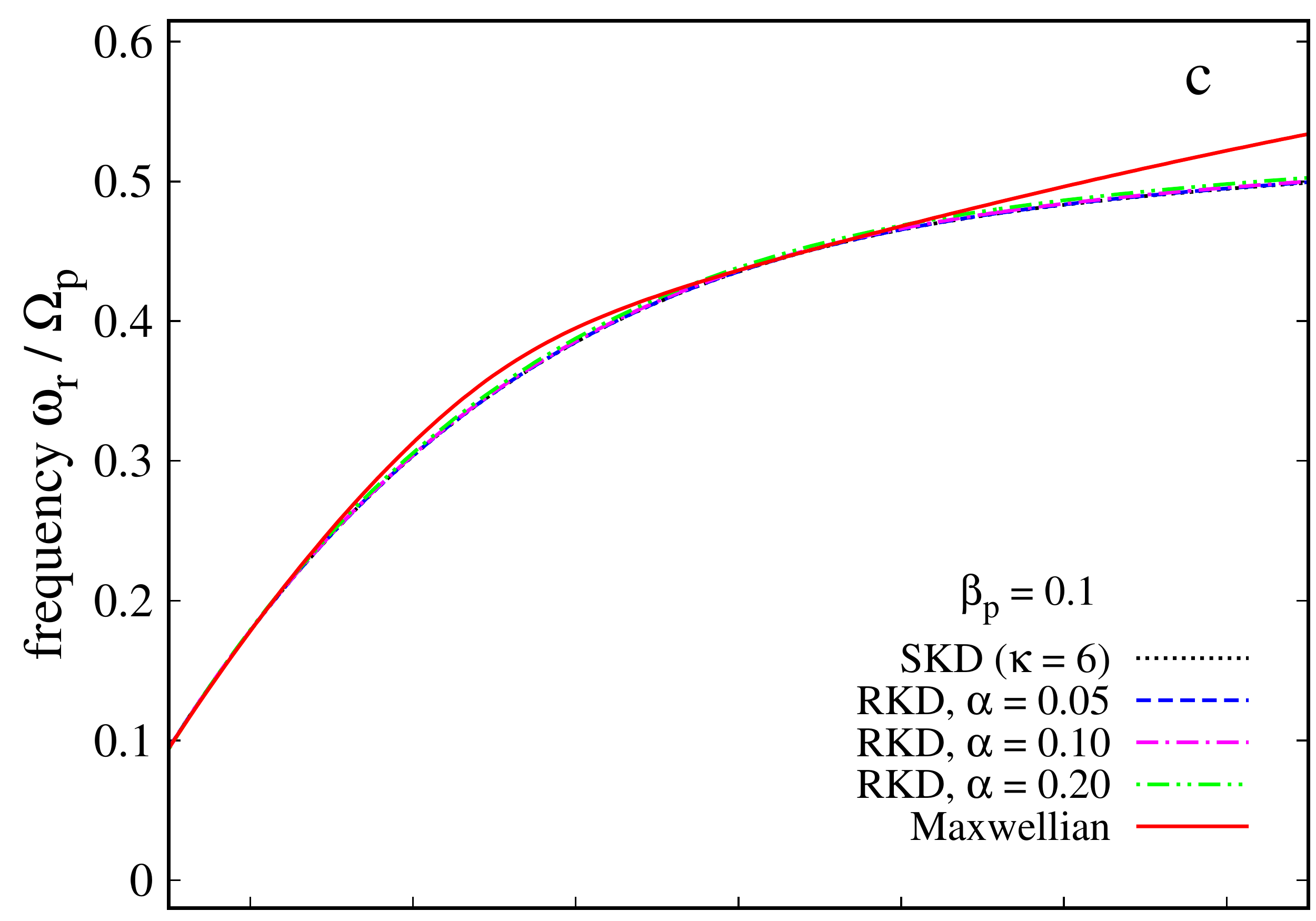}
  \hspace{0.1cm}
  \includegraphics[width=.35\textwidth]{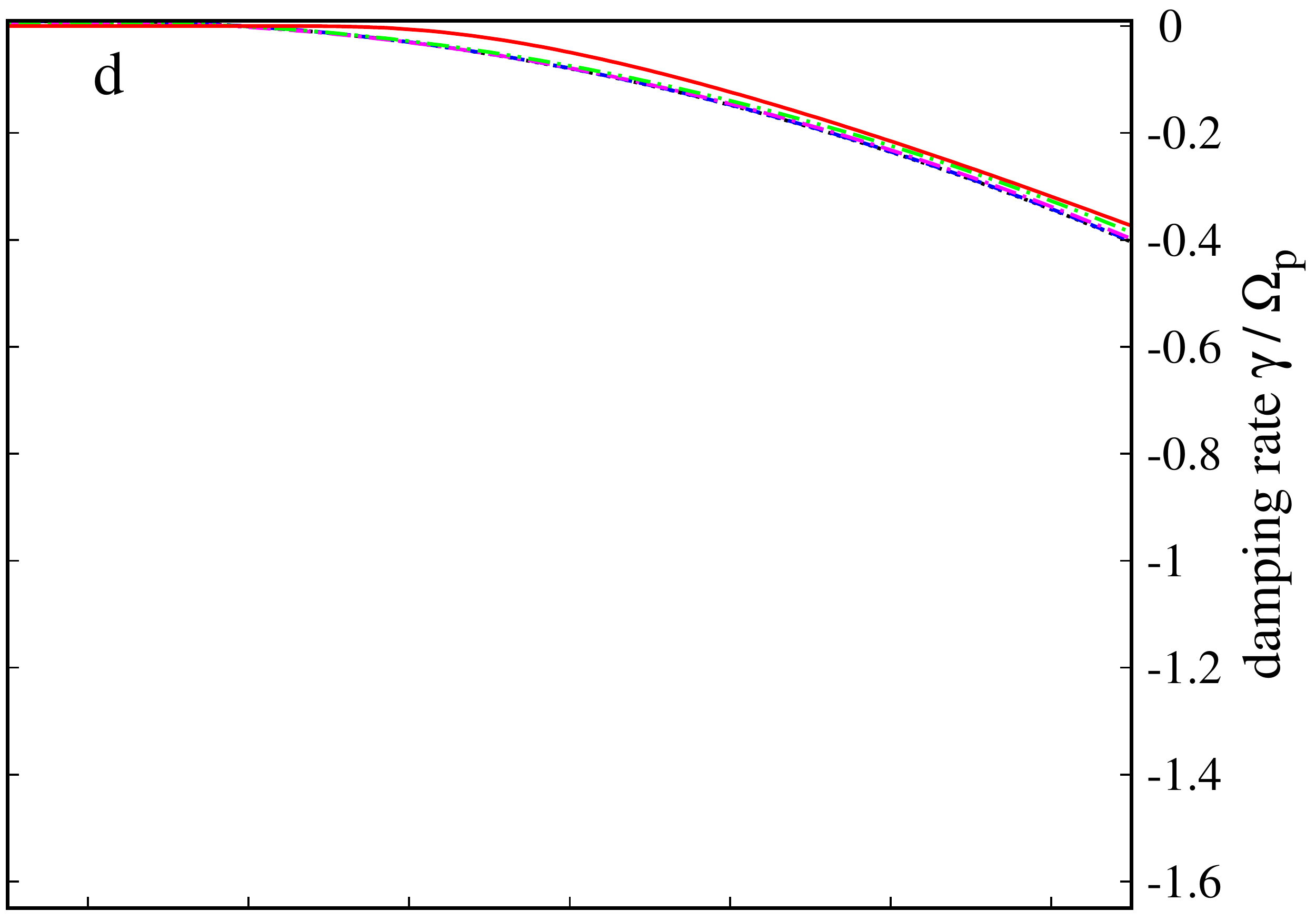}

  \vspace{0.1cm}

  \includegraphics[width=.35\textwidth]{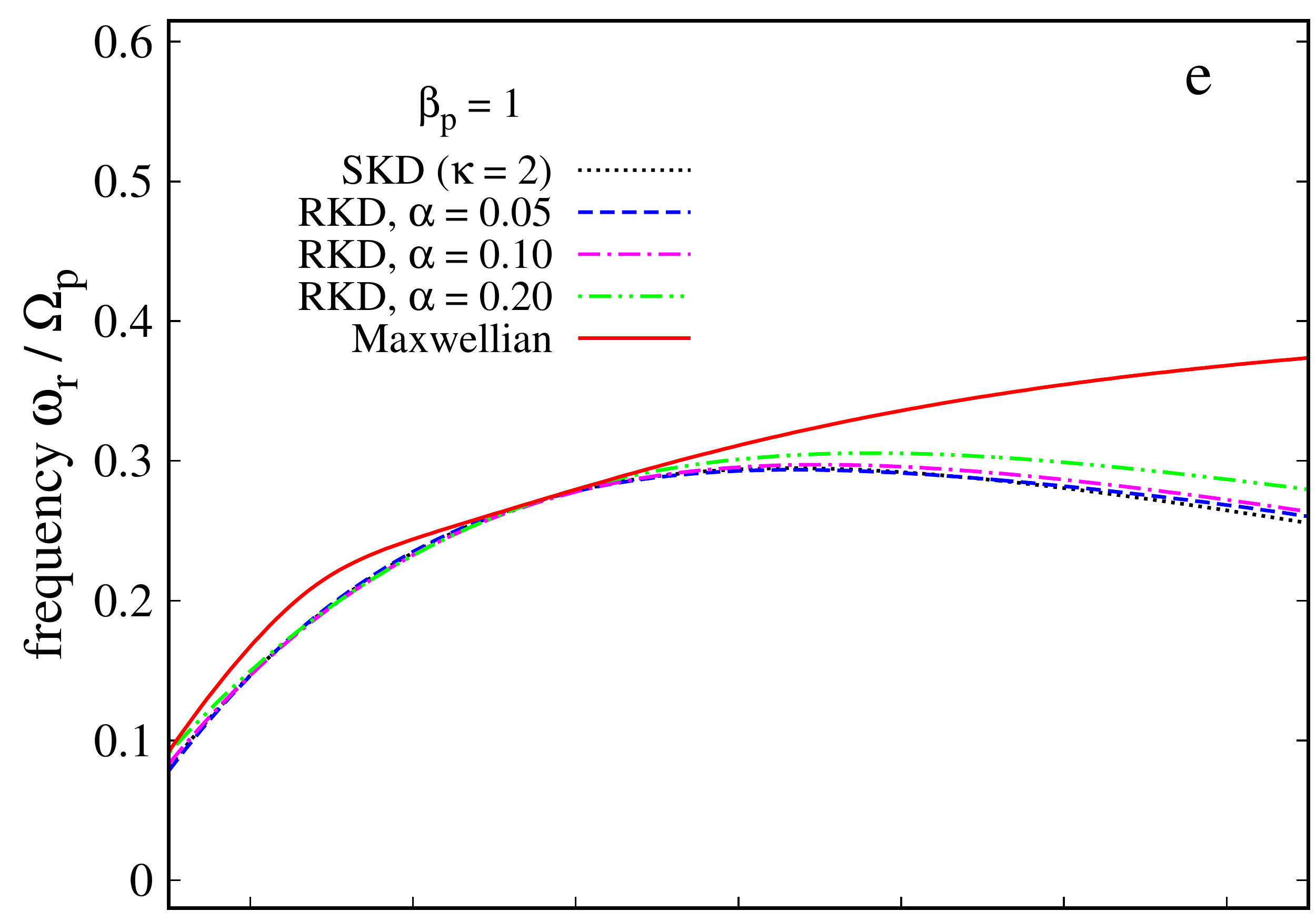}
  \hspace{0.1cm}
  \includegraphics[width=.35\textwidth]{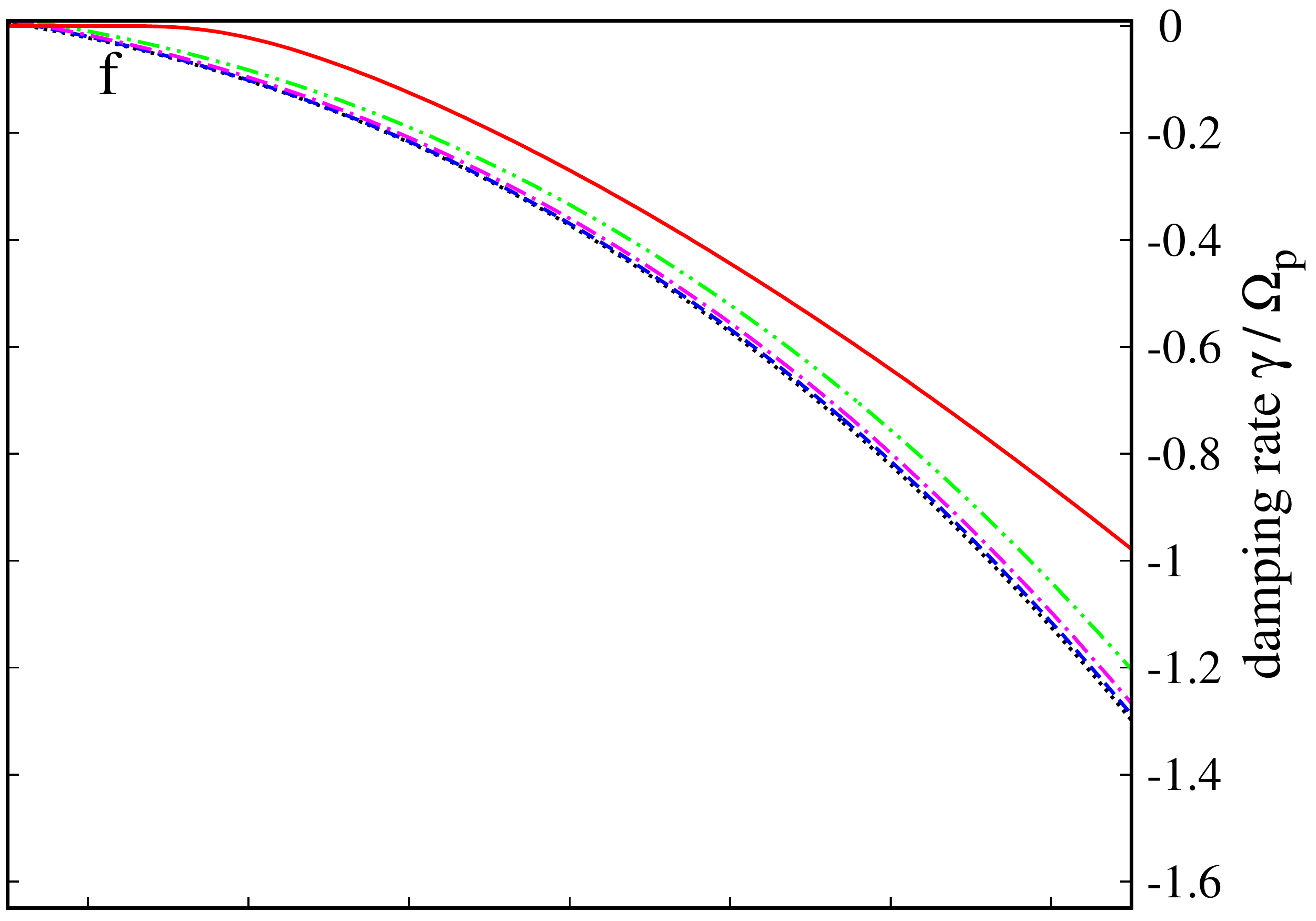}

  \vspace{0.1cm}

  \includegraphics[width=.35\textwidth]{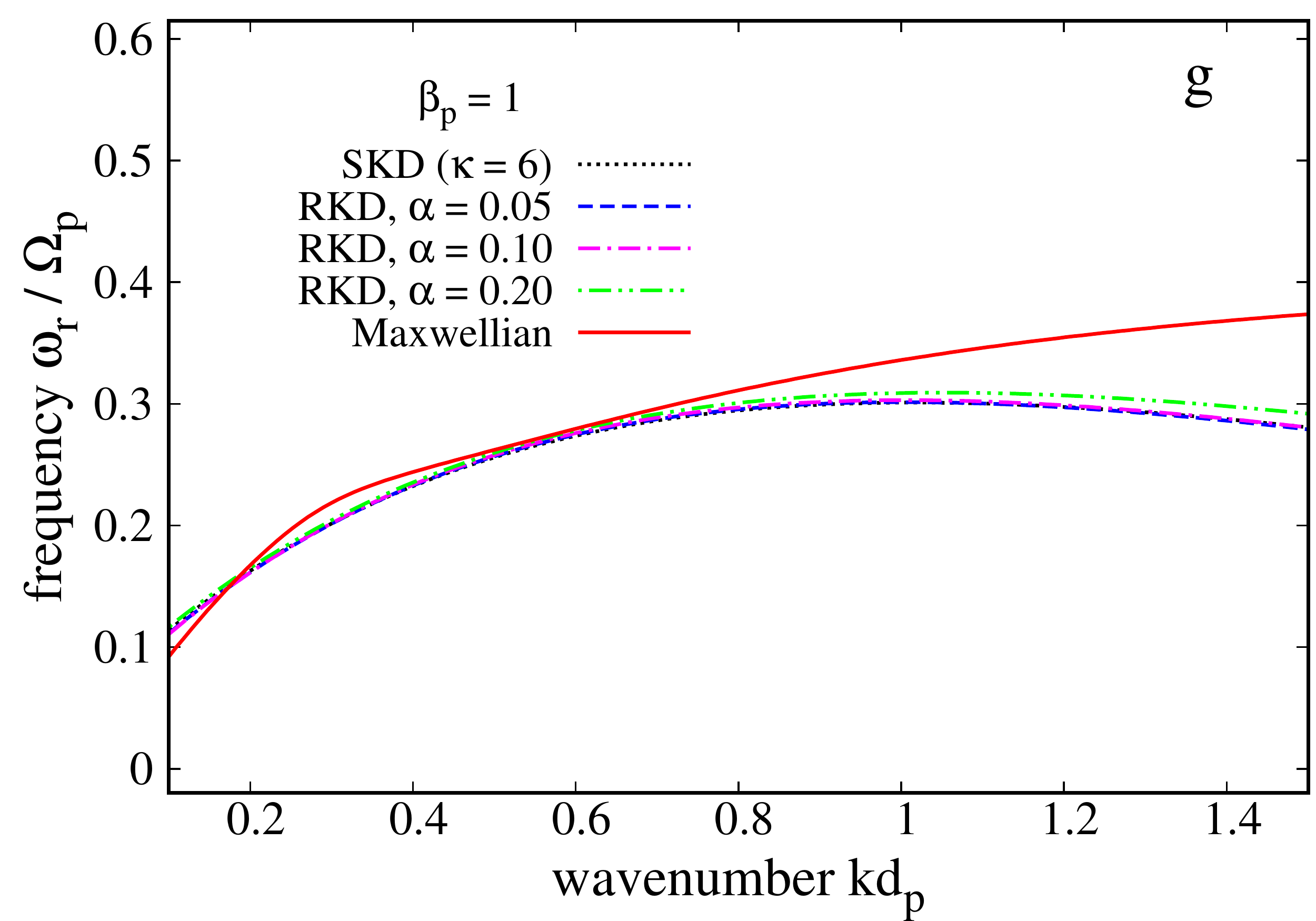}
  \hspace{0.1cm}
  \includegraphics[width=.35\textwidth]{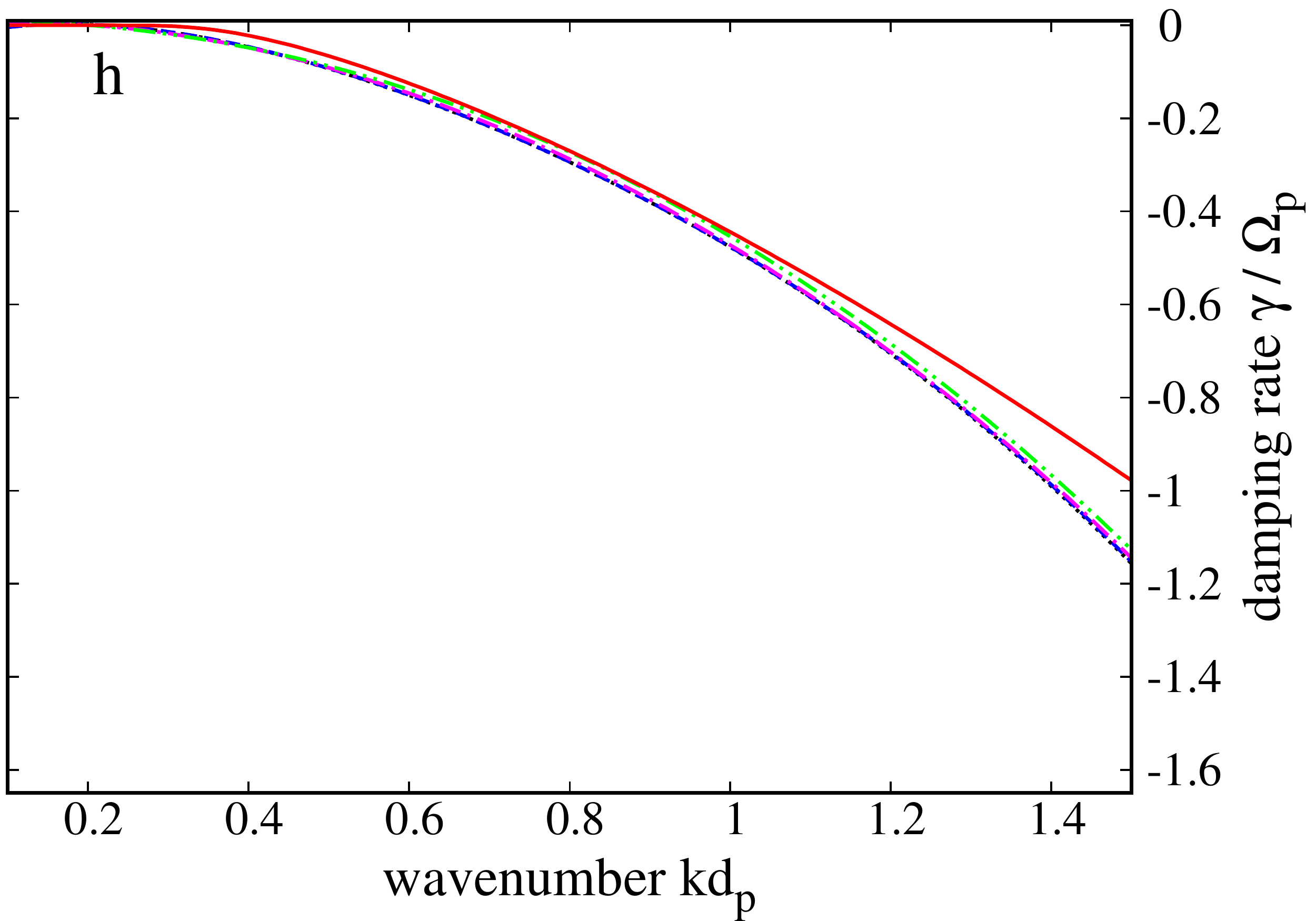}

  \caption{Wavenumber dispersion for the EMIC frequency (left column) and 
           damping rate (right column). Parameters are given in the legends.}\label{fig:emic_rkd}
\end{figure*}

\begin{figure*}[t!]
  \centering
  \includegraphics[width=.35\textwidth]{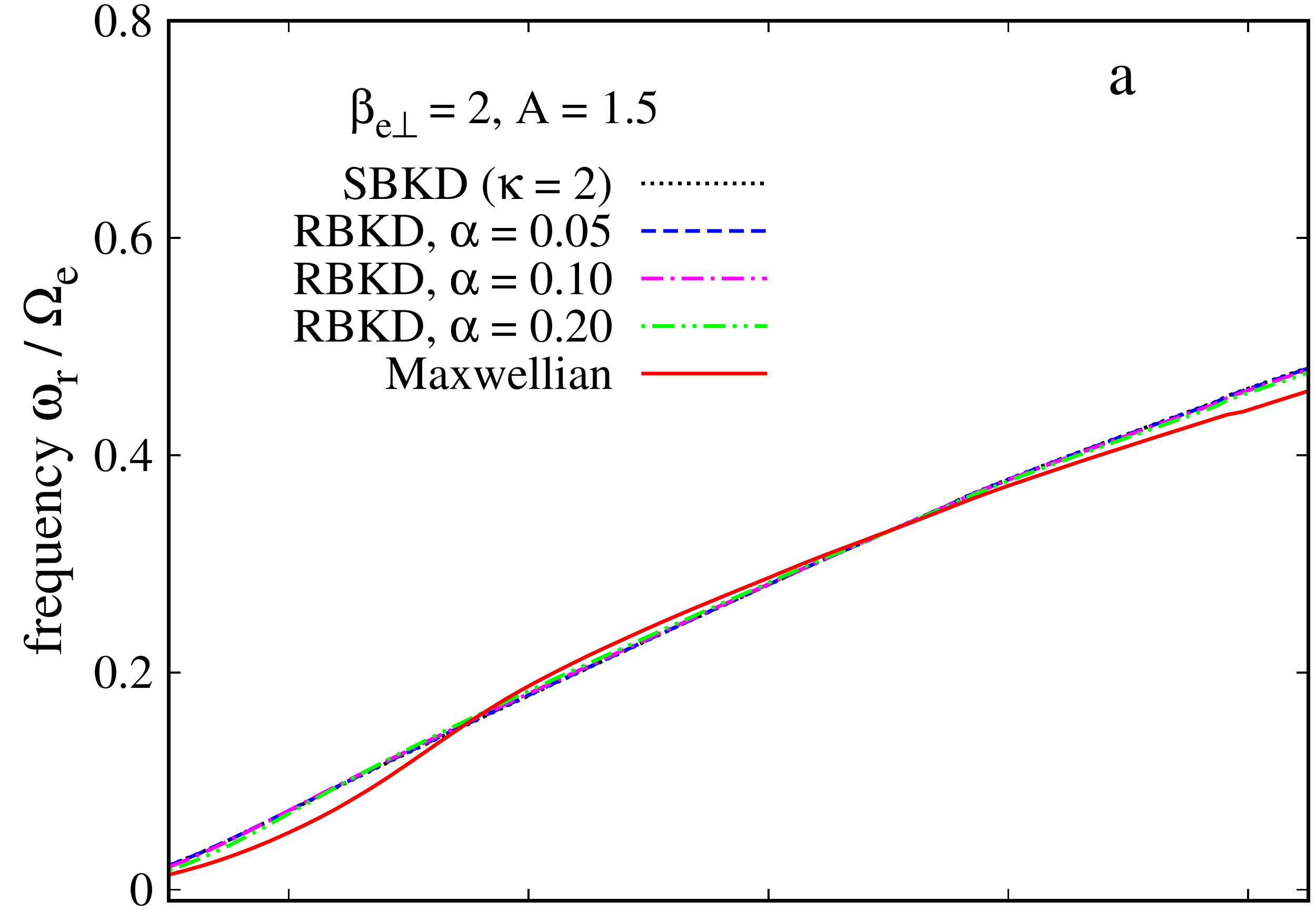}
  \hspace{0.1cm}
  \includegraphics[width=.35\textwidth]{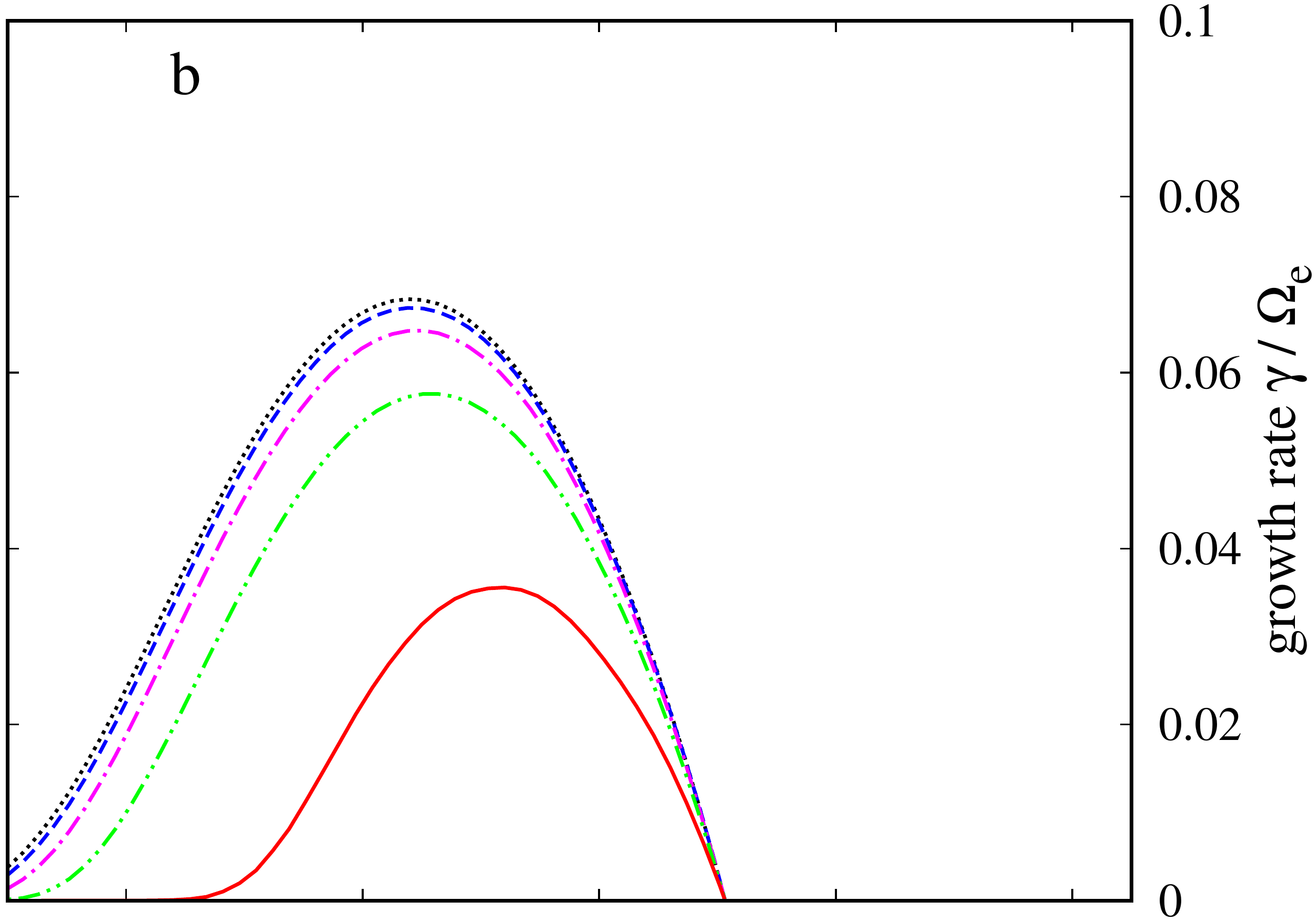}

  \vspace{0.1cm}

  \includegraphics[width=.35\textwidth]{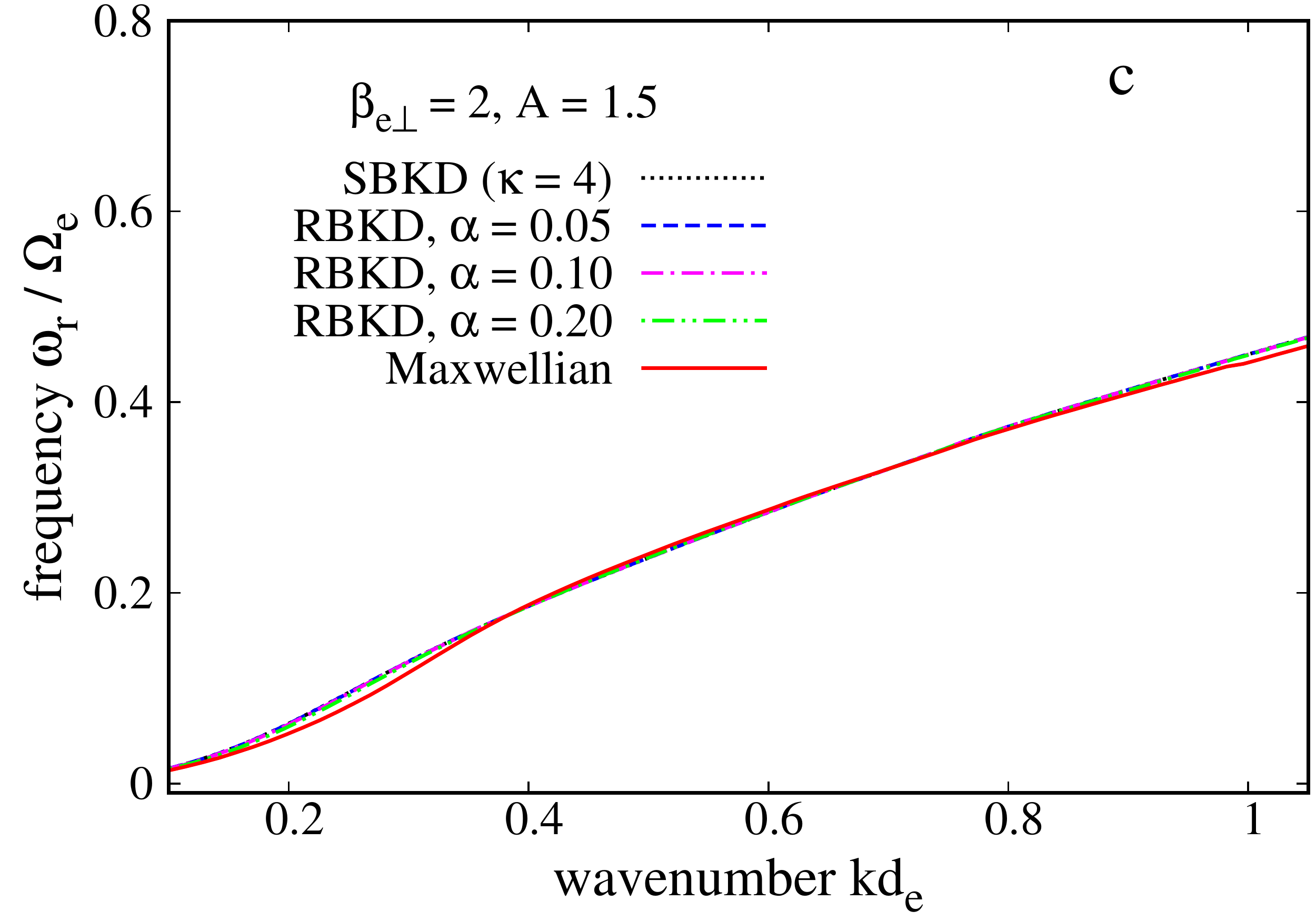}
  \hspace{0.1cm}
  \includegraphics[width=.35\textwidth]{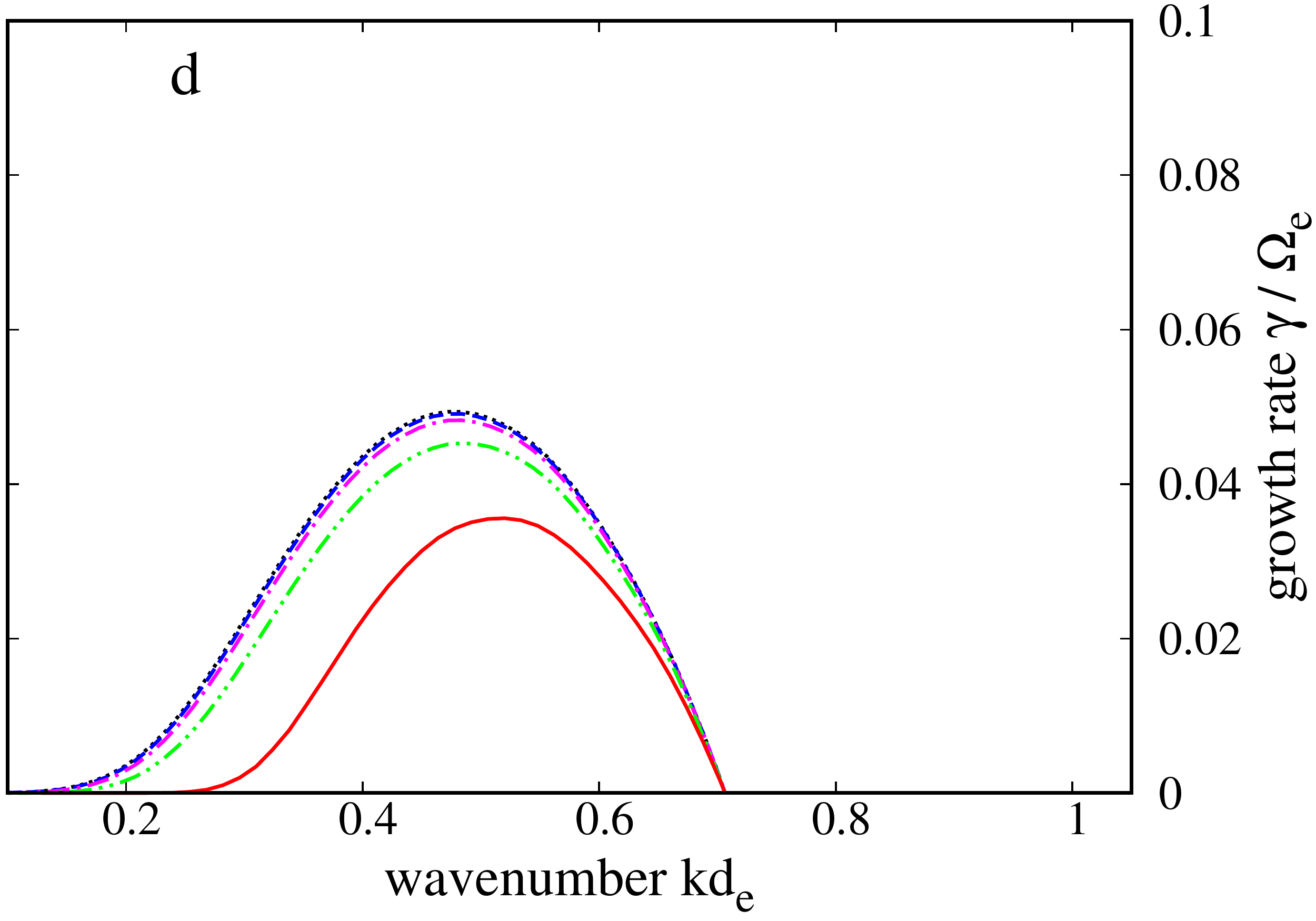}

  \vspace{0.1cm}

  \includegraphics[width=.35\textwidth]{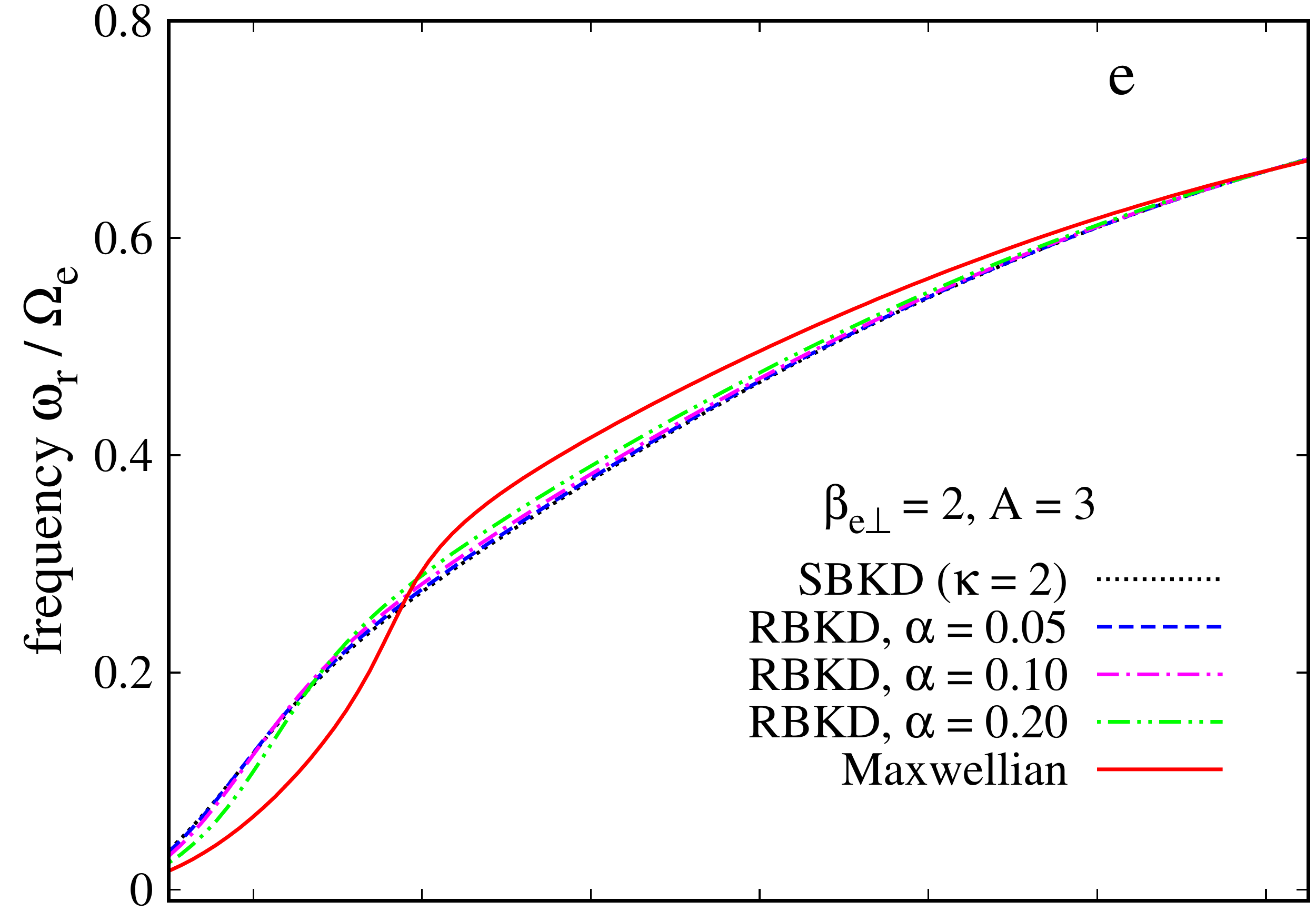}
  \hspace{0.1cm}
  \includegraphics[width=.35\textwidth]{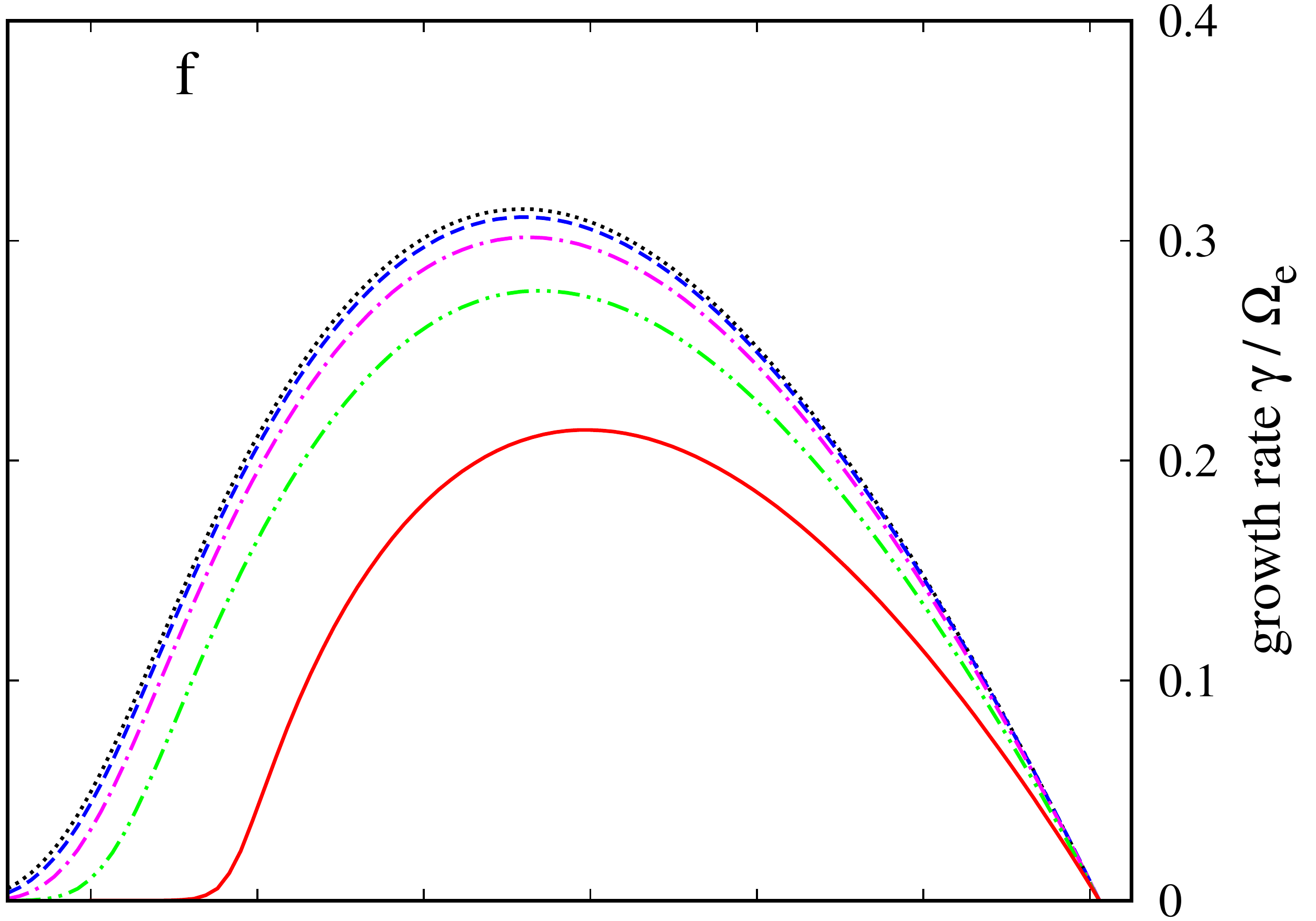}
  
    \vspace{0.1cm}

  \includegraphics[width=.35\textwidth]{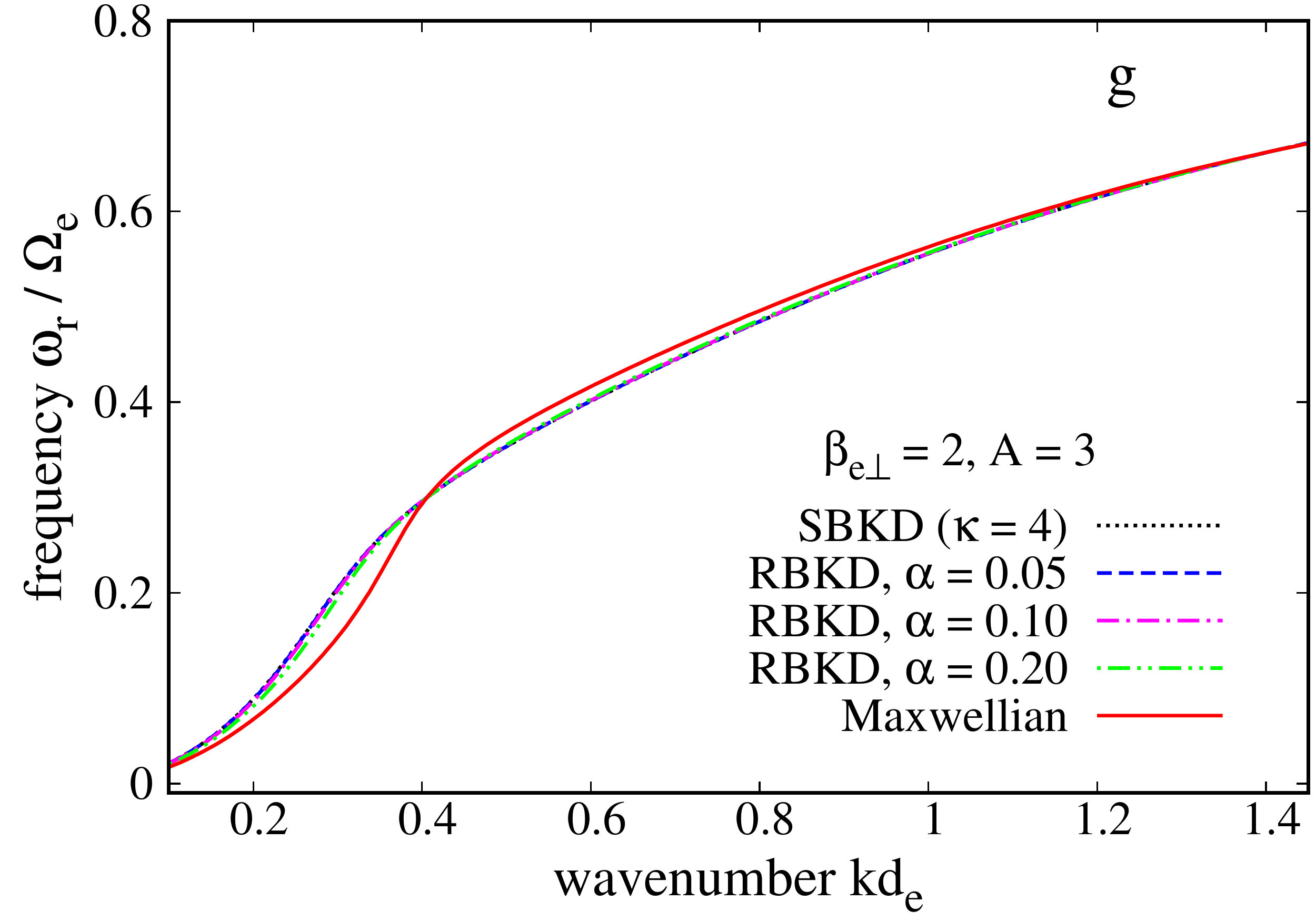}
  \hspace{0.1cm}
  \includegraphics[width=.35\textwidth]{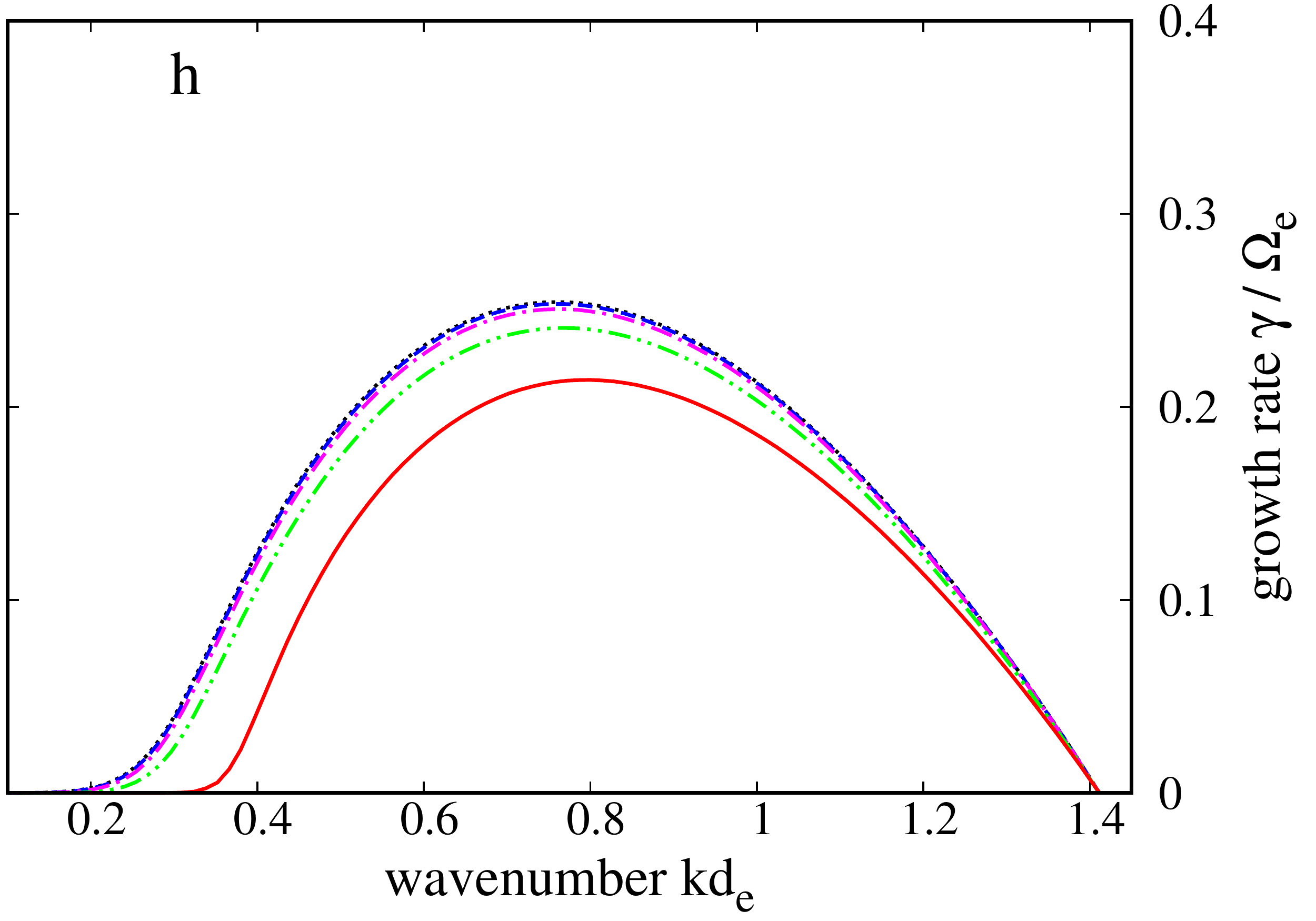}

  \caption{Dispersion curves for wave frequencies (left column) and growth rates 
           (right column) of the EMEC instability with 
           $\alpha_\parallel = \alpha_\perp$. Parameters are 
           explained in the legends.}\label{fig:emec_rbkd}
\end{figure*}

\begin{figure*}[t!]
  \centering
  \includegraphics[width=.35\textwidth]{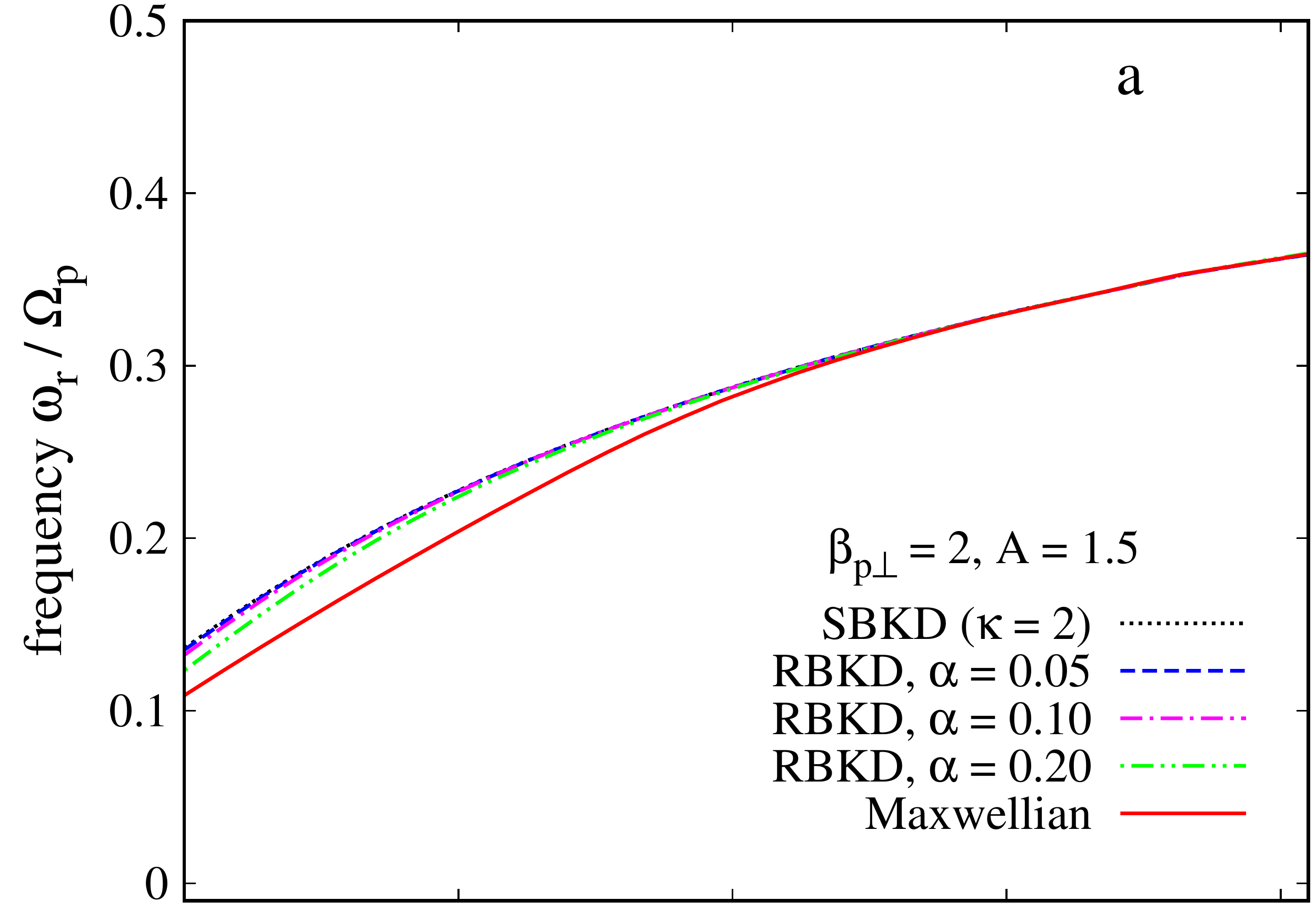}
  \hspace{0.1cm}
  \includegraphics[width=.35\textwidth]{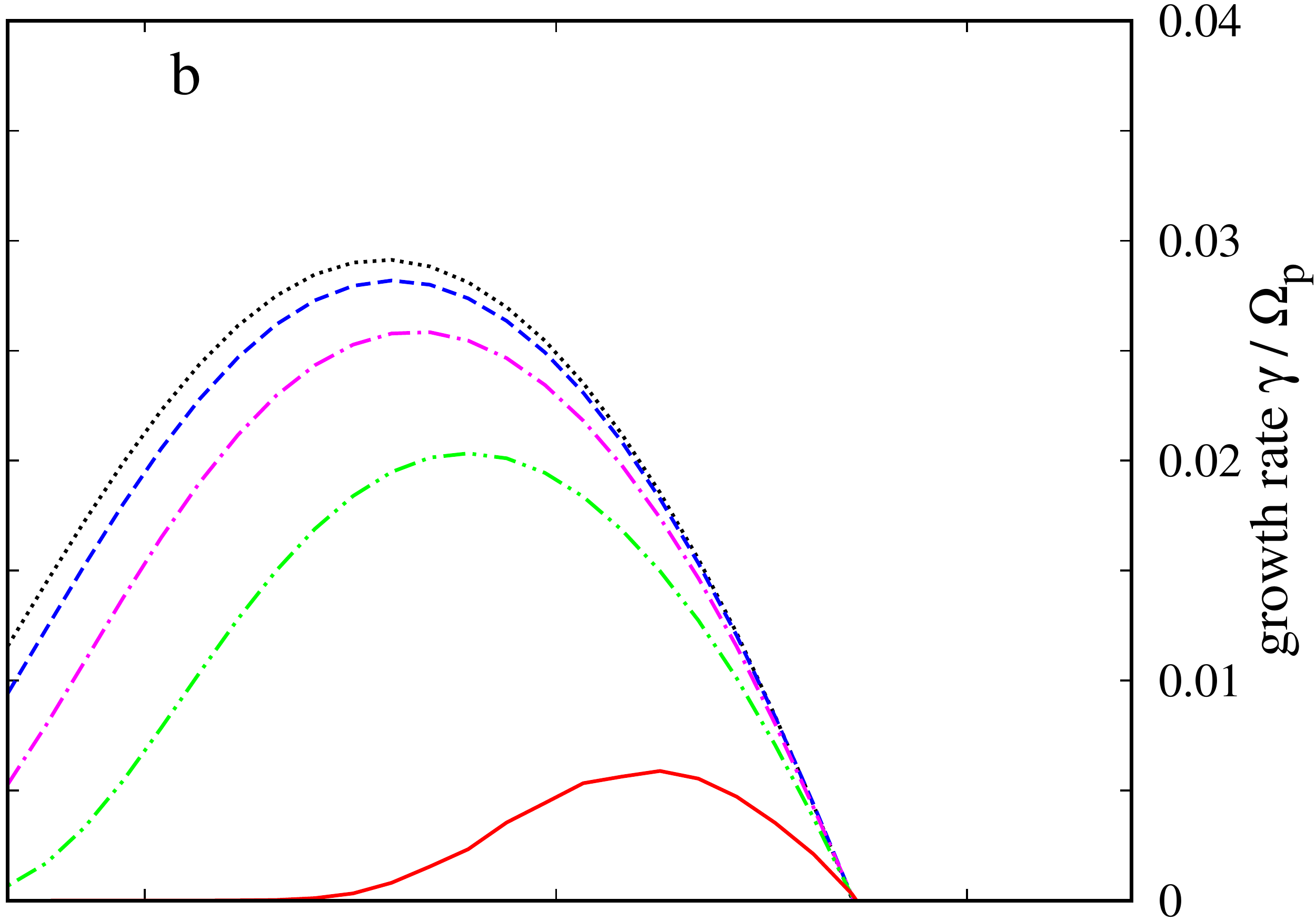}

  \vspace{0.1cm}

  \includegraphics[width=.35\textwidth]{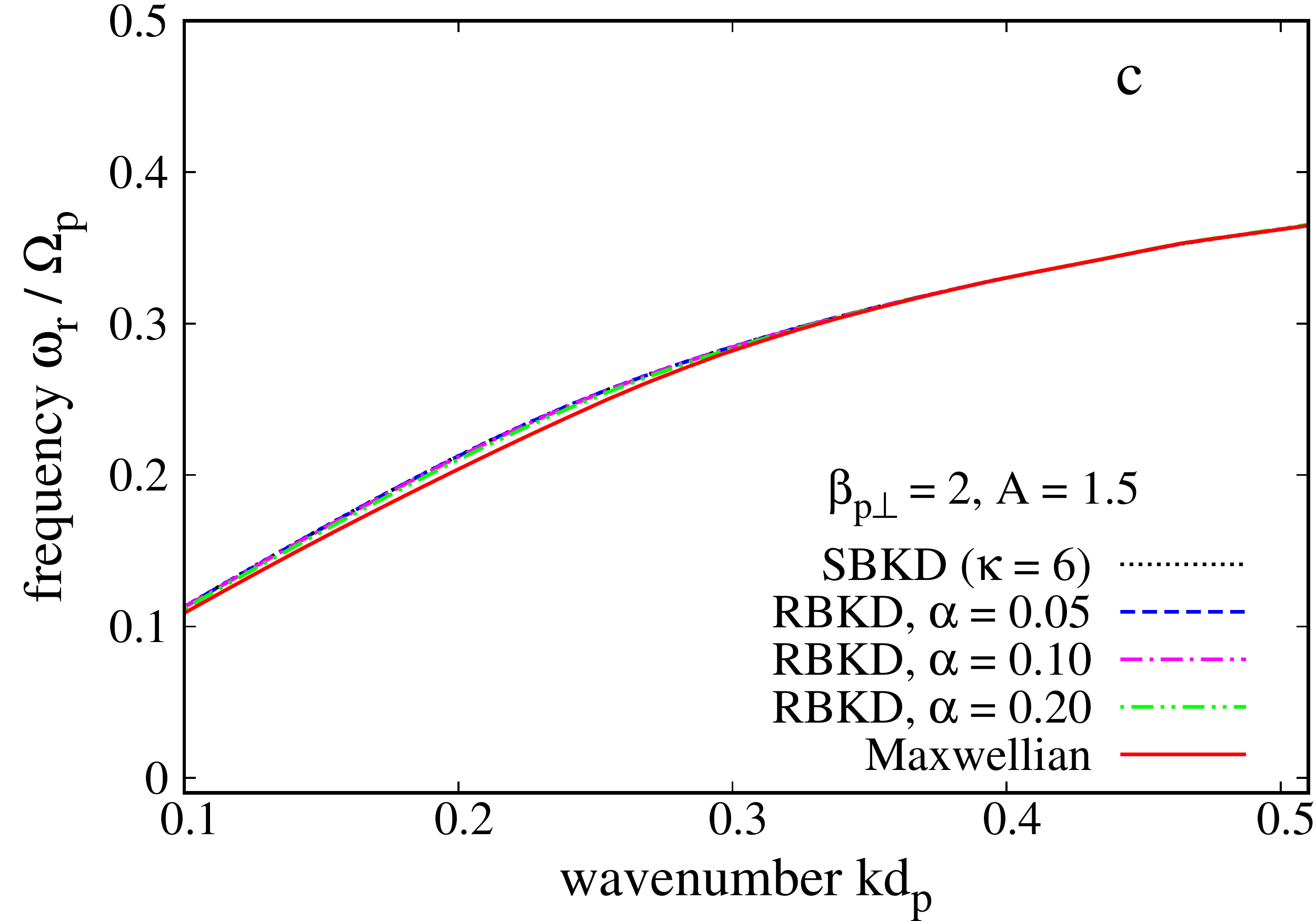}
  \hspace{0.1cm}
  \includegraphics[width=.35\textwidth]{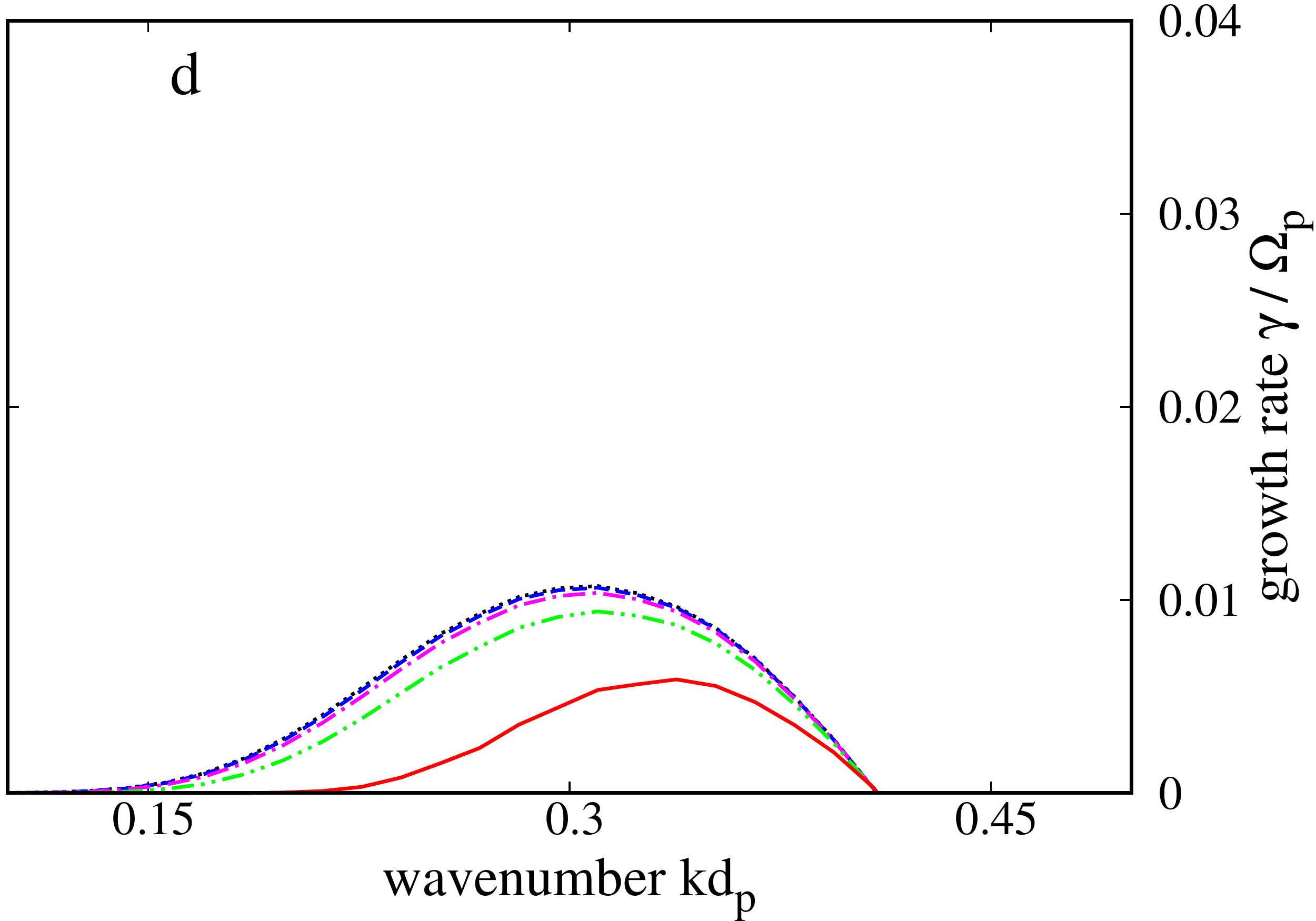}

  \vspace{0.1cm}

  \includegraphics[width=.35\textwidth]{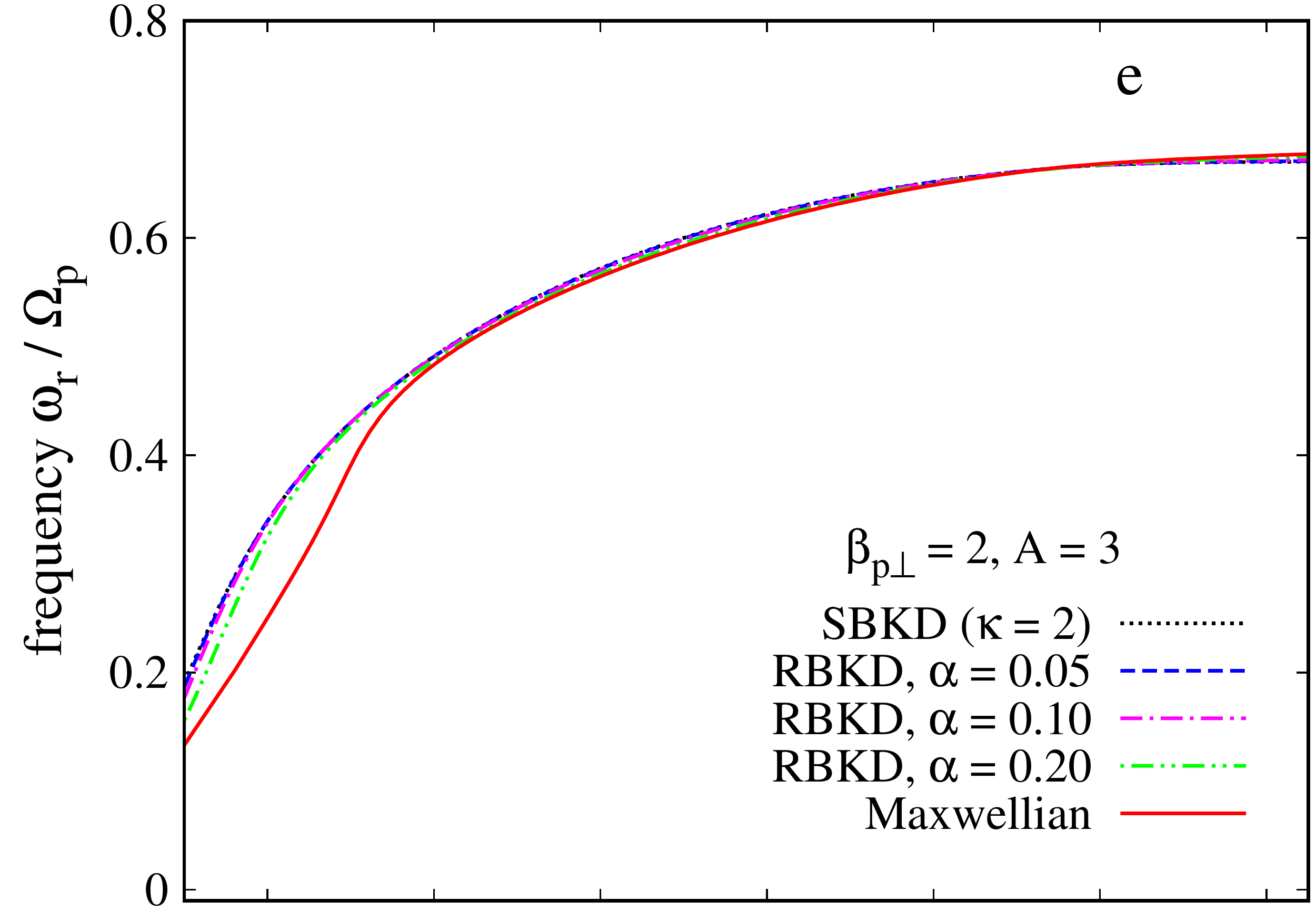}
  \hspace{0.1cm}
  \includegraphics[width=.35\textwidth]{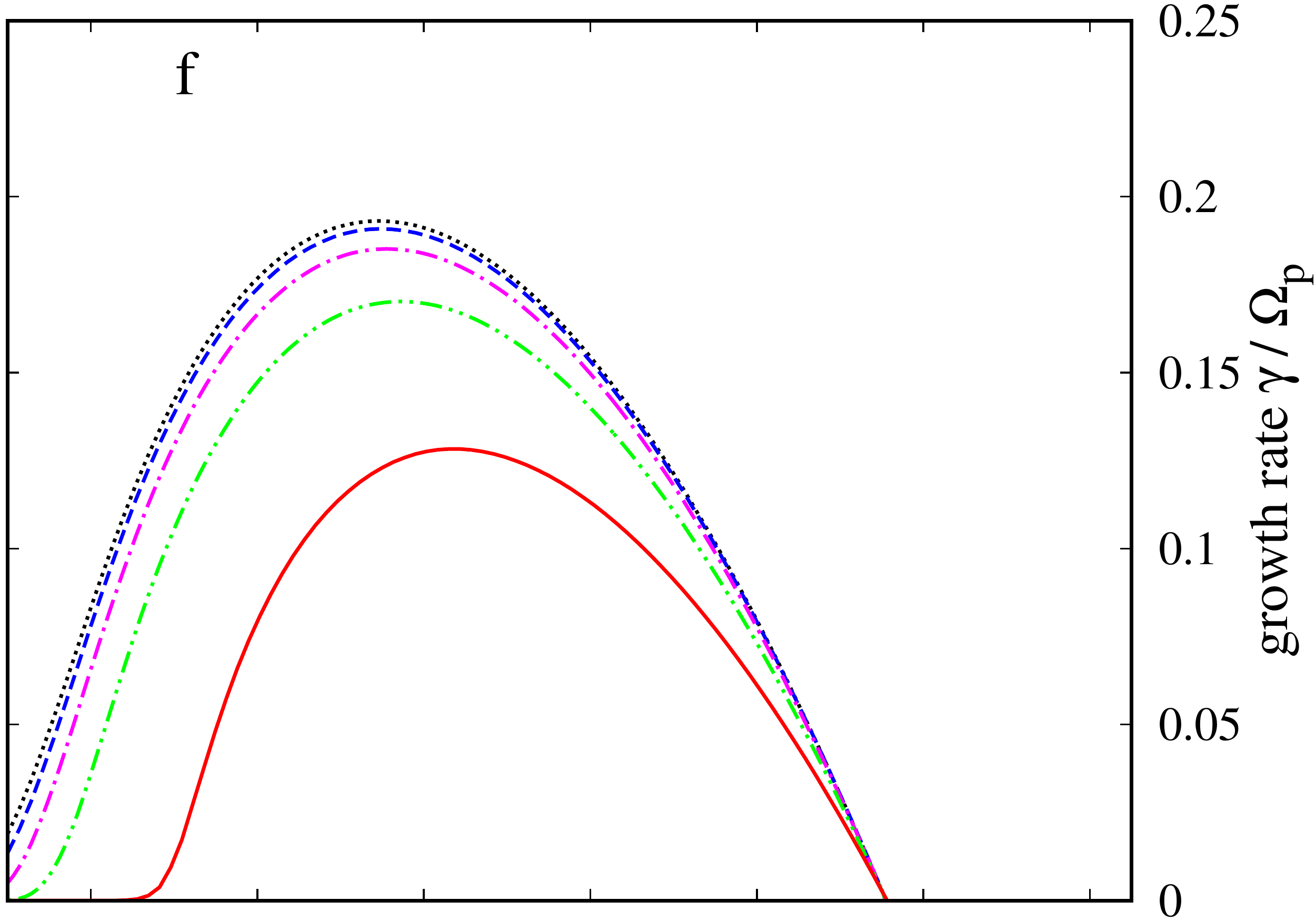}
  
    \vspace{0.1cm}

  \includegraphics[width=.35\textwidth]{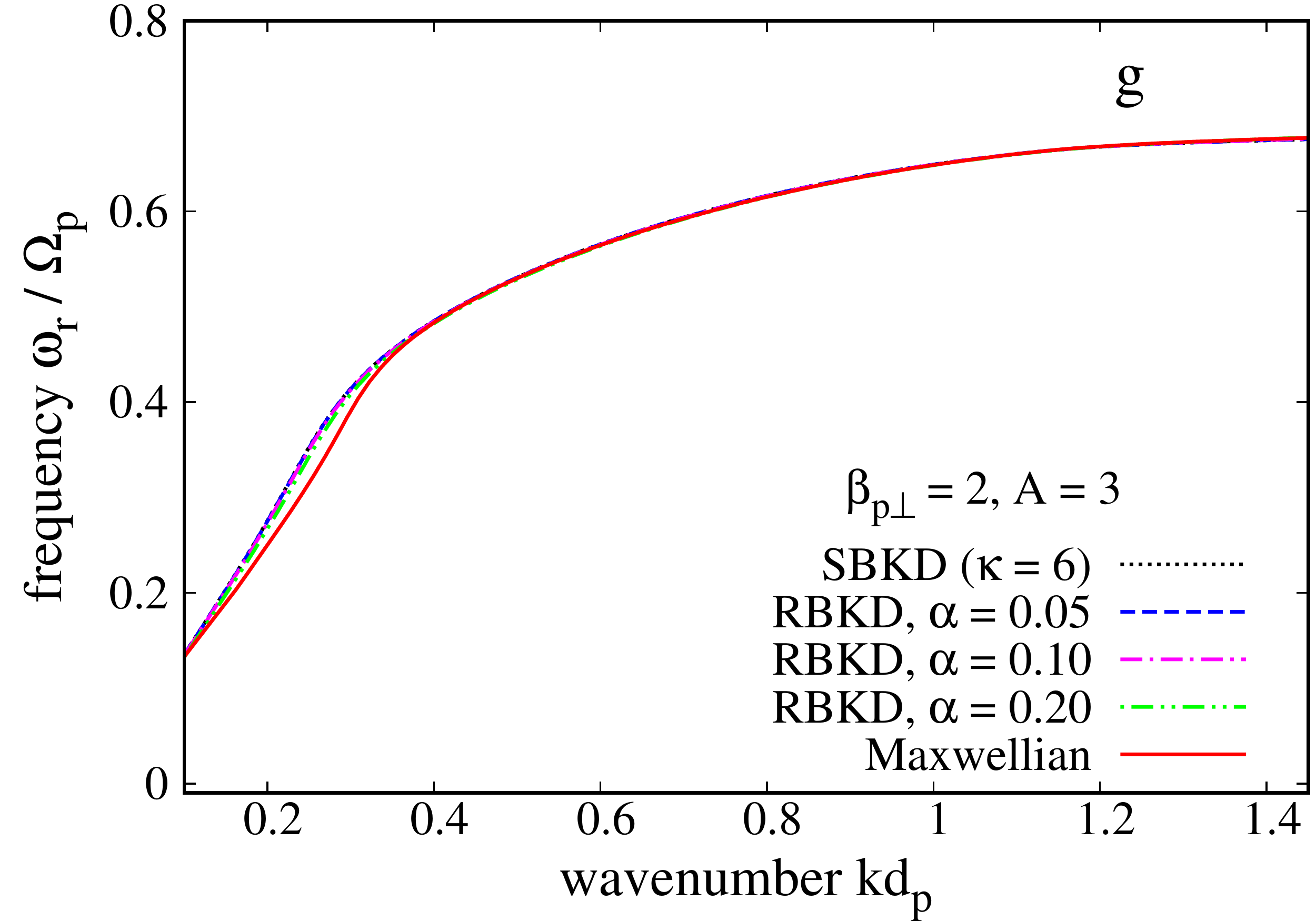}
  \hspace{0.1cm}
  \includegraphics[width=.35\textwidth]{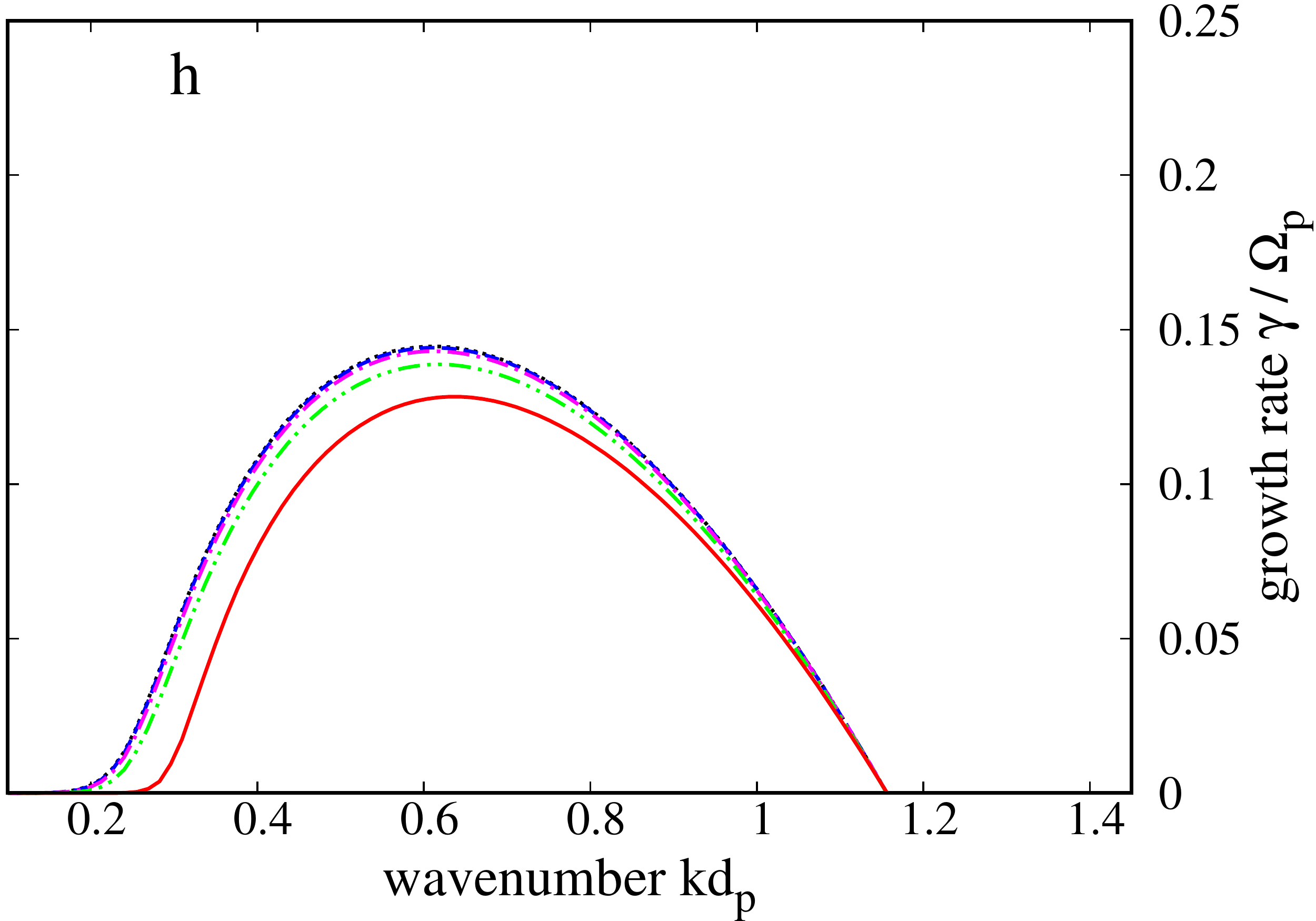}

  \caption{Dispersion curves for wave frequencies (left column) and 
           growth rates (right column) of the EMIC instability
           with $\alpha_\parallel = \alpha_\perp$. Parameters are 
           explained in the legends.}\label{fig:emic_rbkd}
\end{figure*}

\begin{figure*}[t!]
  \centering
  \includegraphics[width=.35\textwidth]{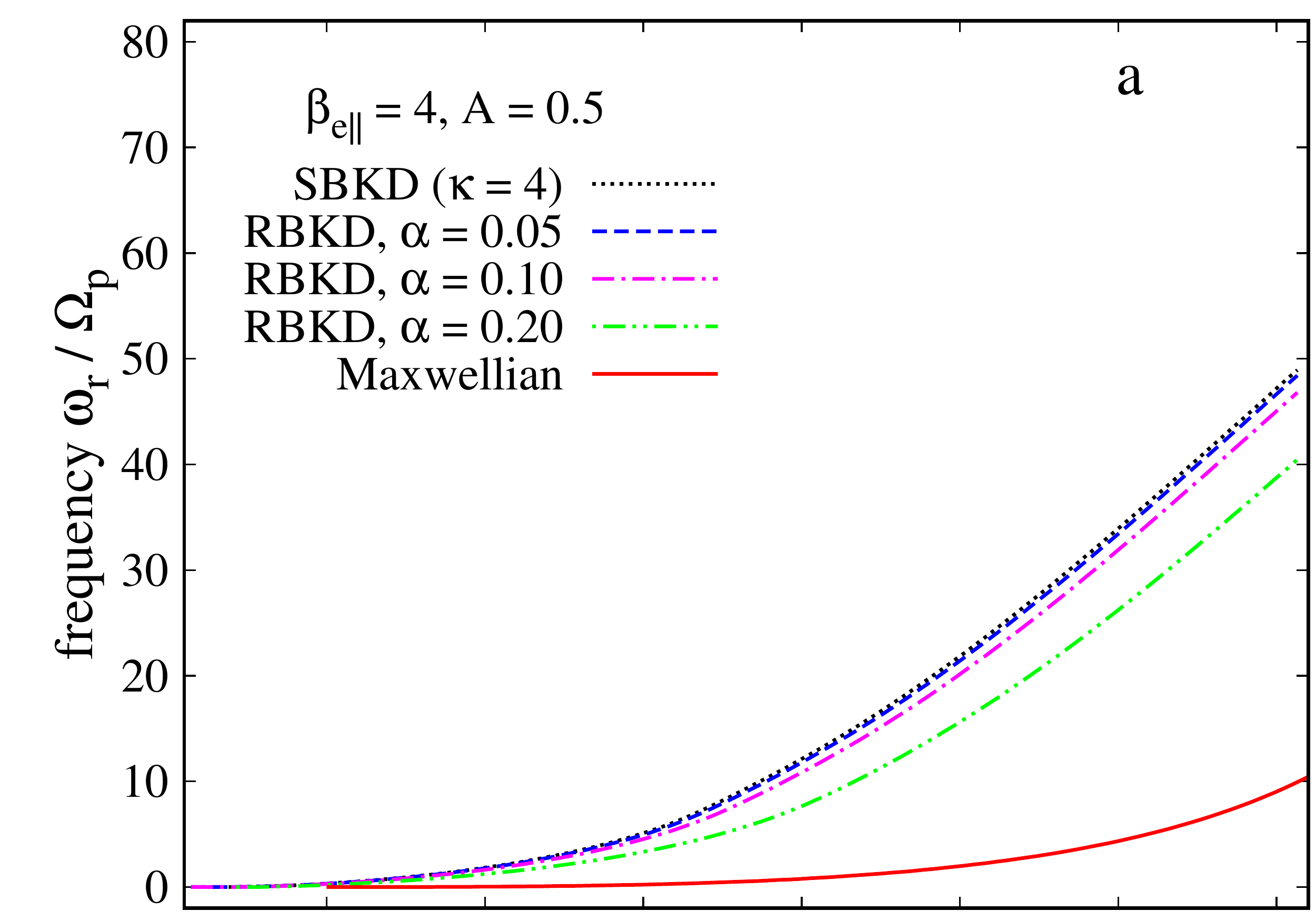}
  \hspace{0.1cm}
  \includegraphics[width=.35\textwidth]{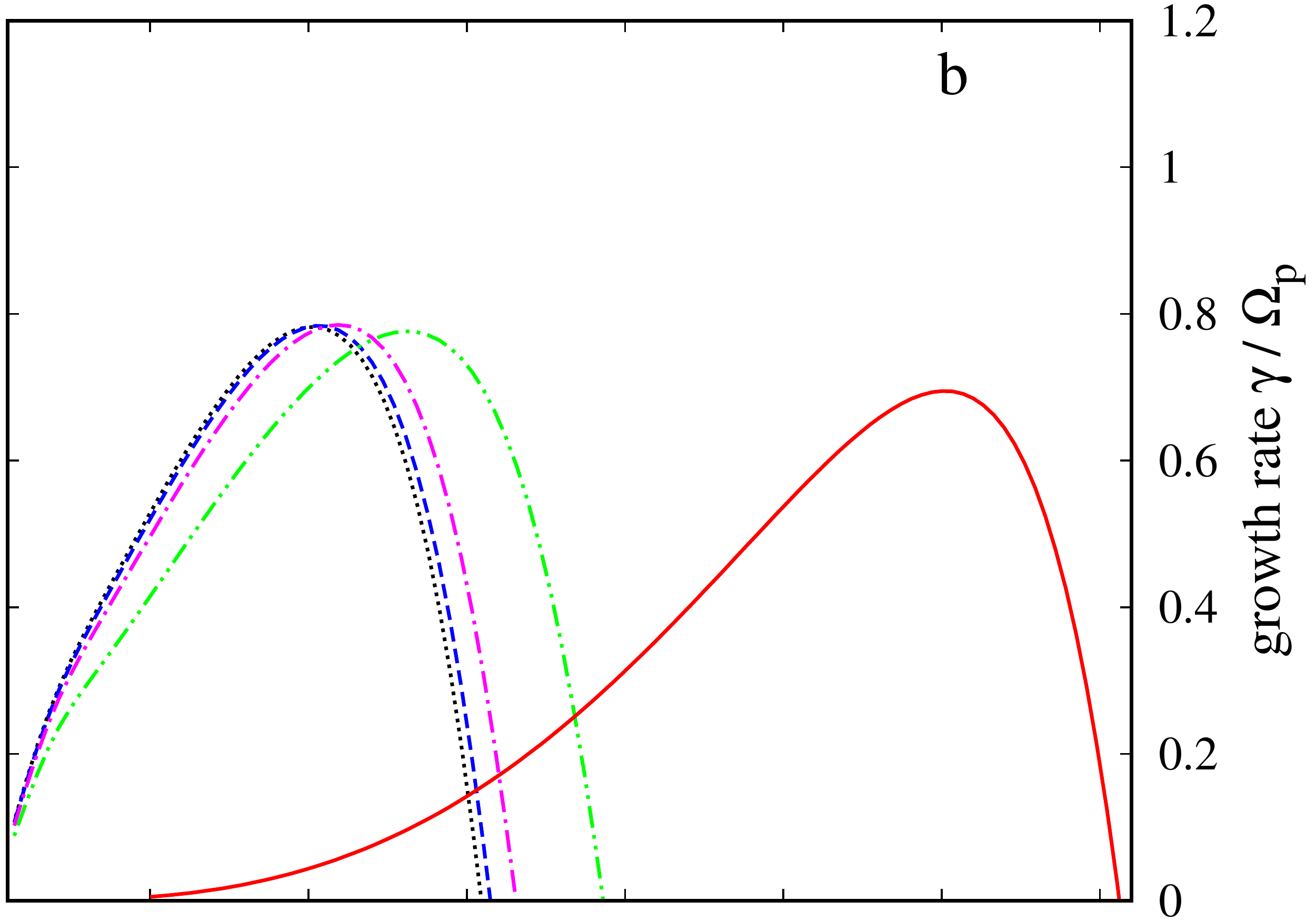}

  \vspace{0.1cm}

  \includegraphics[width=.35\textwidth]{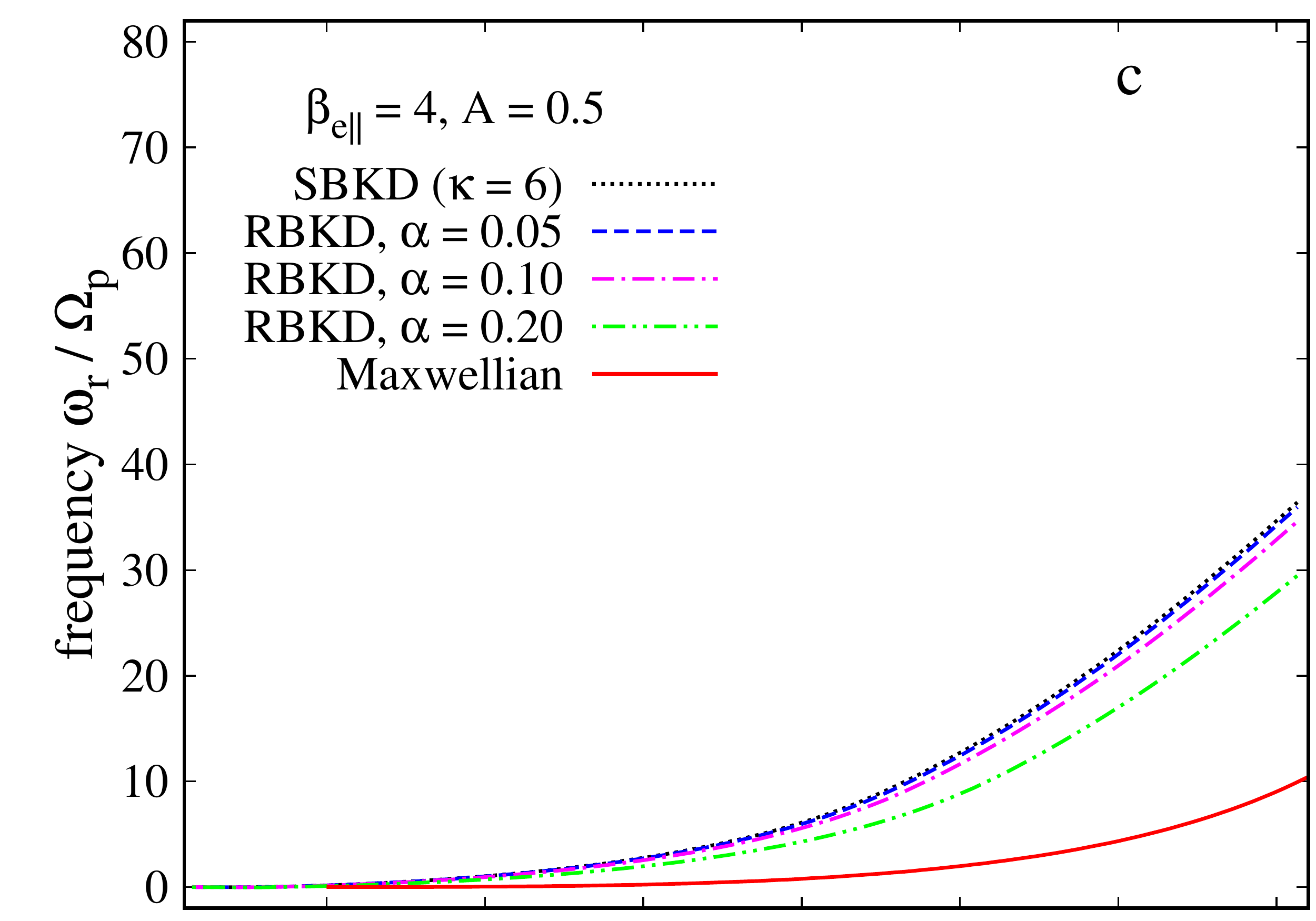}
  \hspace{0.1cm}
  \includegraphics[width=.35\textwidth]{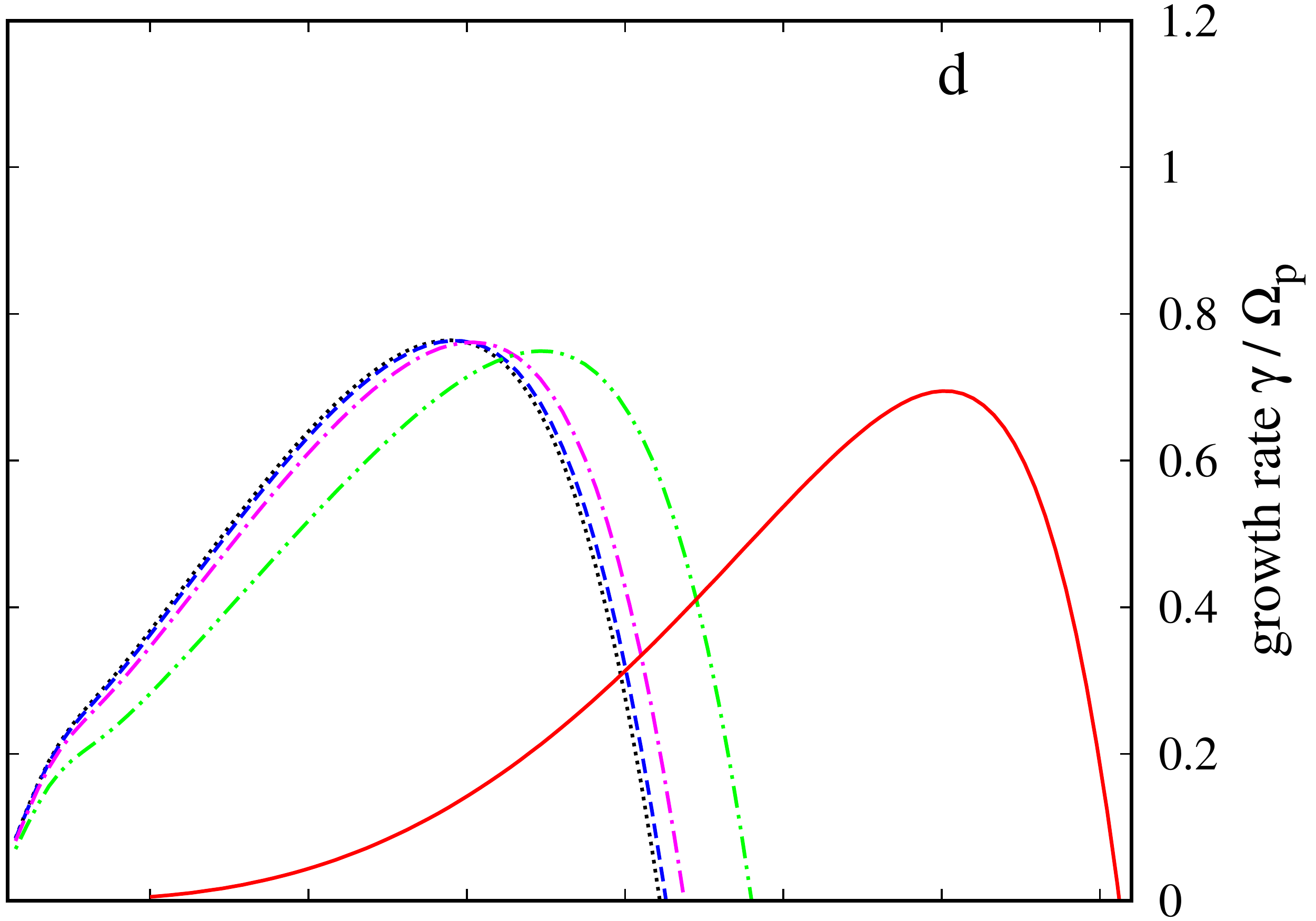}

  \vspace{0.1cm}

  \includegraphics[width=.35\textwidth]{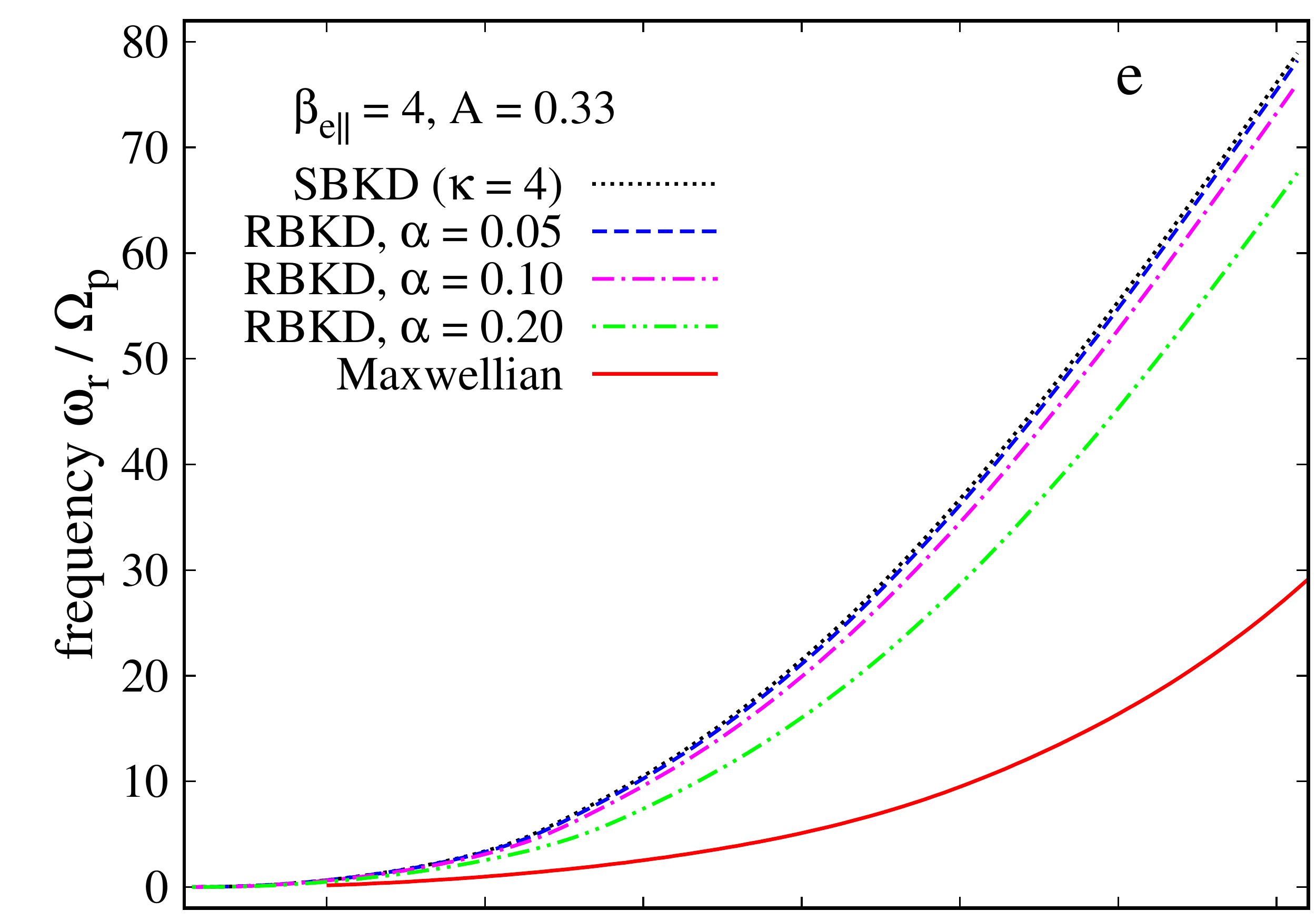}
  \hspace{0.1cm}
  \includegraphics[width=.35\textwidth]{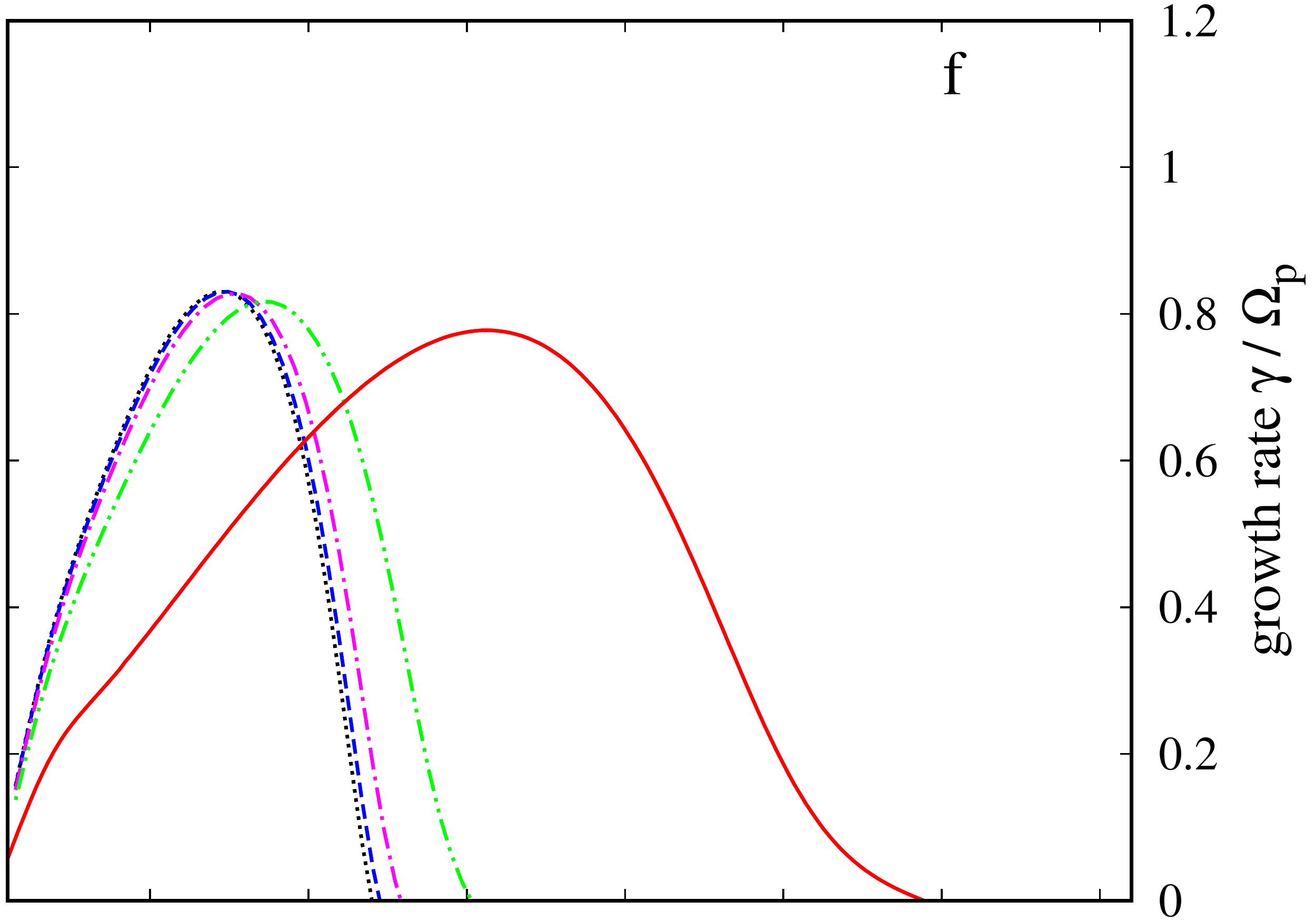}
  
    \vspace{0.1cm}

  \includegraphics[width=.35\textwidth]{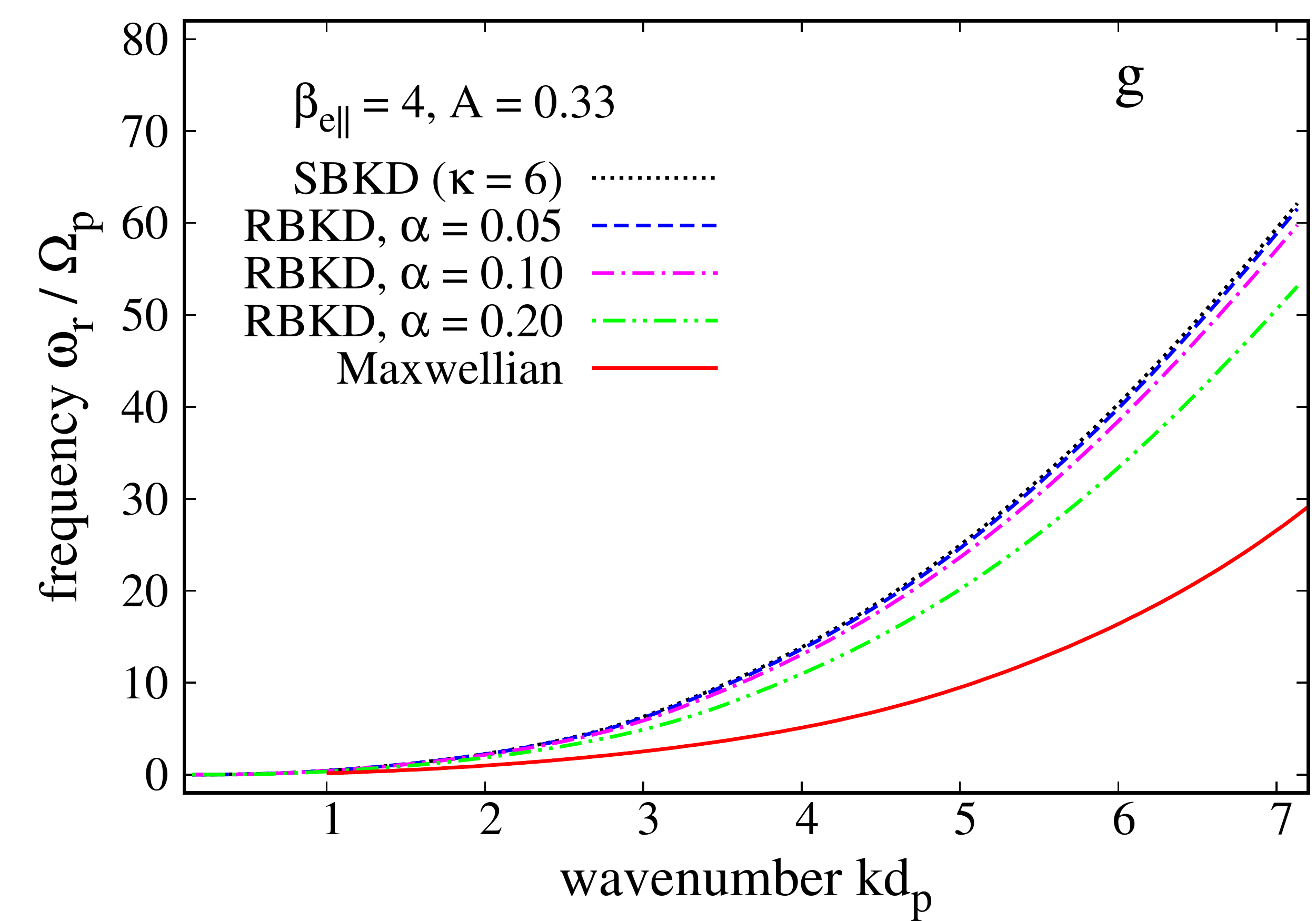}
  \hspace{0.1cm}
  \includegraphics[width=.35\textwidth]{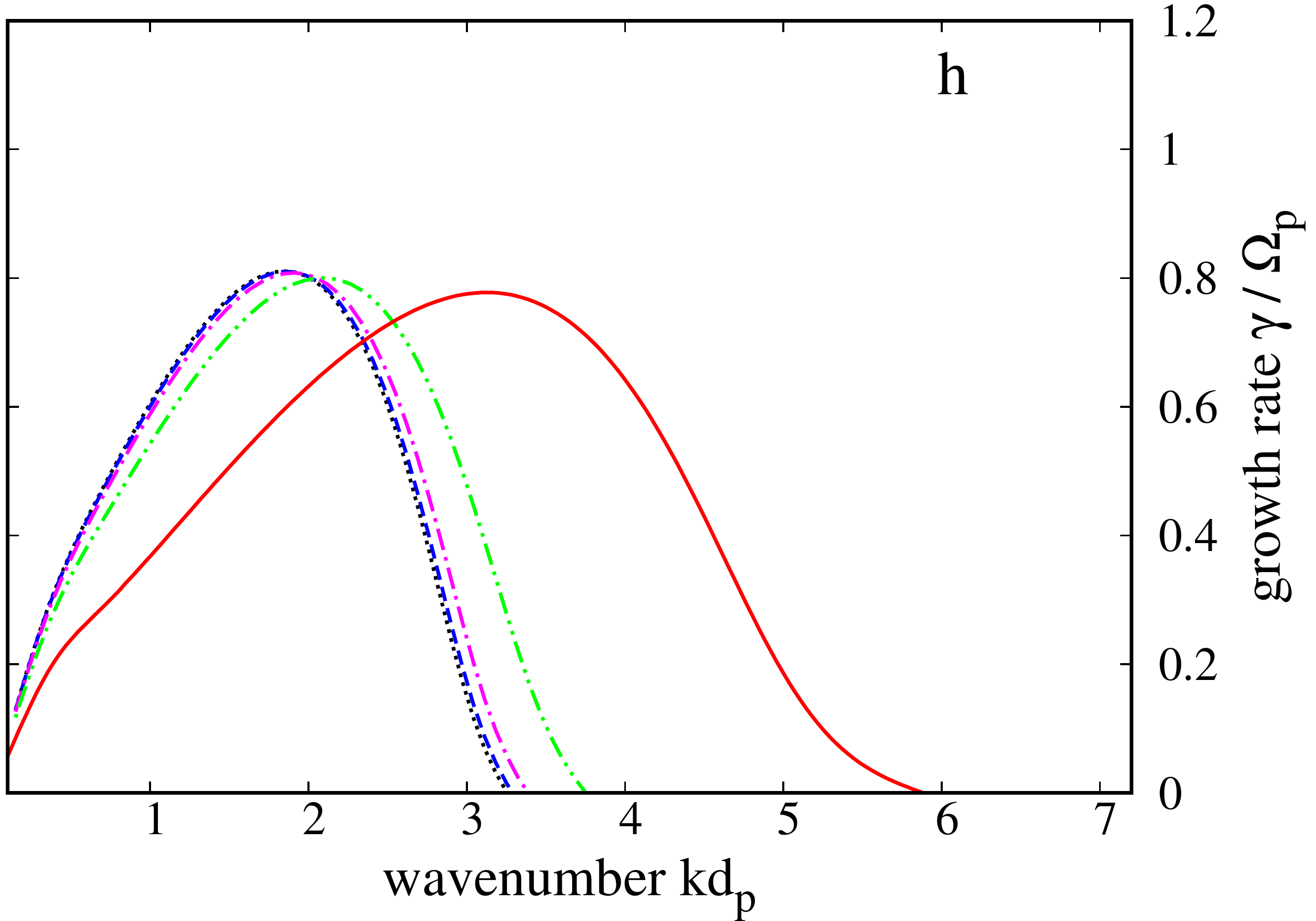}

  \caption{Dispersion curves for wave frequencies (left column) and 
           growth rates (right column) of the EFH instability
           with $\alpha_\parallel = \alpha_\perp$. Parameters are 
           explained in the legends.}\label{fig:pefh_rbkd}
\end{figure*}

\begin{figure*}[t!]
  \centering
  \includegraphics[width=.35\textwidth]{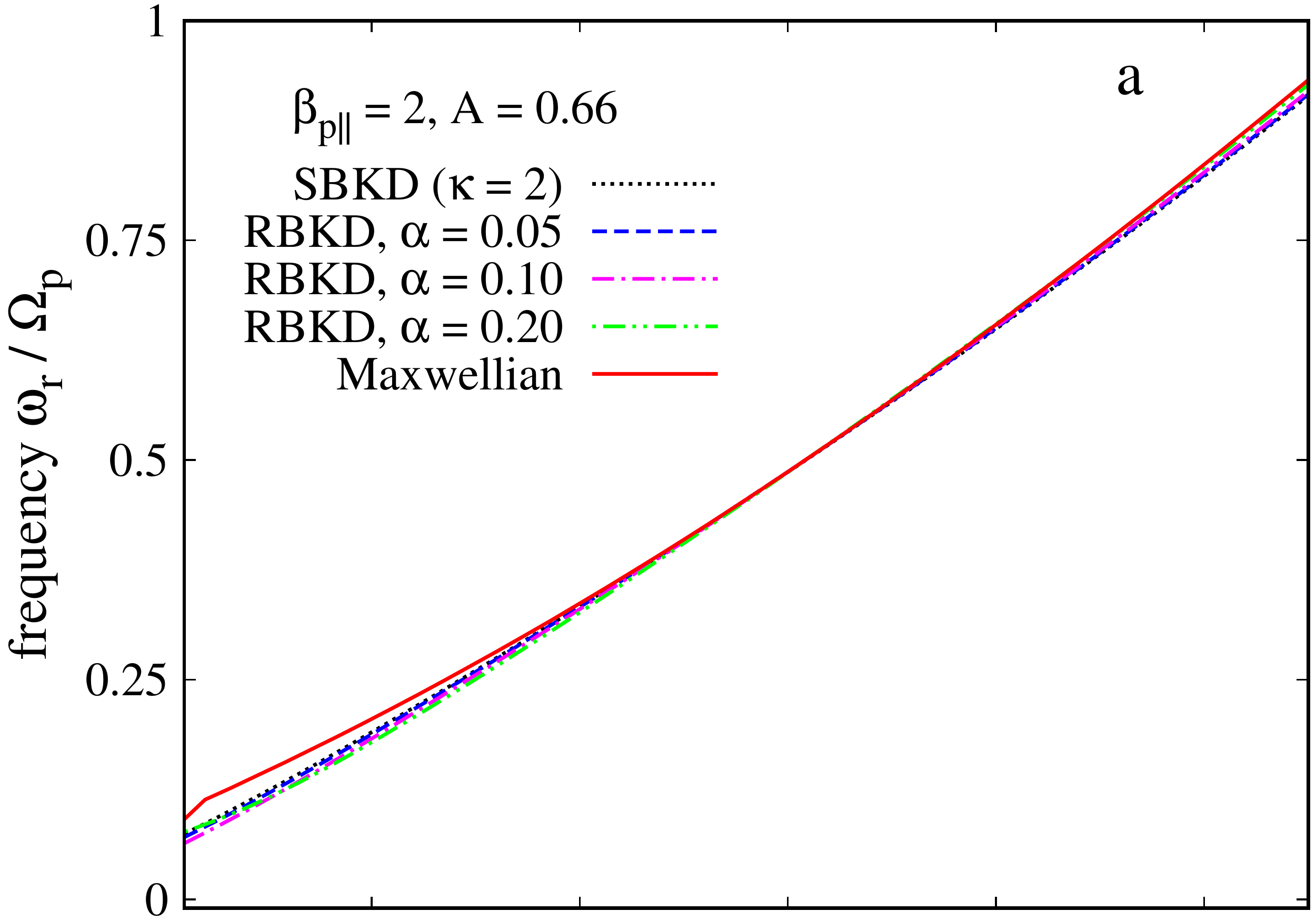}
  \hspace{0.1cm}
  \includegraphics[width=.35\textwidth]{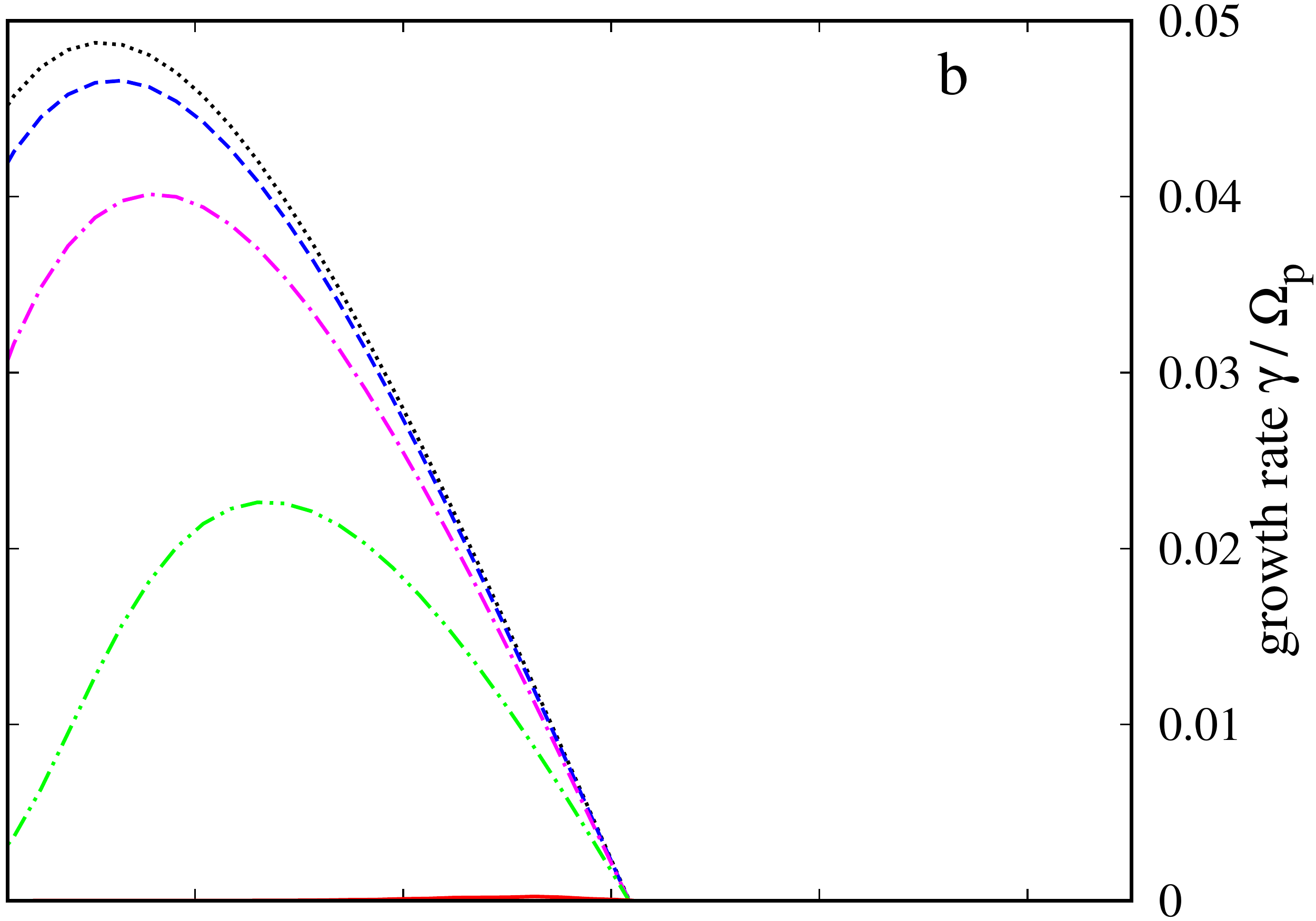}

  \vspace{0.1cm}

  \includegraphics[width=.35\textwidth]{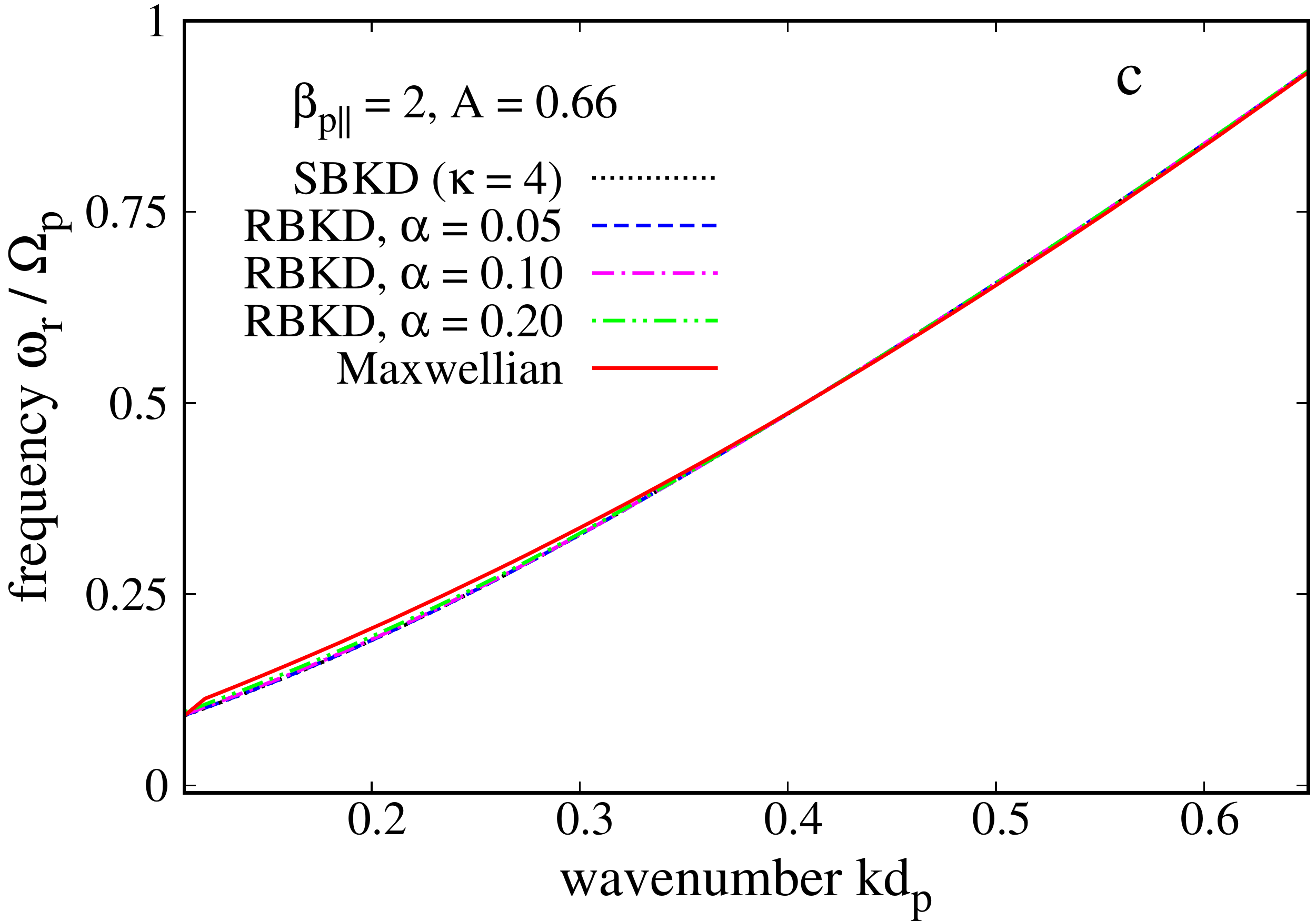}
  \hspace{0.1cm}
  \includegraphics[width=.35\textwidth]{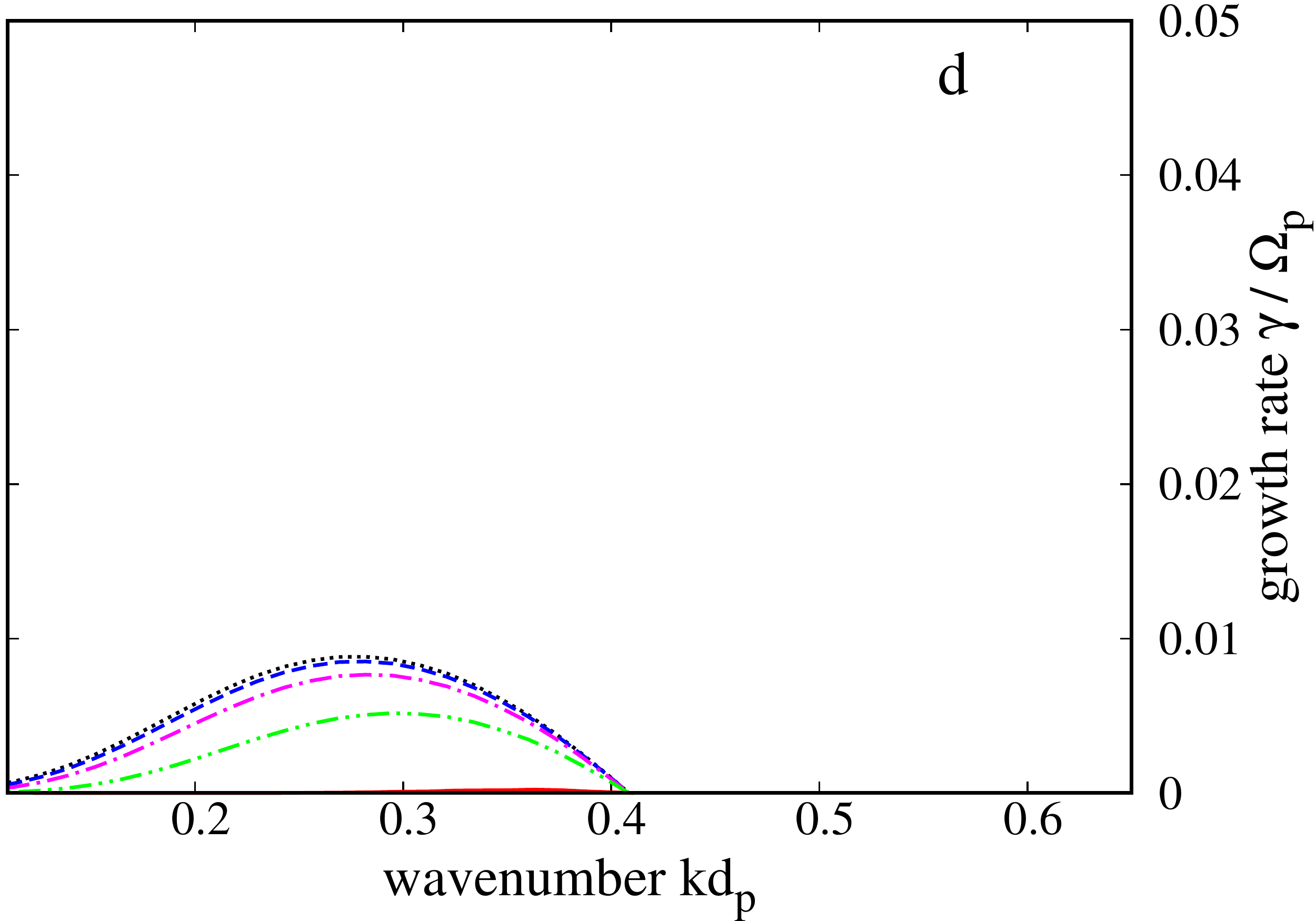}

  \vspace{0.1cm}

  \includegraphics[width=.35\textwidth]{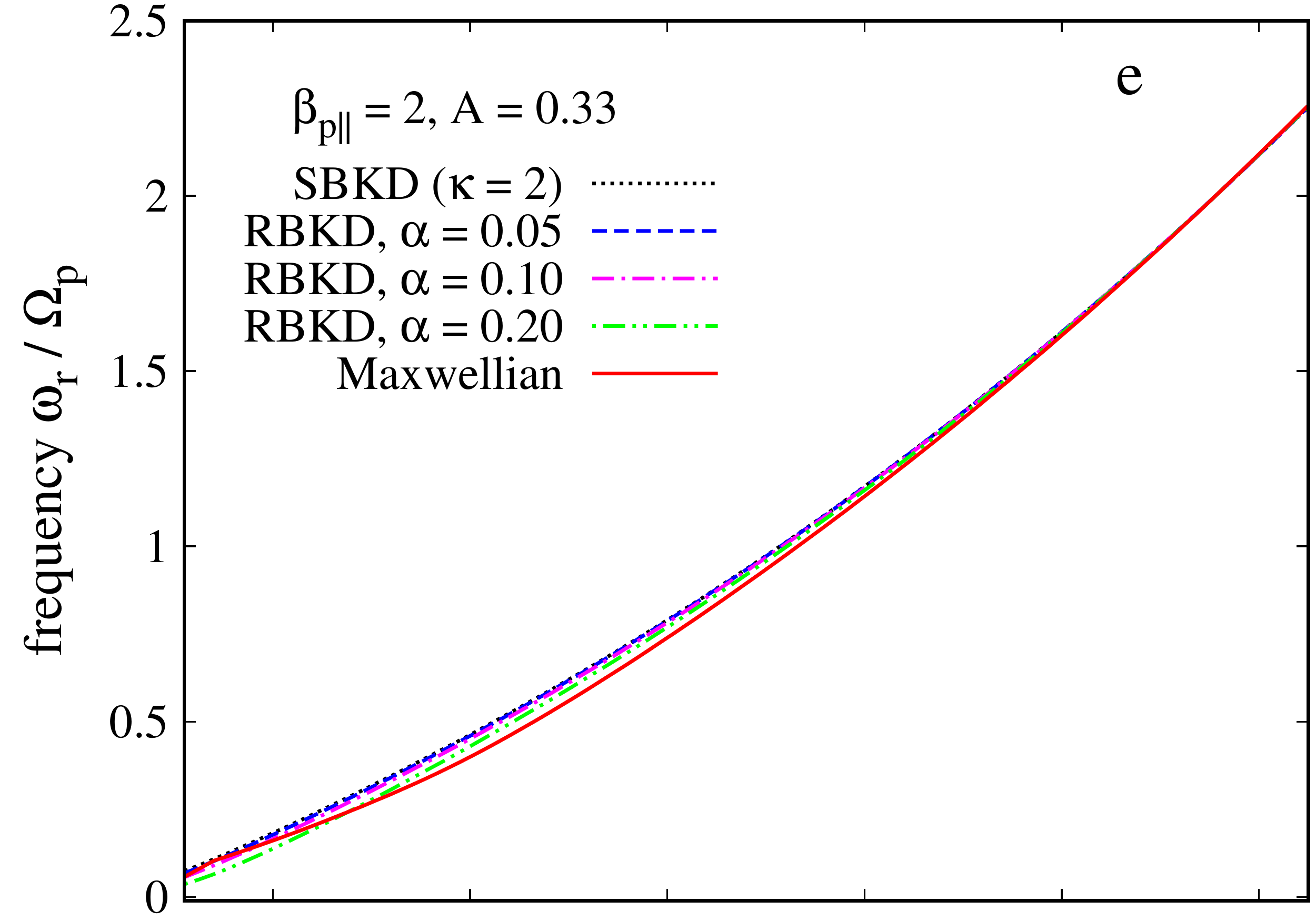}
  \hspace{0.1cm}
  \includegraphics[width=.35\textwidth]{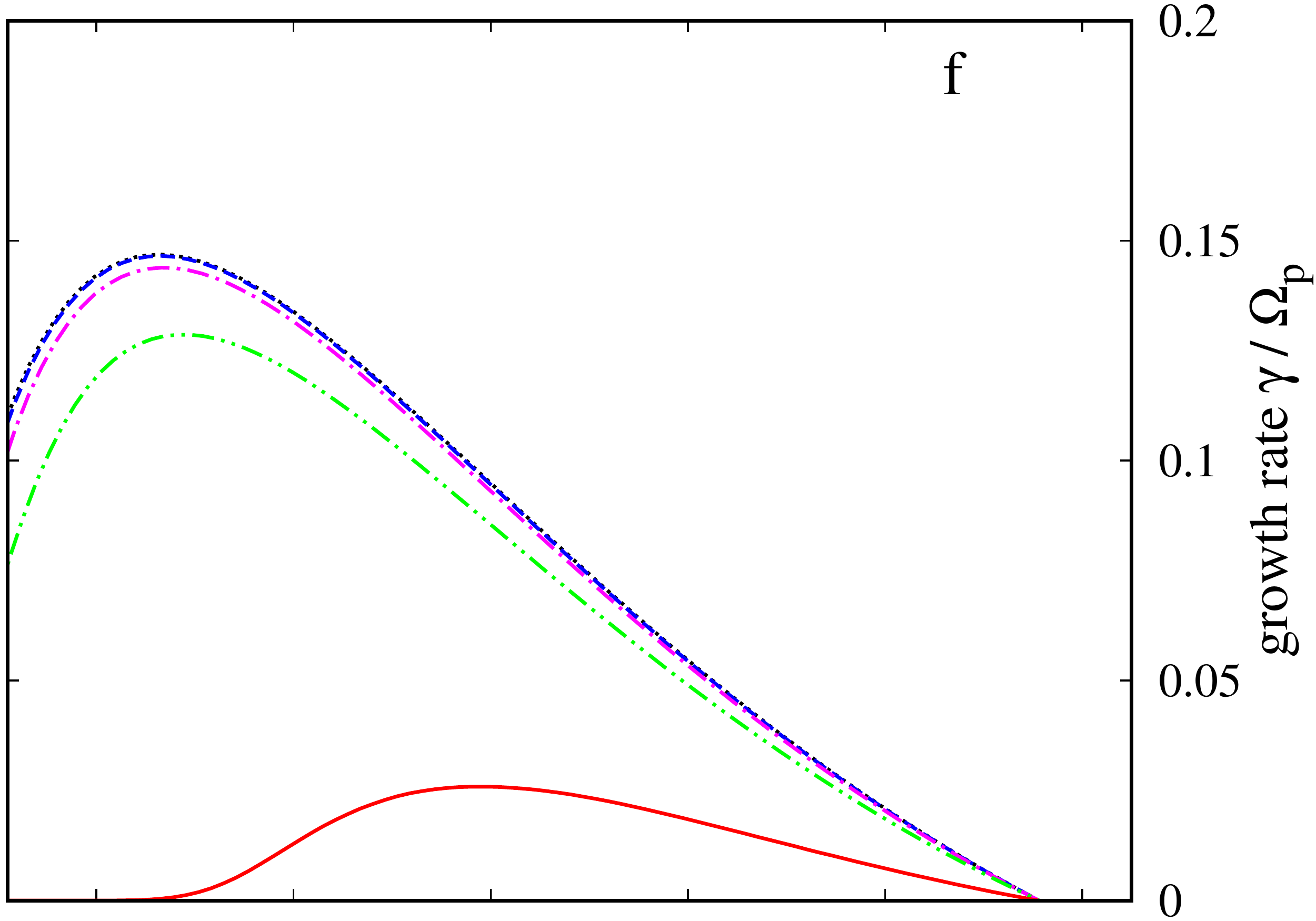}
  
    \vspace{0.1cm}

  \includegraphics[width=.35\textwidth]{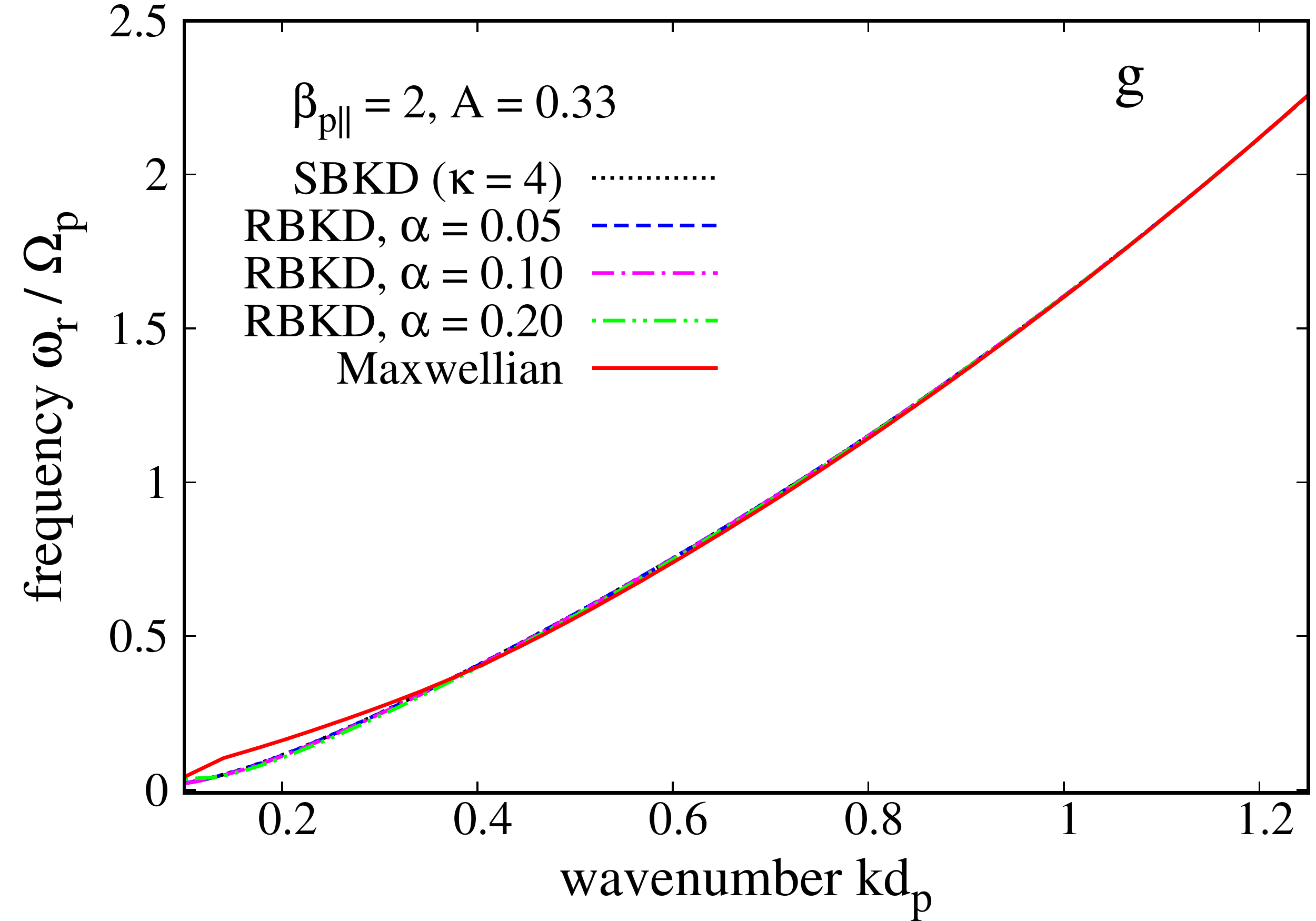}
  \hspace{0.1cm}
  \includegraphics[width=.35\textwidth]{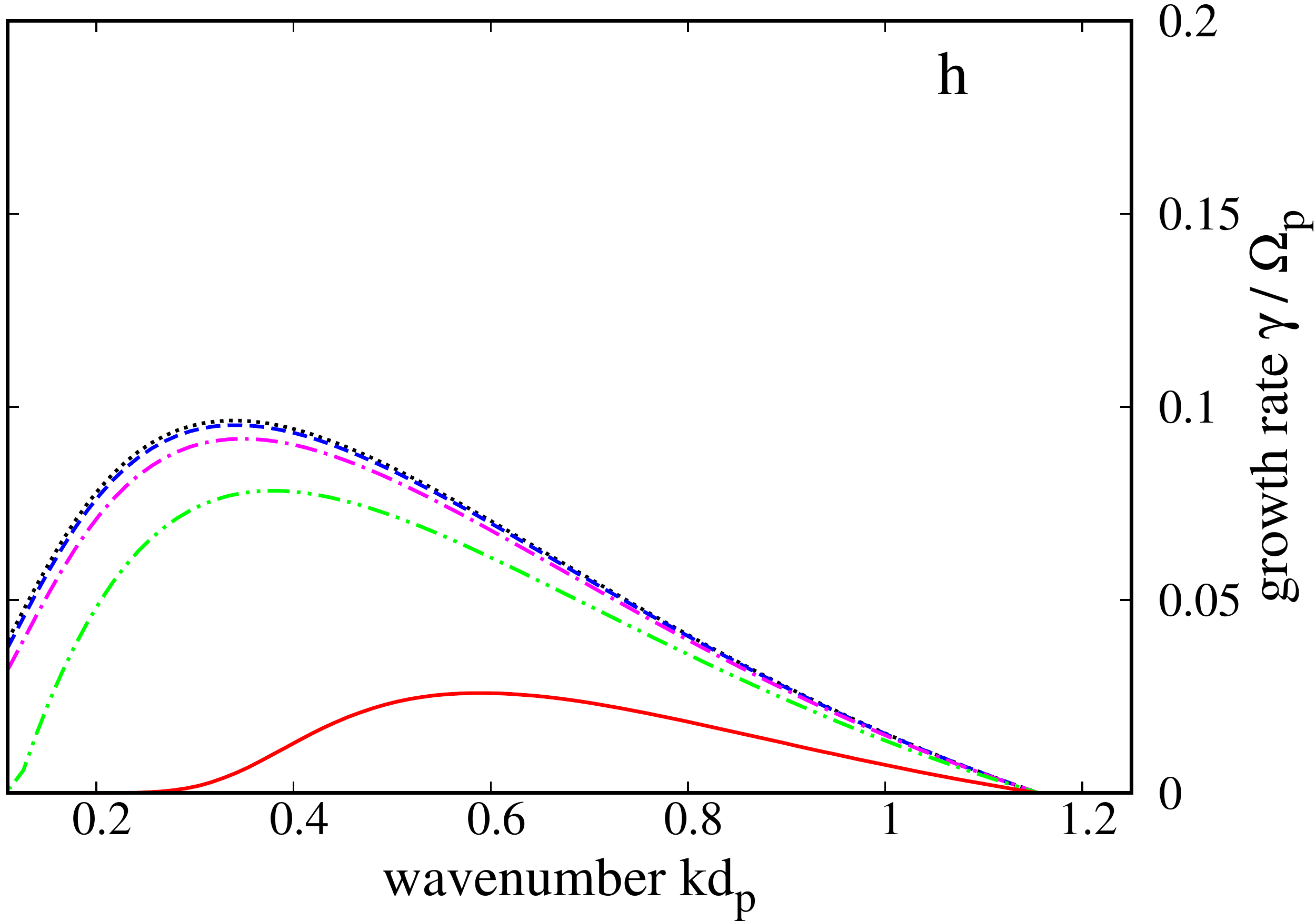}

  \caption{Dispersion curves for wave frequencies (left column) and 
           growth rates (right column) of the PFH instability
           with $\alpha_\parallel = \alpha_\perp$. Parameters are 
           explained in the legends.}\label{fig:ppfh_rbkd}
\end{figure*}

\begin{figure*}[t!]
  \centering
  \includegraphics[width=.35\textwidth]{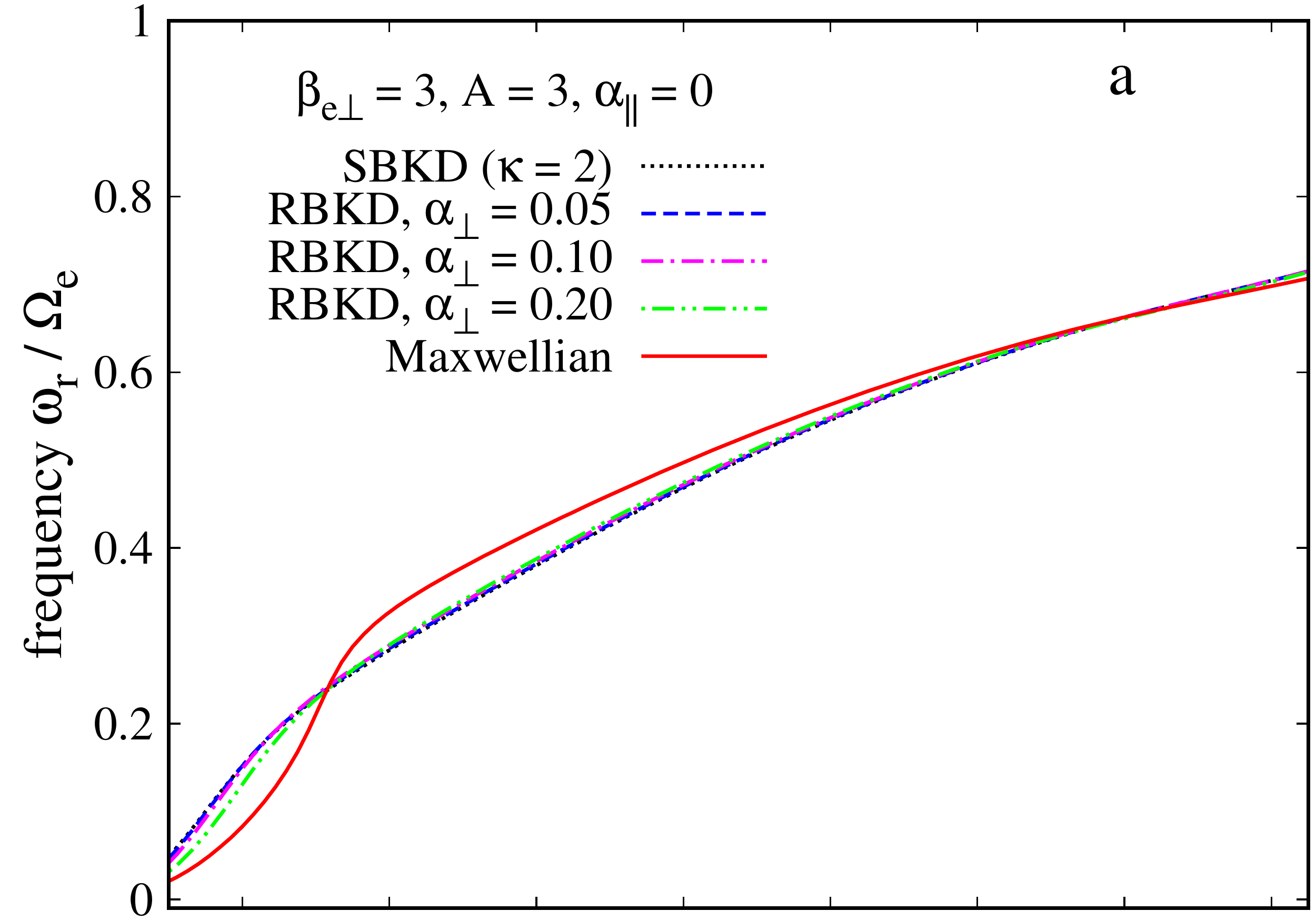}
  \hspace{0.1cm}
  \includegraphics[width=.35\textwidth]{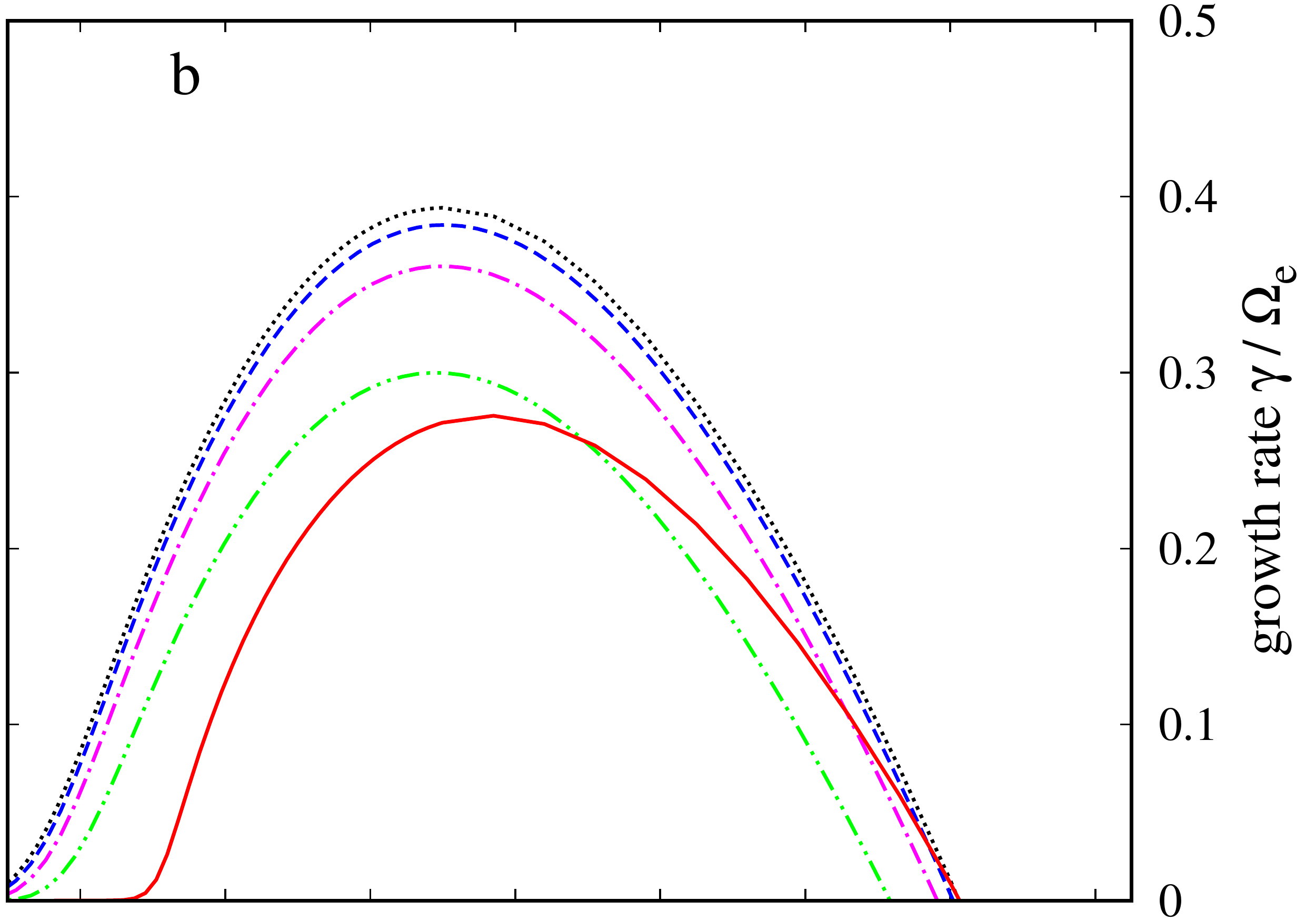}

  \vspace{0.1cm}

  \includegraphics[width=.35\textwidth]{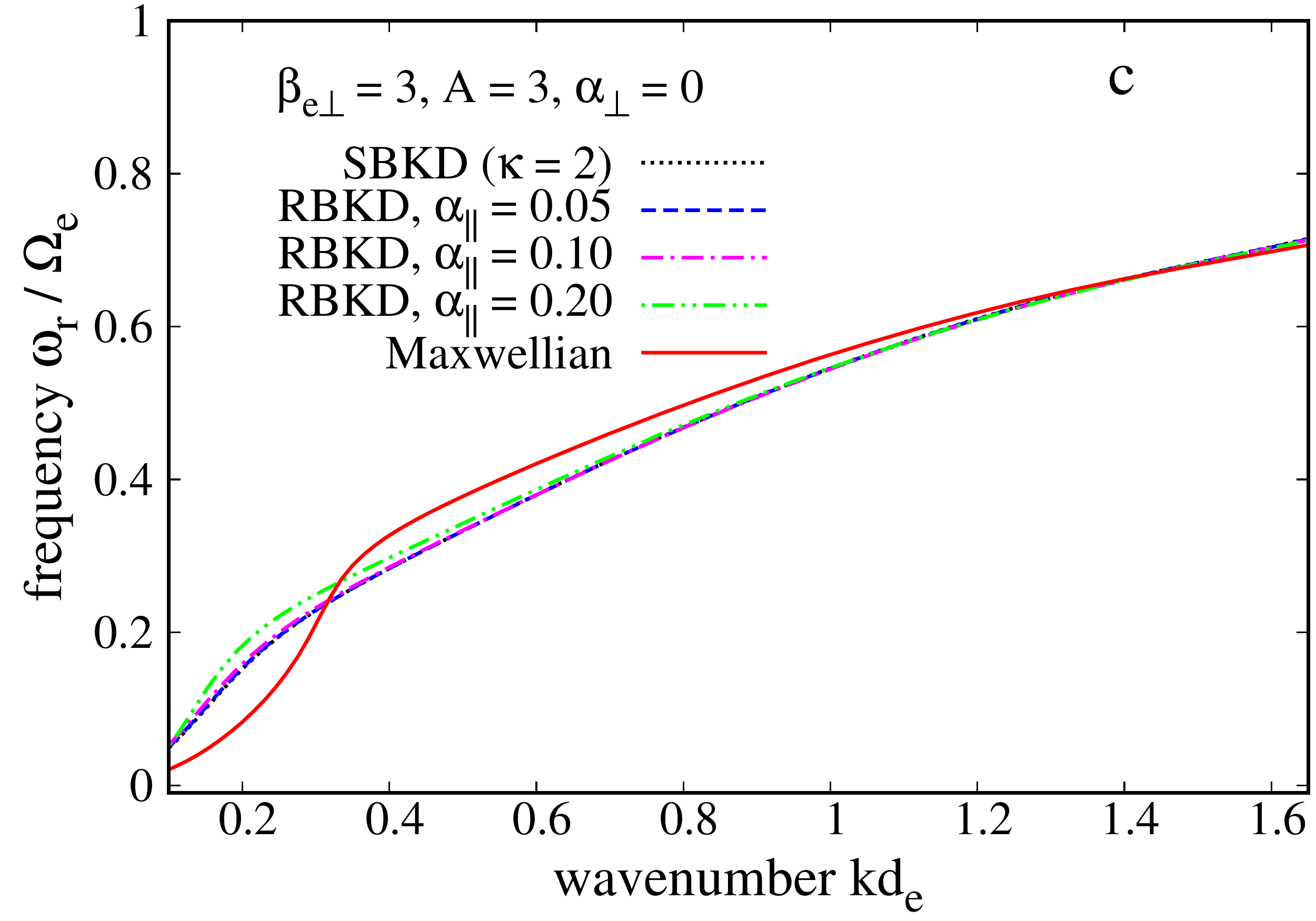}
  \hspace{0.1cm}
  \includegraphics[width=.35\textwidth]{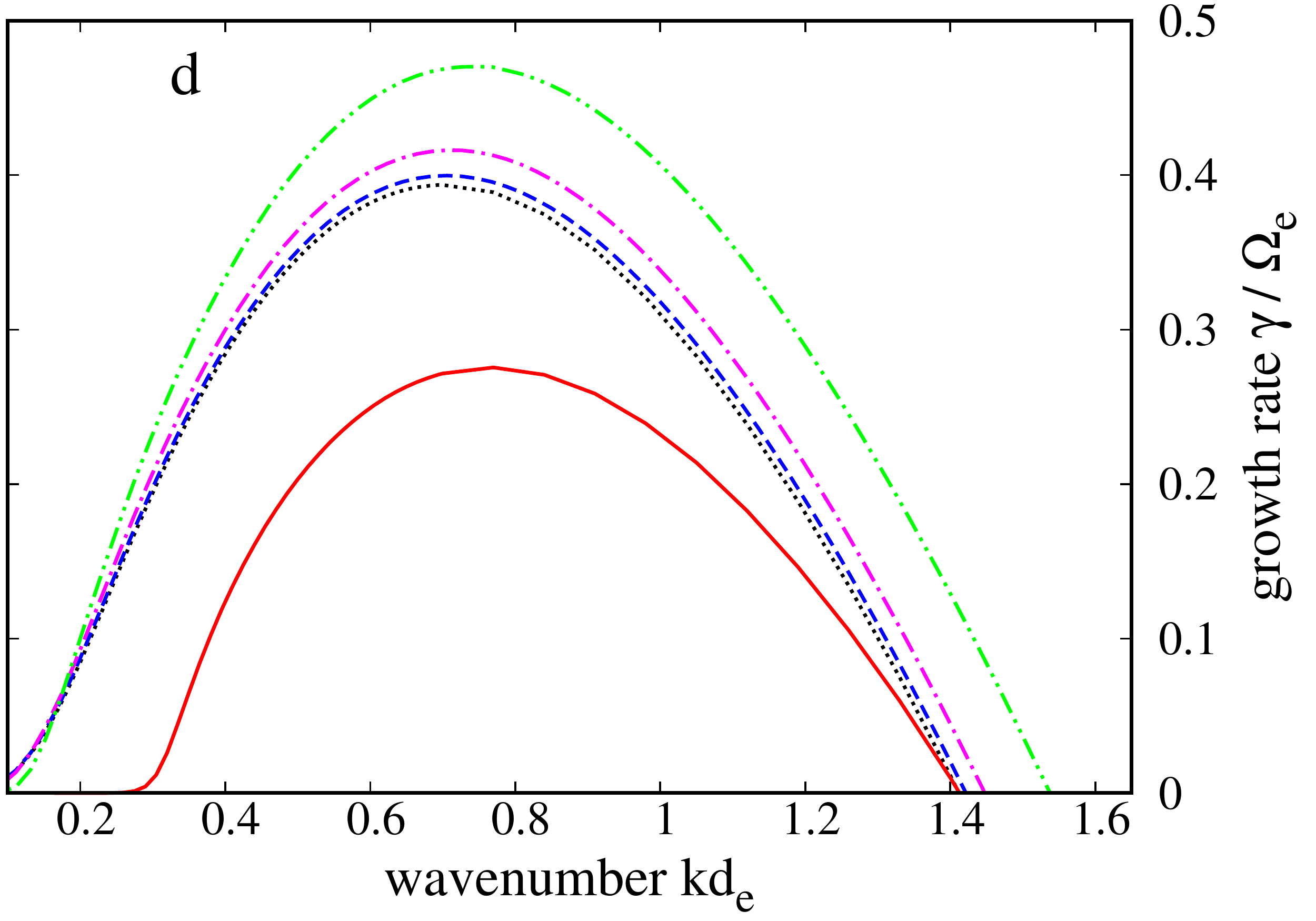}
  
  \includegraphics[width=.35\textwidth]{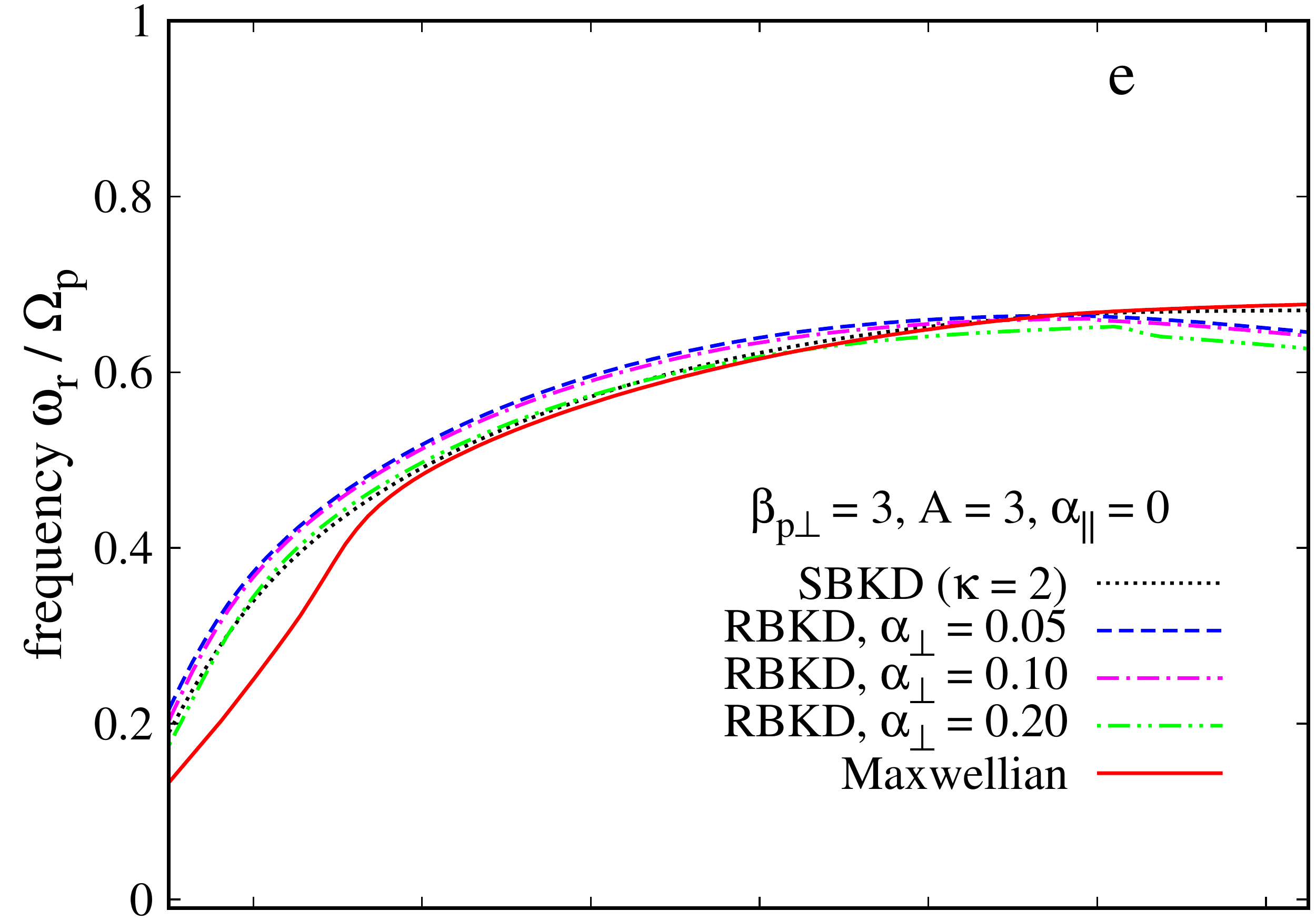}
  \hspace{0.1cm}
  \includegraphics[width=.35\textwidth]{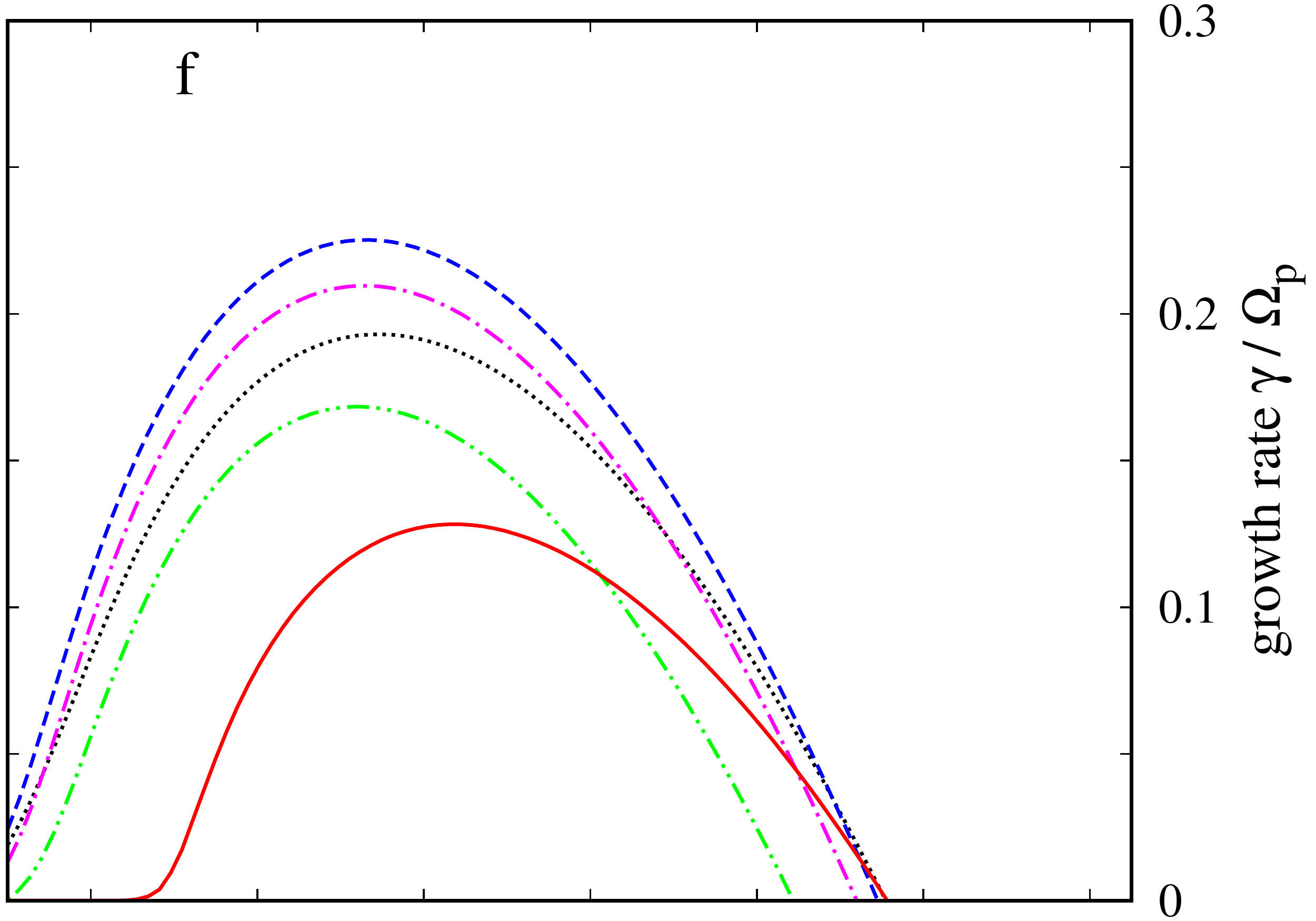}

  \vspace{0.1cm}

  \includegraphics[width=.35\textwidth]{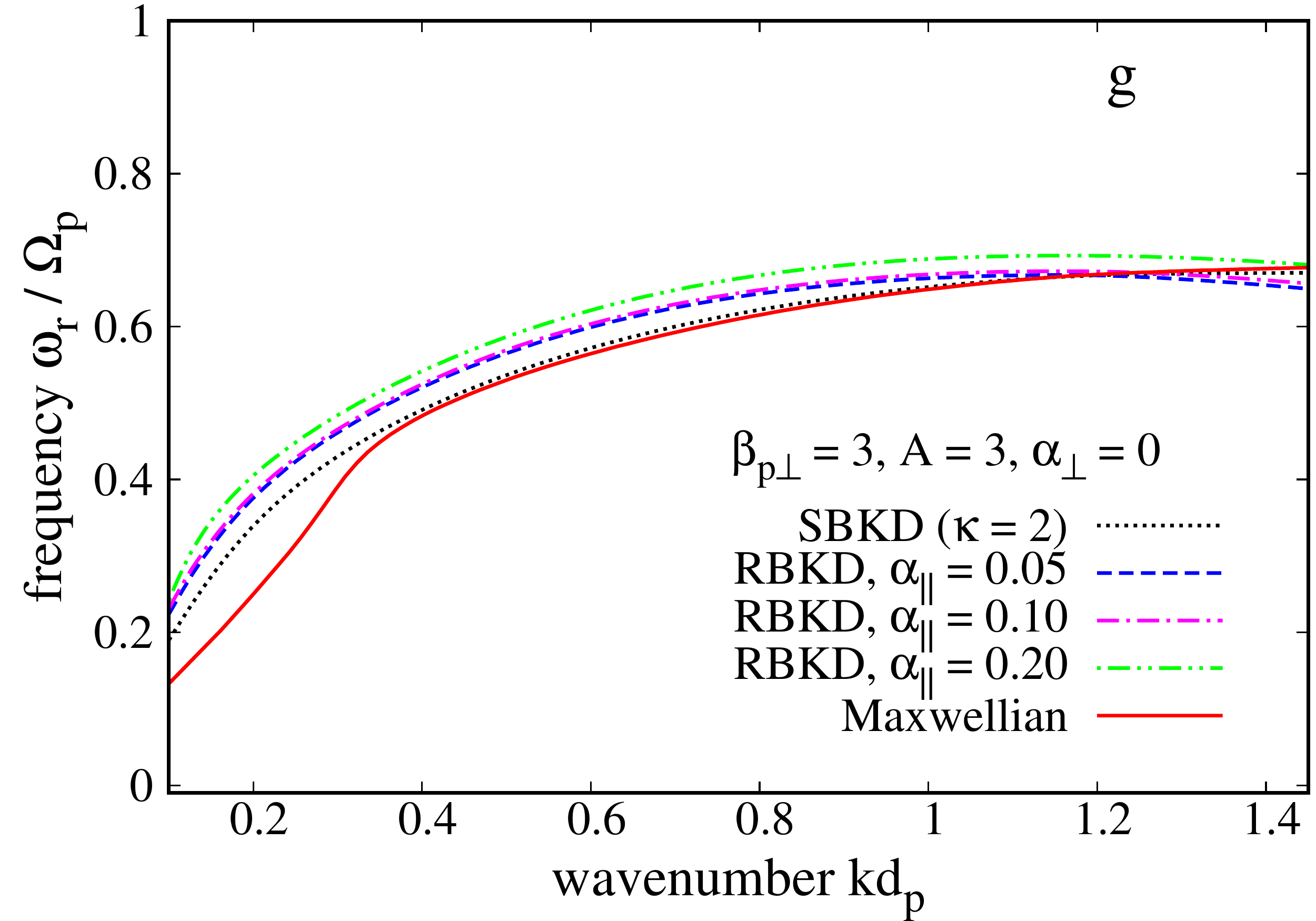}
  \hspace{0.1cm}
  \includegraphics[width=.35\textwidth]{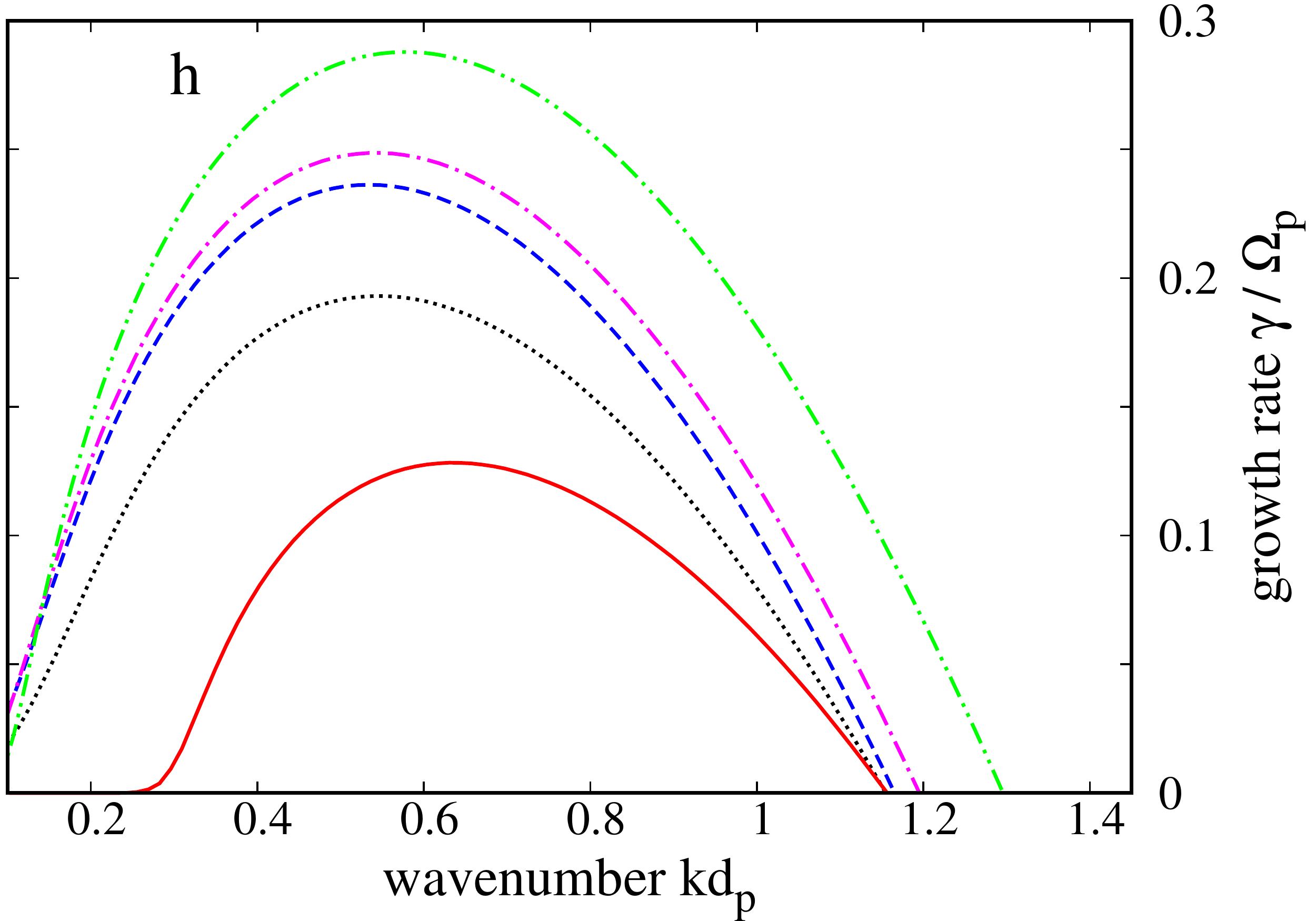}

  \caption{Dispersion curves for wave frequencies (left panels) and 
           growth rates (right panels) of the EMEC (top two rows) and 
           EMIC (bottom two rows) instability with 
           $\alpha_\parallel \neq \alpha_\perp$. Parameters are explained 
           in the legends.}\label{fig:emec_emic_rbkd_apa_ne_ape}
\end{figure*}

\begin{figure*}[t!]
  \centering
  \includegraphics[width=.35\textwidth]{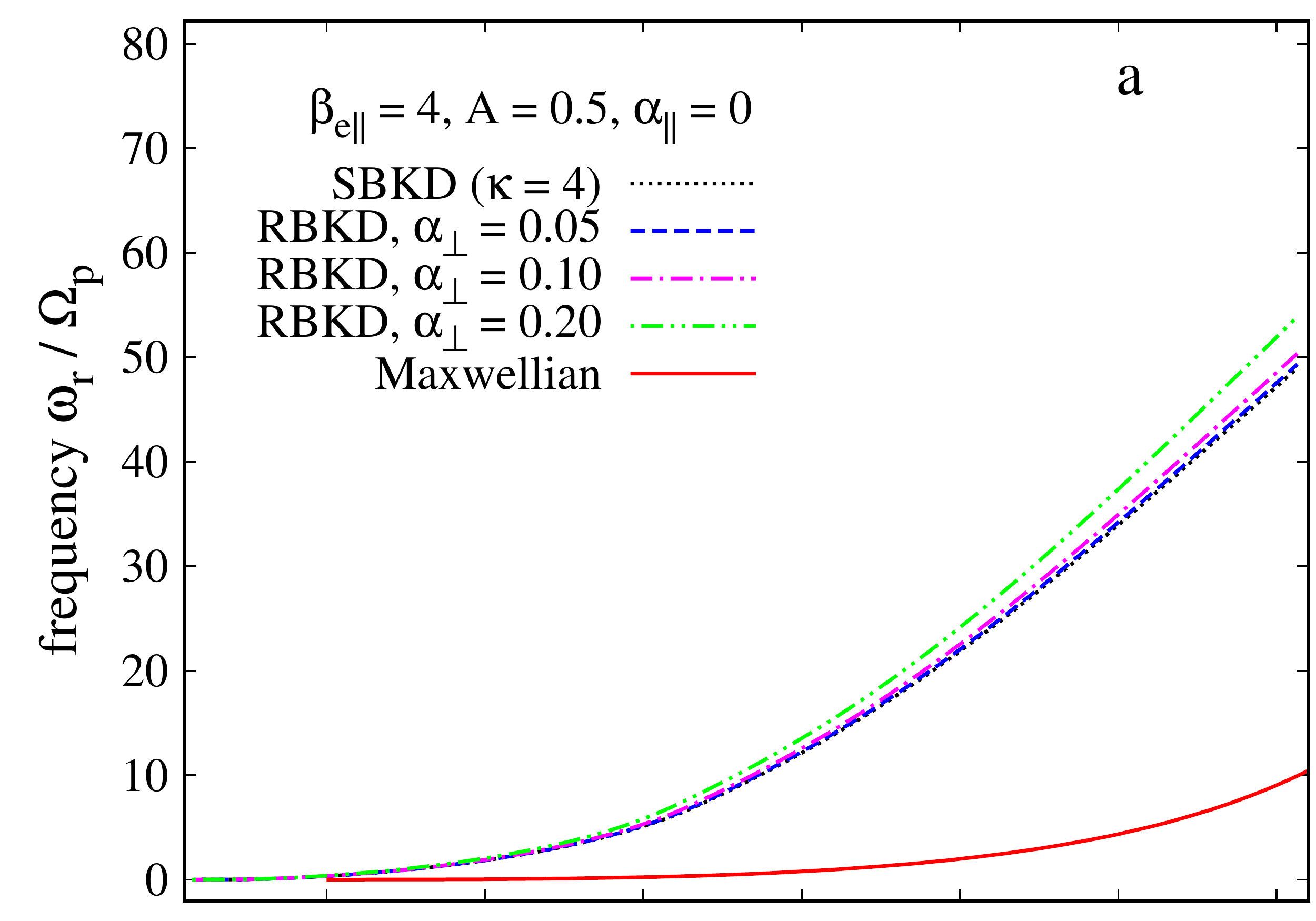}
  \hspace{0.1cm}
  \includegraphics[width=.35\textwidth]{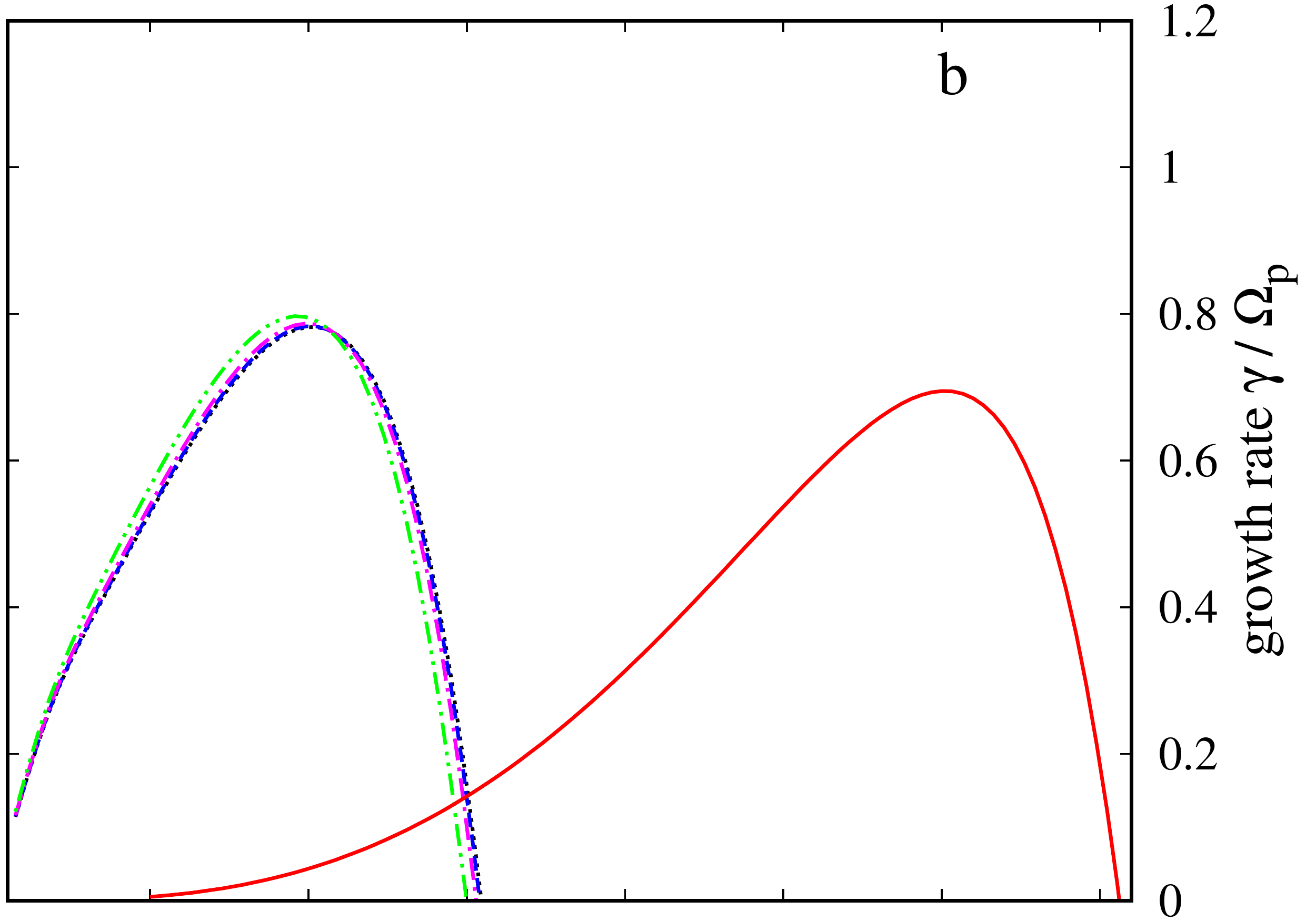}

  \vspace{0.1cm}

  \includegraphics[width=.35\textwidth]{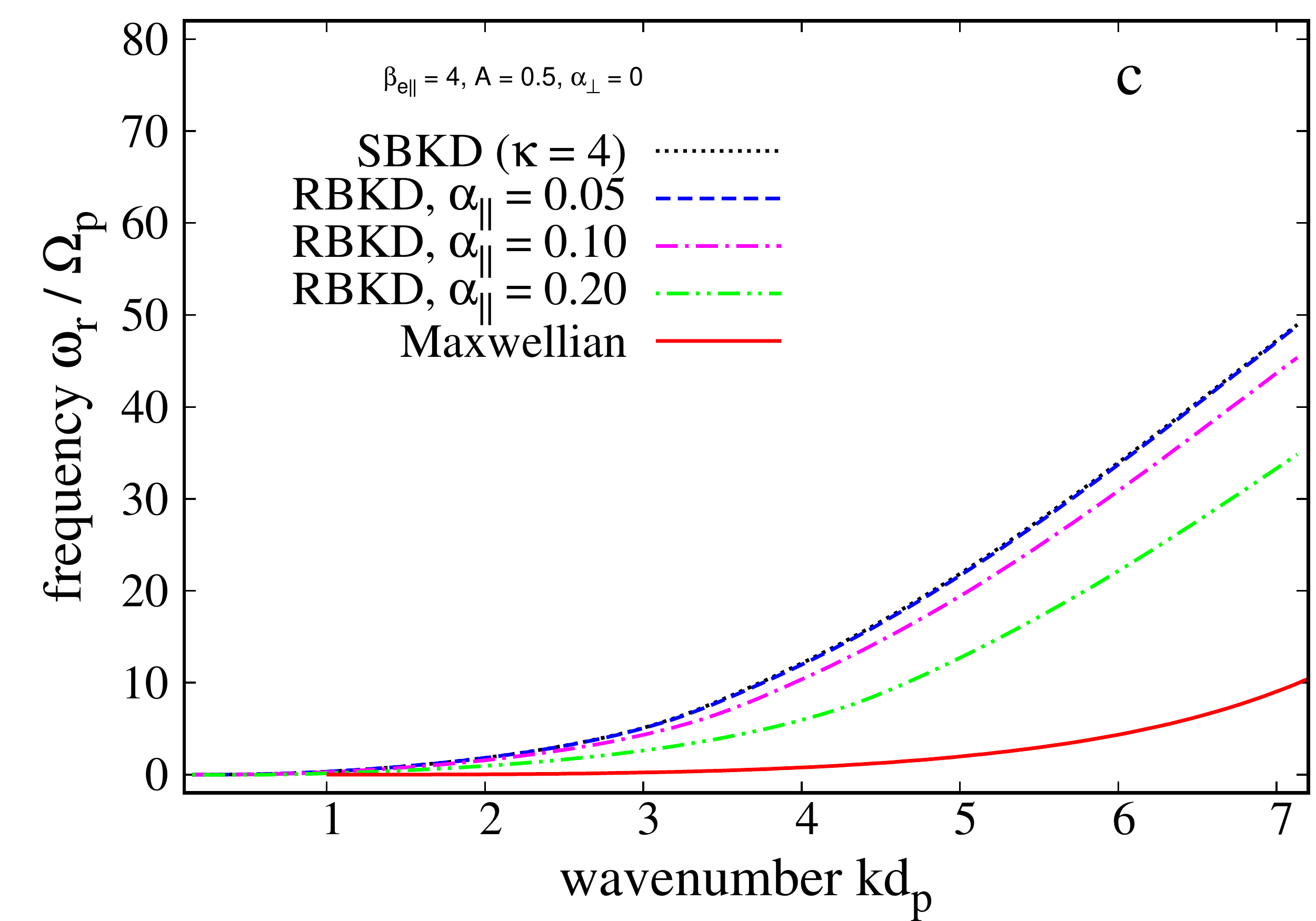}
  \hspace{0.1cm}
  \includegraphics[width=.35\textwidth]{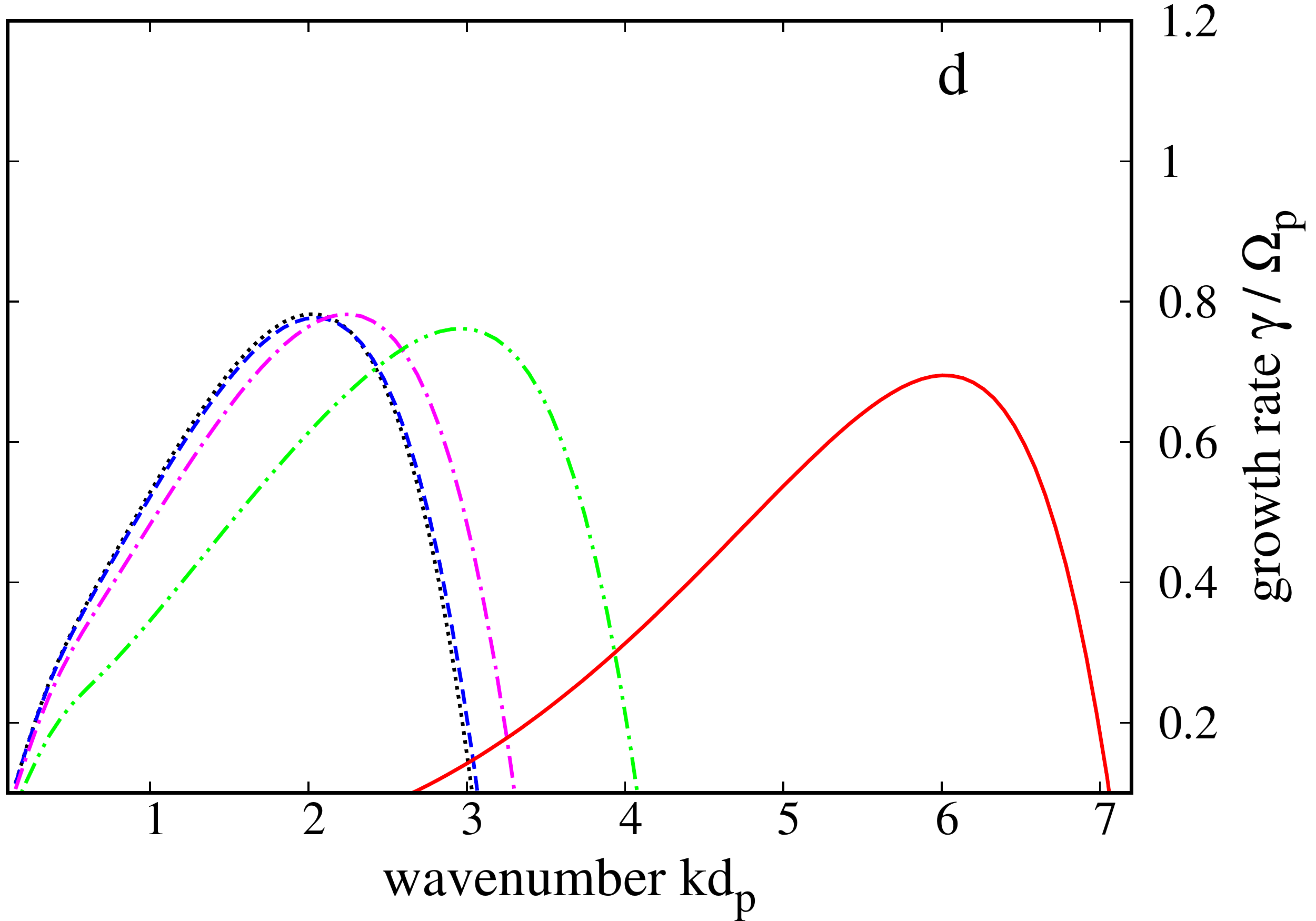}

  \vspace{0.1cm}

  \includegraphics[width=.35\textwidth]{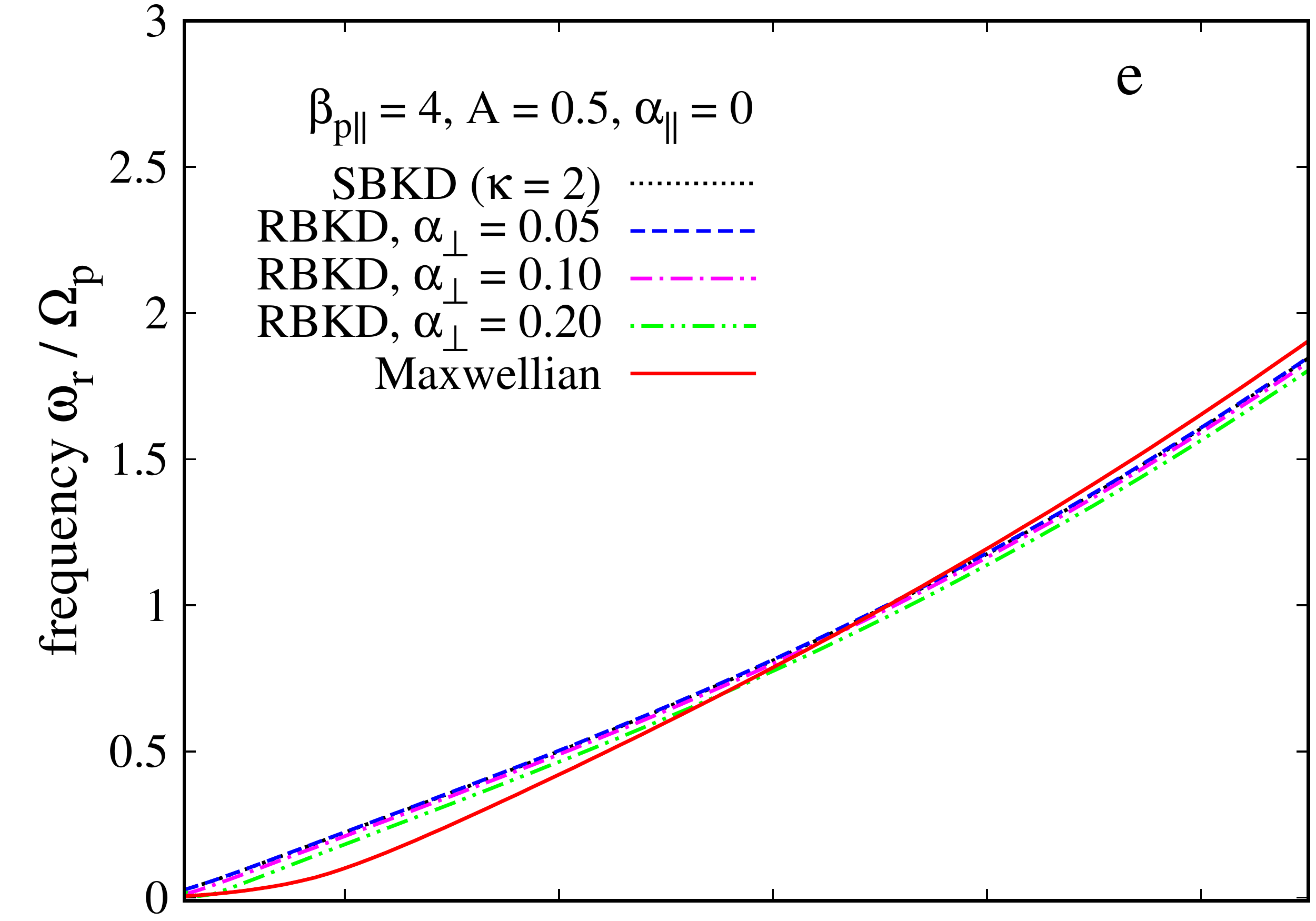}
  \hspace{0.1cm}
  \includegraphics[width=.35\textwidth]{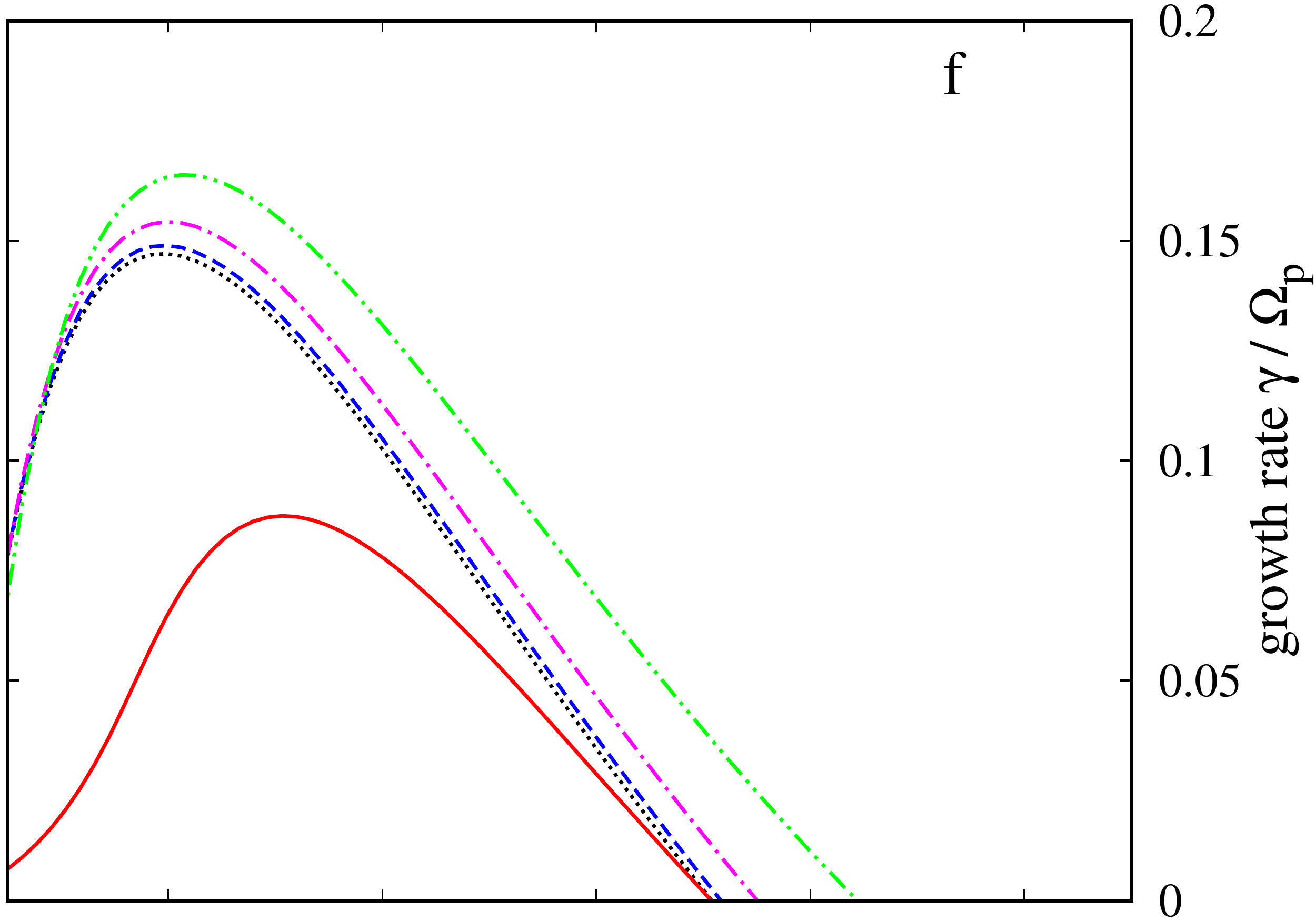}

  \vspace{0.1cm}

  \includegraphics[width=.35\textwidth]{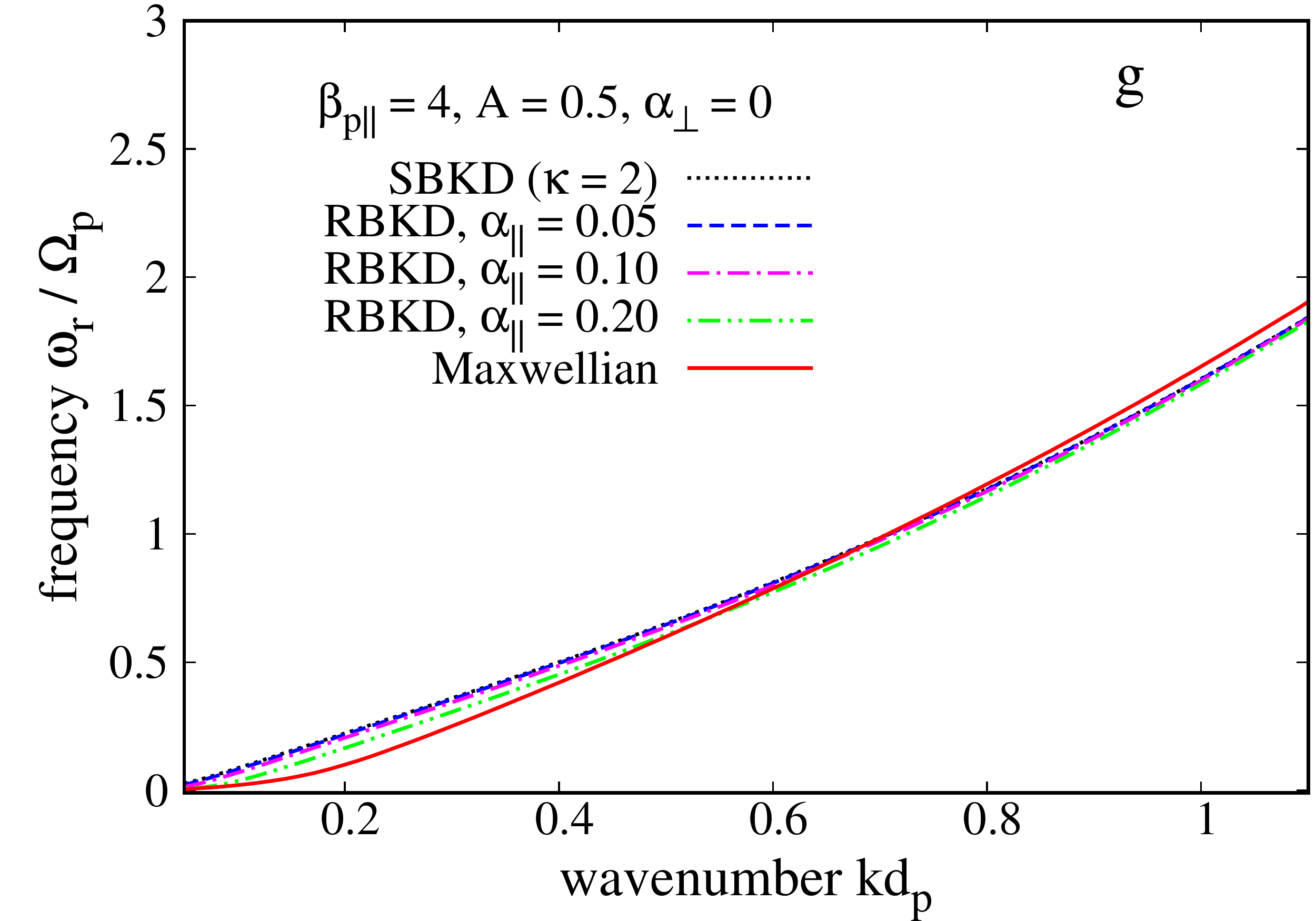}
  \hspace{0.1cm}
  \includegraphics[width=.35\textwidth]{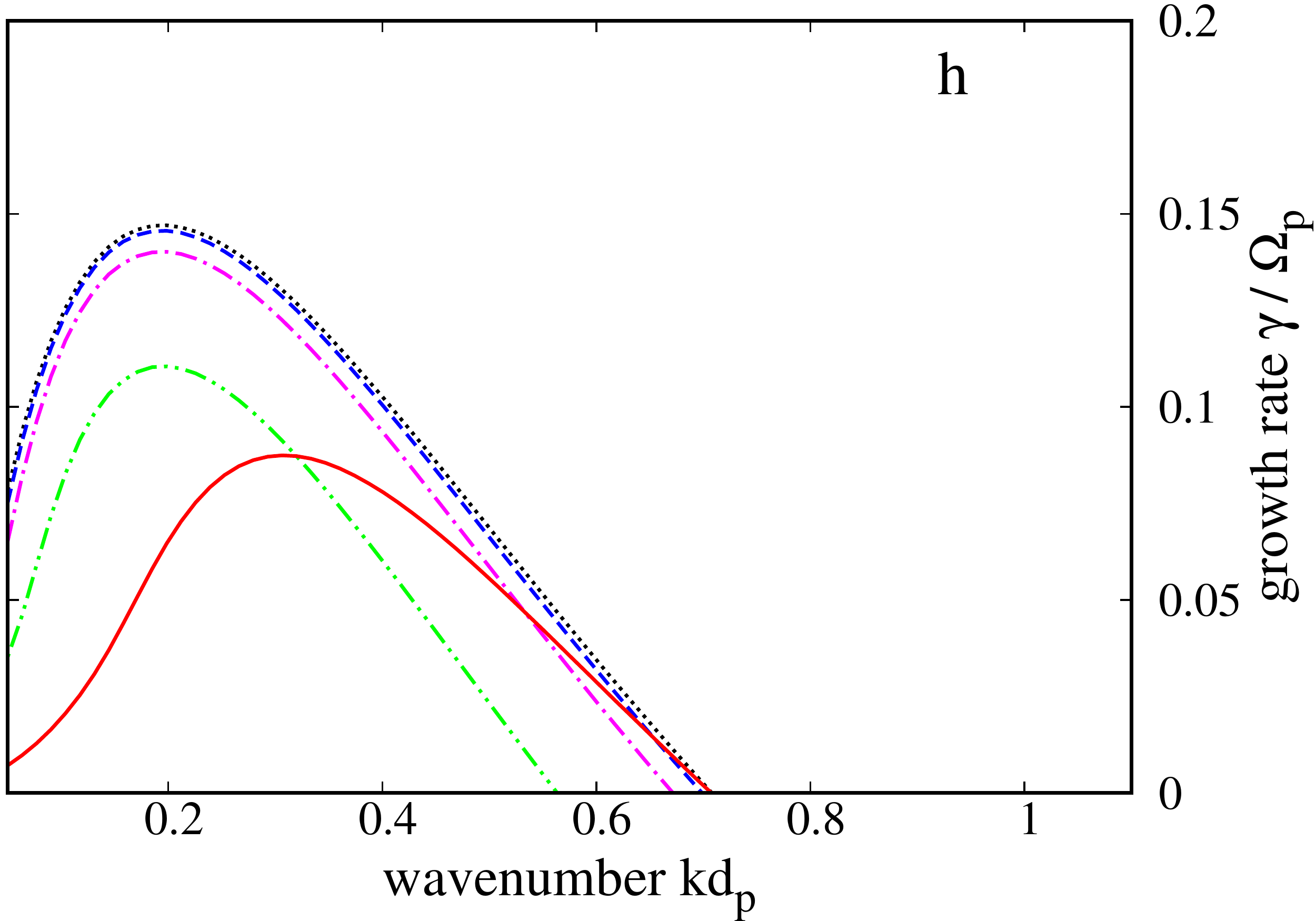}

  \caption{Dispersion curves for wave frequencies (left panels) and 
           growth rates (right panels) of the EFH (top two rows) and 
           PFH (bottom two rows) instability with 
           $\alpha_\parallel \neq \alpha_\perp$. Parameters are explained 
           in the legends.}\label{fig:pefh_ppfh_rbkd_apa_ne_ape}
\end{figure*}

Figure~\ref{fig:emec_rkd} shows the frequencies (left panels) and 
damping rates (right panels) of stable EMEC modes for different 
values of $\kappa$ and $\beta_\rd{e}$ (for exact values see legends). 
In all plots the protons are assumed to be Maxwellian ($\beta_\rd{p} = 1$), 
where the electrons are described by RKDs of different values of $\alpha$.
As test cases\cite{Gary-1993} the results of the Maxwellian (RKD with 
$\alpha = 0$ and $\kappa \to \infty$) are computed and shown in each plot. 
The results for the SKD (with $\alpha = 0$) are also displayed as references 
to be compared with the RKD. A comparison to the corresponding SKD reveals 
that the wave frequency dispersion curves remain basically unaffected even 
for higher values of the $\alpha$-parameter. We quantify the difference in 
the damping rates (or growth rates discussed in 
Secs.~\ref{subsec:emec_emic}-\ref{subsec:apara_neq_aperp}) obtained 
for RKDs by comparison to the corresponding SKD, by the following deviation 
\begin{equation}\label{eq:gamma_diff}
    R = \mid1 - \frac{\gamma_\rd{max,RKD}}{\gamma_\rd{max,SKD}}\mid\,.
\end{equation}
For the damping rates of the EMEC modes we estimate a 13\% deviation at a 
given value $k\,d_\rd{e} = 1$, and for $\kappa = 2$, $\beta_\rd{e} = 0.1$  
and $\alpha = 0.2$. For the lower $\alpha$-values this deviation decreases 
to about 5\%. The results for $\beta_\rd{e} = 1$ show for all values of 
$\alpha$ a deviation below 9\%. All deviations are presented in 
Tab.~\ref{tab:gamma_diff_sec_a}. For $\kappa = 6$ the deviations in the 
damping rates are overall below 8\%.
\begin{table}[ht]
  \begin{center}
    \begin{tabular}{ccc|ccc|ccc|ccc}
      \multicolumn{3}{c}{EMEC (0.1)} & \multicolumn{3}{c}{EMEC (1)} & \multicolumn{3}{c}{EMIC (0.1)} & \multicolumn{3}{c}{EMIC(1)} \\
      $\kappa$ & $\alpha$ & $R$ & $\kappa$ & $\alpha$ & $R$ 
      & $\kappa$ & $\alpha$ & $R$ & $\kappa$ & $\alpha$ & $R$ \\
      \hline
      2 & 0.05 & 0.01 & 2 & 0.05 & $<1\%$ & 2 & 0.05 & $<1\%$ & 2 & 0.05 & $<1\%$ \\
      2 & 0.1 & 0.04 & 2 & 0.1 & 0.02 & 2 & 0.1 & 0.02 & 2 & 0.1 & 0.02 \\
      2 & 0.2 & 0.13 & 2 & 0.2 & 0.08 & 2 & 0.2 & 0.08 & 2 & 0.2 & 0.07 \\
      6 & 0.05 & $<1\%$ & 6 & 0.05 & $<1\%$ & 6 & 0.05 & $<1\%$ & 6 & 0.05 & $<1\%$ \\
      6 & 0.1 & 0.02 & 6 & 0.1 & 0.01 & 6 & 0.1 & 0.01 & 6 & 0.1 & 0.01 \\
      6 & 0.2 & 0.07 & 6 & 0.2 & 0.03 & 6 & 0.2 & 0.04 & 6 & 0.2 & 0.02 \\
      \hline
    \end{tabular}
      \caption{The deviations for the damping rates of EMEC 
               (at $k\,d_\rd{e} = 1$) and EMIC (at $k\,d_\rd{p} = 1.4$) 
               modes, as provided by Eq. \eqref{eq:gamma_diff}. 
               The numbers in the brackets denote 
               $\beta$.}\label{tab:gamma_diff_sec_a}
  \end{center}
\end{table}
Differences may increase with increasing wavenumber, where the mode 
is highly damped, i.e., shifted to lower wave frequencies and lower 
damping rates with an increase of $\alpha$. For $\kappa = 6$ the 
differences are much less significant as the distribution models are 
closer to Maxwellian limits. Damping rates for the SKD plasma remain 
below the Maxwellian, and for the RKD it shifts toward the Maxwellian 
with increasing $\alpha$. This trend occurs due to the increasing value 
of $\alpha$, by which more resonant particles that can gain energy from 
the electromagnetic waves, are removed 
(see, e.g., Ref.~\onlinecite{Verscharen-et-al-2019}).

The electromagnetic ion-cyclotron modes (EMIC) are low-frequency 
modes, left-hand circularly polarized and have been detected in 
the solar wind\cite{Jian-et-al-2009} and in planetary 
magnetospheres.\cite{Nguyen-et-al-2007} In Fig.~\ref{fig:emic_rkd} 
the computed frequencies and damping rates are presented for stable 
EMIC modes. The setup and organization of the panels is the same as 
for Fig.~\ref{fig:emec_rkd}, except that the protons are described by 
RKDs now and electrons are Maxwellian and isotropic with 
$\beta_\rd{e} = 1$. We can see that the wave frequency curves for the 
SKD and RKDs reach a plateau or maximum around $k\,d_\rd{p} \gtrsim 0.8$, 
while the same curves in the Maxwellian-distributed plasmas continue 
to increase\cite{Gary-1993}. At higher temperatures the wave frequency 
decreases with increasing the wavenumber. The effect is more significant 
for low values of the $\kappa$-index, i.e., $\kappa = 2$, since with the 
increase of $\kappa$ the distributions approach the Maxwellian (and with 
the increase of $\alpha$ the dispersion curves overall approach the 
Maxwellian). Depending on the kinetic temperature of SKD particles, for 
low values of $\kappa$ one may critically need to invoke RKD models, which 
reduce the unrealistic contribution of superluminal particles present in 
the SKD models, see the $\kappa=2$ case in 
Ref.~\onlinecite{Scherer-et-al-2019b}. The influence on the damping rates 
is similar to the EMEC modes: comparing to Maxwellian limits, in 
Kappa-distributed plasmas the waves are stronger damped, but with the 
increase of the $\alpha$-parameter more resonant particles are removed and 
the damping curves approach the Maxwellian. The deviation of the RKD-based 
damping rates from the one based on the SKD is overall below 9\%. All 
deviations are calculated at $k\,d_\rd{p} = 1.4$ and displayed 
in Tab.~\ref{tab:gamma_diff_sec_a}.

\subsection{Electromagnetic Cyclotron Instabilities}\label{subsec:emec_emic}

In space plasmas, the free energy of anisotropic electron 
populations, e.g., an electron temperature anisotropy of 
$A_{\rd{e}} = T_{\rd{e}\perp} / T_{\rd{e}\parallel} > 1$, 
may lead to an instability of the EMEC (or whistler-cyclotron) 
modes, with maximum growth rates in parallel 
direction.\cite{Lazar-et-al-2013} By convention, the growth is 
described via a positive rate $\gamma > 0$. In the low-collisional 
plasmas from space, the enhanced whistler fluctuations 
selfconsistently induced by the electron anisotropy can be very 
effective in constraining large deviations from isotropy, and the 
observations appear to confirm 
that.\cite{Stverak-et-al-2008, Lazar-et-al-2017b}  
The frequency curves (left panels) and growth rates (right panels) 
of the EMEC instability computed for an RBKD-plasma are displayed 
in Fig.~\ref{fig:emec_rbkd}. Each plot contains the results for 
three RBKDs, which are compared to the corresponding SBKD and 
Bi-Maxwellian distribution.\cite{Lazar-et-al-2013} The exact values 
of $\kappa$, $\beta_{\rd{e}\perp}$ and the anisotropy are displayed 
in the legends. The protons are assumed to be isotropic and 
Maxwellian with $\beta_\rd{p} = 1$. In the left column of 
Fig.~\ref{fig:emec_rbkd} the wave frequency dispersion curves remain 
basically unchanged for small anisotropies, e.g., $A_\rd{e} = 1.5$, and 
with the increase of the anisotropy, e.g., for $A_\rd{e} = 3$, all 
Kappa-curves deviate stronger from the Maxwellian. This difference is 
reduced for higher spectral indices, e.g., for $\kappa = 4$, and the 
influence of the cutoff parameter $\alpha$ is minor. By comparing the 
maximums of the growth rates obtained for the RBKD with the one obtained 
for the SBKD we find the greatest deviation with 16\% for 
$\kappa = 2$, $A = 1,5$ and $\alpha = 0.2$, while the other curves 
mostly deviate to about 1 to 5\% in their maximums. All deviations are 
presented in Tab.~\ref{tab:gamma_diff_sec_bc}.
\begin{table}[ht]
  \begin{center}
    \begin{tabular}{ccc|ccc|ccc|ccc}
      \multicolumn{3}{c}{EMEC ($1.5$)} & \multicolumn{3}{c}{EMIC ($1.5$)} &
      \multicolumn{3}{c}{EFH ($0.5$)} & \multicolumn{3}{c}{PFH ($0.66$)} \\
      $\kappa$ & $\alpha$ & $R$ & $\kappa$ & $\alpha$ & $R$ 
      & $\kappa$ & $\alpha$ & $R$ & $\kappa$ & $\alpha$ & $R$\\ 
      \hline
      2 & 0.05 & 0.01 & 2 & 0.05 & 0.03 & 4 & 0.05 & $<1\%$ & 2 & 0.05 & 0.04 \\
      2 & 0.1 & 0.05 & 2 & 0.1 & 0.11 & 4 & 0.1 & $<1\%$ & 2 & 0.1 & 0.17 \\
      2 & 0.2 & 0.16& 2 & 0.2 & 0.30 & 4 & 0.2 & $<1\%$ & 2 & 0.2 & 0.53 \\
      4 & 0.05 & $<1\%$ & 6 & 0.05 & $<1\%$ & 6 & 0.05 & $<1\%$ & 4 & 0.05 & 0.03 \\
      4 & 0.1 & 0.02 & 6 & 0.1 & 0.03 & 6 & 0.1 & $<1\%$ & 4 & 0.1 & 0.13 \\
      4 & 0.2 & 0.08 & 6 & 0.2 & 0.12 & 6 & 0.2 & 0.02 & 4 & 0.2 & 0.41 \\
      \hline
      \multicolumn{3}{c}{EMEC ($3$)} & \multicolumn{3}{c}{EMIC ($3$)} & 
      \multicolumn{3}{c}{EFH ($0.33$)} & \multicolumn{3}{c}{PFH ($0.33$)} \\
      \hline
      2 & 0.05 & 0.01 & 2 & 0.05 & 0.01 & 4 & 0.05 & $<1\%$ & 2 & 0.05 & $<1\%$ \\
      2 & 0.1 & 0.04 & 2 & 0.1 & 0.04 & 4 & 0.1 & $<1\%$ & 2 & 0.1 & 0.02 \\
      2 & 0.2 & 0.12 & 2 & 0.2 & 0.11 & 4 & 0.2 & 0.01 & 2 & 0.2 & 0.12 \\
      4 & 0.05 & $<1\%$ & 6 & 0.05 & $<1\%$ & 6 & 0.05 & $<1\%$ & 4 & 0.05 & 0.01 \\
      4 & 0.1 & 0.01 & 6 & 0.1 & 0.01 & 6 & 0.1 & $<1\%$ & 4 & 0.1 & 0.04 \\
      4 & 0.2 & 0.05 & 6 & 0.2 & 0.04 & 6 & 0.2 & 0.01 & 4 & 0.2 & 0.18 \\
      \hline
    \end{tabular}
      \caption{The deviations for the maximum growth rates of the 
               EMEC, EMIC, EFH and PFH instabilities, as provided 
               by Eq.~\eqref{eq:gamma_diff}. 
               The numbers in the brackets denote the 
               anisotropy $A$.}\label{tab:gamma_diff_sec_bc}
  \end{center}
\end{table}
The growth rates, displayed in the right column, are in general 
highly modified under the influence of the $\kappa$- and 
$\alpha$-parameters. Profiles are the same for all cases, showing 
peaking (maximum) values increasing with the increase of the 
anisotropy that drives the instability. With the increase of 
$\alpha$ the growth rates decrease with the peak shifting toward 
higher wavenumbers, and this effect can be understood in the 
following way. With increasing $\alpha$ the distribution function 
receives a greater cutoff in the high-velocity range, and the 
number of particles that resonate with, and lose energy to the 
waves is diminished. Consequently, the growth rates are reduced. 
Again, the influence of $\alpha$ on the growth rates is reduced 
for a higher $\kappa$.

We should mention that for the cyclotron resonant instabilities 
not only the number of resonant particles is responsible to 
determine the magnitude of the growth rate, but also the pitch 
angle anistropy of the velocity distribution in the area of 
resonance in the distribution that has to be sufficiently 
anisotropic, see Ref.~\onlinecite{Astfalk-and-Jenko-2016} 
and \onlinecite{Singhal-2009}.

The EMIC instability arises due to an excess of kinetic energy of 
the protons in perpendicular direction, i.e., 
$T_{\rd{p}\perp} > T_{\rd{p}\parallel}$. The maximum growth rates 
of the EMIC waves are again obtained for parallel propagation, so 
that one can take $k \simeq k_\parallel$. Figure~\ref{fig:emic_rbkd} 
shows the wave frequency curves and growth rates. The composition of 
the plots is similar as for the EMEC instability, where we have 
computed as test cases the results for the 
SBKD\cite{Lazar-and-Poedts-2013} and the 
Bi-Maxwellian\cite{Lazar-and-Poedts-2013}, assuming this time 
$\beta_\rd{e} = 1$, while the protons are described by RBKDs. 
For the details of the parameter setup see the legends. 
While the dispersion curves are only slightly affected again 
by the change in the $\alpha$-parameter, the growth rates 
decrease with the increase in $\alpha$ (and hence a removal 
of resonant particles that lose energy to the waves) and the 
peak of the growth rates shifts toward higher wavenumbers. 
For $\kappa = 6$ these effects are less prominent. When comparing 
the maximum growth rates we find for the EMIC instability for 
$\kappa = 2$, $A = 1,5$ and $\alpha = 0.2$ the greatest deviation 
with about 30\%, while all other results show a deviation around 10\% 
for the curves with $\alpha = 0.2$, and mostly below 5\% for the 
curves with lower values of $\alpha$. All results are displayed 
in Tab.\ref{tab:gamma_diff_sec_bc}.

\subsection{The Electron and Proton Firehose Instabilities}\label{subsec:pfh_efh}

For opposite anisotropies of electrons, i.e., $A_{\rd{e}} < 1$, 
the theory predicts the so-called electron firehose (EFH) instability, 
developing as a low-frequency left-hand circularly polarized mode in 
parallel direction, at wavenumbers $k\,d_\rd{e} > 1$ and with 
frequencies between the proton cyclotron frequency and the electron 
cyclotron frequency 
($\Omega_\rd{p} < \omega_\rd{r} < \Omega_\rd{e}$ ).\citep{Gary-1993} 
The parallel firehose instability is competed by another aperiodic 
branch propagating only obliquely to the magnetic field, 
see Refs.~\onlinecite{Maneva-et-al-2016, Lopez-et-al-2019} and 
references therein, but our present study intended to testing 
regularized Kappa-distributions is limited to parallel EM modes. 

Figure~\ref{fig:pefh_rbkd} shows the wave frequency curves 
(left column) and growth rates (right column) of the EFH 
instability. The legends summarize the values used for the 
model parameters and the protons are described by an isotropic 
Maxwellian of $\beta_\rd{p} = 1$. Profiles of the wavenumber 
dispersion curves remain unchanged, as already known for the 
standard bi-Maxwellian and SBKD\cite{Lazar-et-al-2017b}, i.e., 
with the decrease of the temperature anisotropy the frequencies 
are significantly enhanced, while the peak of the growth rates 
increases and shifts toward lower wavenumbers. Decreasing $\kappa$ 
has similar effect, but with increasing $\alpha$ the growth rates 
shift toward higher wavenumbers towards the Maxwellian-based curves.
The comparison of the maximum growth rates shows beside the shift of 
the peak almost no deviation between the RBKD- and SBKD-based results, 
where the deviation is for almost all results below 1\% 
(see, Tab.~\ref{tab:gamma_diff_sec_bc} for detailed results).

The proton firehose (PFH) instabilities, either parallel or the 
aperiodic oblique branches, arise when protons are characterized 
by a temperature anisotropy $A_\rd{p} < 1$ and a sufficiently high 
$\beta_{\rd{p}\parallel} > 1$.\cite{Astfalk-and-Jenko-2016}
Different observations suggest that these firehose instabilities 
limit any further increase of the proton temperature in parallel 
direction. If the temperature anisotropy exceeds the instability 
threshold, the fluctuations excited by the firehose modes grow and 
act back on the protons contributing to their relaxation and 
constraining their anisotropy.\cite{Lazar-et-al-2011} 
For certain conditions maximum growth may occur at parallel 
propagation, i.e., $\vec{k} \times \vec{B}_0 = \vec{0}$, with 
$\vec{B}_0$ being the vector of the local background magnetic field, 
motivating our limitation to parallel modes in the present analysis.
The proton firehose instability shows right-hand circular polarization 
in parallel direction of propagation and evolves out of the 
magnetosonic/whistler wave as the proton temperature 
anisotropy increases.\citep{Gary-1993} Figure~\ref{fig:ppfh_rbkd} 
presents the dispersion curves for the wave frequencies (left column) 
and growth rates (right column) of the PFH instability computed for 
the RBKD-distributed plasmas with different values of $\alpha$, 
together with the test case results for the SBKD- and Maxwellian 
distributed plasmas. All results are shown for two different values 
of $\kappa$ and $A_\rd{p}$, while the electrons are now described by 
an isotropic Maxwellian of $\beta_\rd{e} = 1$. As in previous 
studies\cite{Lazar-et-al-2011} comparing SBKD with Maxwellian, 
the frequency changes with $A_\rd{p}$, but does not vary much with 
$\kappa$. It remains also unaffected by the variation of the 
$\alpha$-parameter. However, the growth rates and their peaking 
values are in general much higher than Maxwellian limit. For small 
$A_\rd{p}$ and small $\kappa$, e.g., Fig.\ref{fig:ppfh_rbkd} panel b 
with $\kappa = 2$ and $A_\rd{p} = 0.66$, the growth rates decrease 
significantly (the maximum decreases by about 53\%) with increasing 
$\alpha$ and shift toward higher wavenumbers. Another rather significant 
decrease by 41\% in the maximun growth rate occurs in panel d with 
$\kappa = 4$ and $A_\rd{p} = 0.66$, while the differences for all other 
curves are below $20\%$ (see, Tab.~\ref{tab:gamma_diff_sec_bc}.)
For higher values of $\kappa$, e.g., panel d, or higher 
anisotropies, e.g, panel f, these effects become less prominent.

\subsection{Anisotropic regularization parameters}\label{subsec:apara_neq_aperp}

In Sections.~\ref{subsec:emec_emic} and \ref{subsec:pfh_efh} 
we examined the effects of an isotropic regularization using 
the same values for parameter $\alpha$, i.e., 
$\alpha_\parallel = \alpha_\perp$, in both parallel and 
perpendicular directions. As for a more general case in this 
section we allow for $\alpha$ to be anisotropic, i.e., 
$\alpha_\parallel \neq \alpha_\perp$ and show the effects of 
this regularization on the wavenumber dispersion properties of 
the EMEC, EMIC, EFH and PFH instabilities.

Figure~\ref{fig:emec_emic_rbkd_apa_ne_ape} displays the results 
for the EMEC (top two rows) and EMIC (bottom two rows) instabilities, 
where we examine the effect of changing the regularization parameter 
only in one direction, while keeping it constant in the other direction 
(see the legends). The plots for the EMEC instability show the 
frequencies (left column) and growth rates (right column) for the case of 
a constant $\alpha_\parallel$ and varied $\alpha_\perp$ (panels a and b) 
and the case of a constant $\alpha_\perp$ and different values of 
$\alpha_\parallel$ (panels c and d). We assume $\kappa = 2$, 
$\beta_{\rd{e}\perp} = 3$, $\beta_\rd{p} = 1$ and $A_\rd{e} = 3$ 
for all plots. Although similar to the previous case with 
$\alpha_\parallel = \alpha_\perp$ in Fig.~\ref{fig:emec_rbkd}, 
the wave frequency is slightly lowered with increasing $\alpha_\perp$ or 
slightly enhanced with increasing $\alpha_\parallel$.
On the other hand, the growth rates are significantly affected by the 
change of the regularization parameters. In the case of constant 
$\alpha_\parallel$ (panel b), the growth rates markedly decrease with 
the increase of $\alpha_\perp$ (peak decreases by $23\%$ compared to the 
SBKD-based result), and the peak shifts toward lower wavenumbers. 
All differences in the maximums of the growth rates between RBKD- and 
SBKD-based results are shown in Tab.\ref{tab:gamma_diff_sec_d}. 
Physically, this can be explained by the variations of the effective 
anisotropy, which drives the EMEC instability. In the case of 
$\alpha_\parallel \neq \alpha_\perp$ the anisotropy becomes a function 
of the two regularization parameters and thus a change in either 
$\alpha_\parallel$ or $\alpha_\perp$ alters the anistropy. With the 
increase of $\alpha_\perp$ (and a constant $\alpha_\parallel$), the 
distribution loses more high-energy particles from perpendicular 
direction that can resonate with the waves, and the effective 
anisotropy decreases, which is reflected by lower peaks of the 
growth rates. However, the same effective anisotropy increases if 
$\alpha_\perp$ is constant but $\alpha_\parallel$ increases, leading 
in this case to a stimulation of the instability. For the EMIC 
instability we find a similar behavior of the results that can be 
explained analogously, i.e., the dispersion curves for the wave 
frequency remain basically unchanged with increasing components of 
the $\alpha$-parameter, while the growth rates change in the way 
described for the EMEC instability. Rather big differences occur in 
panel h, where the maximum of the curve with $\alpha_\parallel = 0.05$ 
deviates by 22\% from the SBKD-curve, $\alpha_\parallel = 0.1$ by 28\%, 
and $\alpha_\parallel = 0.2$ by almost 50\%. Detailed results for the 
EMIC instability are shown in Tab.~\ref{tab:gamma_diff_sec_d}.

\begin{table}[ht]
  \begin{center}
    \begin{tabular}{ccc|ccc|ccc|ccc}
      \multicolumn{3}{c}{EMEC} & \multicolumn{3}{c}{EMIC} 
      & \multicolumn{3}{c}{EFH} & \multicolumn{3}{c}{PFH} \\
      $\alpha_\parallel$ & $\alpha_\perp$ & $R$ 
      & $\alpha_\parallel$ & $\alpha_\perp$ & $R$ 
      & $\alpha_\parallel$ & $\alpha_\perp$ & $R$ 
      & $\alpha_\parallel$ & $\alpha_\perp$ & $R$ \\
      \hline
      0 & 0.05 & 0.02 & 0 & 0.05 & 0.16 & 0 & 0.05 & $<1\%$ & 0 & 0.05 & 0.01 \\
      0 & 0.1 & 0.08 & 0 & 0.1 & 0.08 & 0 & 0.1 & $<1\%$ & 0 & 0.1 & 0.05 \\
      0 & 0.2 & 0.23 & 0 & 0.2 & 0.12 & 0 & 0.2 & 0.02 & 0 & 0.2 & 0.12 \\
      0.05 & 0 & 0.01 & 0.05 & 0 & 0.22 & 0.05 & 0 & $<1\%$ & 0.05 & 0 & 0.01 \\
      0.1 & 0 & 0.05 & 0.1 & 0 & 0.28 & 0.1 & 0 & $<1\%$ & 0.1 & 0 & 0.04 \\
      0.2 & 0 & 0.19 & 0.2 & 0 & 0.49 & 0.2 & 0 & 0.02 & 0.2 & 0 & 0.24 \\
      \hline
    \end{tabular}
      \caption{The deviations for the maximum growth rates of the 
               EMEC and EMIC ($A = 3$, $\kappa = 2$), 
               EFH ($A = 0.5$, $\kappa = 4$) and PFH ($A = 0.5$, $\kappa = 2$) 
               instabilities, as provided by Eq.~\eqref{eq:gamma_diff}.
               }\label{tab:gamma_diff_sec_d}
  \end{center}
\end{table}

In Fig.~\ref{fig:pefh_ppfh_rbkd_apa_ne_ape} we show the effect 
of different $\alpha_\parallel$ and $\alpha_\perp$ on the EFH 
(top two rows) and the PFH (bottom two rows) instabilities.
Again, we can see that both the frequency (left panels) and 
growth rates (right panels) vary with the increase of the 
$\alpha$-components, especially for the EFH instability. 
However, comparing to the EMEC and EMIC instabilities, here 
we observe opposite effects, as for instance, for the EFH 
instability the frequency is lowered with increasing 
$\alpha_\perp$, but markedly lowered with increasing 
$\alpha_\parallel$. In this case the growth rates are not much 
affected by $\alpha_\perp$ (i.e., maximum growth rates deviate 
by less than 3\%), but are slightly lowered moving to higher 
wavenumbers for higher values of $\alpha_\parallel$. The PFH wave 
frequencies do not change much, but the growth rates increase 
with increasing $\alpha_\perp$ and decrease with increasing 
$\alpha_\parallel$. The largest deviation occurs for a strong 
cut-off parameter $\alpha_\parallel = 0.24$ in panel h, 
where the maximum growth rate deviates by 24\%.

\section{Summary and Outlook}\label{sec:summary}    

There is consistent evidence that shows the existence of 
particle populations in space plasmas, e.g., 
in the solar wind and planetary magnetospheres,
to be out of 
thermal equilibrium and showing enhanced suprathermal 
tails that can be well-fitted by standard Kappa-distributions 
(SKD). Due to unphysical limitations of the SKD, the 
regularized Kappa-distribution (RKD) has been introduced, 
which removes the critical restrictions of the SKD and is 
able to reproduce as limit cases the Maxwellian and the SKD 
as well as its results considering plasma waves and instabilities. 
We have investigated for the first time parallel electromagnetic 
modes in a plasma described by regularized Kappa-distributions 
and used parameters typical for space plasmas, but not restrained 
to these conditions. The results were derived by using the kinetic 
dispersion solver \emph{LEOPARD} able to resolve the dispersion 
properties for plasma particles with arbitrary distribution functions. 
The wavenumber dispersion curves obtained for Maxwellian and SKD 
distribution functions are in perfect agreement with those predicted 
by the numerical solvers dedicated to these models.

In Section~\ref{subsec:stable_modes} we presented the 
dispersion curves for wave frequencies and damping rates 
derived with \emph{LEOPARD} for the stable electromagnetic 
electron-cyclotron (EMEC) and ion-cyclotron (EMIC) modes by 
applying the isotropic RKD. The results show that even for 
higher values of the regularization parameter $\alpha$ the 
dispersion curves show almost no deviation from the curves 
obtained for the SKD. However, for the damping rates a value 
of $\alpha < 0.1$ is necessary to reproduce the data of the 
SKD, i.e., allowing only for small deviations from the 
SKD-based damping rates. 

In Section~\ref{subsec:emec_emic} we used the anisotropic 
regularized Kappa-distribution (RBKD) and computed the 
frequencies and growth rates of the EMEC and EMIC instabilities, 
which are driven by a temperature anisotropy of 
$A = T_\perp / T_\parallel > 1$. Here, we could also see that 
the frequencies even for relative high values of 
$\alpha_\parallel = \alpha_\perp = \alpha$ are close to the 
results of the corresponding anisotropic standard Kappa-distribution 
(SBKD) and that the growth rates decrease with increasing $\alpha$ 
(as with higher values of $\alpha$ there are less resonant particles 
to drive the instability). For the electron firehose (EFH) and proton 
firehose (PFH) instabilities in Section~\ref{subsec:pfh_efh} we found 
similar results, except the increase of the growth rates for increasing 
$\alpha_\perp$ (while keeping $\alpha_\parallel$ constant), as the 
effective temperature anisotropy increases in that case. 
For anisotropic regularization parameters, i.e., 
$\alpha_\parallel \neq \alpha_\perp$, examined in 
Section~\ref{subsec:apara_neq_aperp}, the growth rates of the EMEC and 
EMIC instabilities decrease with increasing $\alpha_\perp$ 
(keeping  $\alpha_\parallel$ constant), and increase with increasing 
$\alpha_\parallel$ (keeping $\alpha_\perp$ constant), which is due 
either to a decrease of the effective (temperature) anisotropy for the 
former case, or a higher anisotropy in the latter case. For the 
EFH and PFH instabilities we found the opposite case, i.e., an increase 
of the growth rates with increasing $\alpha_\perp$ and a decrease with 
increasing $\alpha_\parallel$ (keeping the component in the other 
direction constant, respectively).

The regularized Kappa-distribution, therefore, can reproduce the results 
of a standard Kappa-distribution with the benefit of removing its 
limitations by being defined for all $\kappa > 0$, having no diverging 
velocity moments and no contribution to macroscopic quantities 
(e.g., pressure) by superluminal particles (provided that $\alpha$ 
is choosen properly). As a continuation of the wave dispersion 
properties of plasma systems described by the regularized 
Kappa-distribution, future studies must consider oblique 
modes,\cite{Gaelzer-and-Ziebell-2016} e.g., 
the aperiodic mirror\cite{Shaaban-et-al-2018} and 
firehose\cite{Meneses-et-al-2018, Shaaban-et-al-2019,Lopez-et-al-2019} 
instabilities. However, along with these strengths, we have 
encountered a few issues which need only to be mentioned at 
this stage. The code is very effective in the study of kinetic 
instabilities, such as the EMEC, EMIC and PFH instabilities with 
the assumption of a two-component plasma (electrons and protons) 
with only one anisotropic component, while the other component is 
assumed to be isotropic and Maxwellian. Since the code has the option 
to produce the Maxwellian distribution internally, this reduces the 
computation time significantly. Once the proper initial guesses for 
the real and imaginary part of $\omega$ were found, the code worked 
solidly for these cases. However, the choice of the initial guesses 
is crucial for the code to start on the desired solution branch.
Difficulties occurred when we considered damped modes, i.e., when 
applied isotropic distributions for all particle species. The results in 
Figs.~\ref{fig:emec_rkd} and \ref{fig:emic_rkd} had to be smoothed by 
an interpolation method, because the raw results from \emph{LEOPARD} 
showed many irregularities toward higher wavenumbers in the trend of the 
curves. The reason for this problem could be the method of spline 
interpolation that is applied inside the code. 
Verscharen et al.\cite{Verscharen-et-al-2018} mention that the fit method 
with cubic splines becomes inaccurate for strongly damped solutions. 
In order to solve the dispersion relation the evaluation of the spline at 
a complex value is needed, but the value is distant from the real grid 
points by which the spline is supported.

Further challenges occurred also in the computation of the results 
of the EFH instability. In order to produce the results we had to 
use electrons as the first species, so that all results are obtained 
in electron-based units, and then we transformed the results to 
proton-units as shown in Figs.~\ref{fig:pefh_rbkd} and 
\ref{fig:pefh_ppfh_rbkd_apa_ne_ape}. Therefore, we had to reach very 
low values of $k\,d_\rd{e}$, which caused computational issues. 
Additionally, for very low values of the real and/or imaginary part of 
the frequency the code easily jumps away from the desired branch of 
solutions. In general the code goes through the wavenumber interval only 
forward and one cannot force to the code to stay on a particular 
solution branch, which means that the code will sometimes jump to the 
wrong solution branch. This can be somewhat resolved by starting at 
higher values of $k\,d_\rd{s}$ and slowly working back to lower values. 
These issues will be addressed in our future studies in an attempt to 
provide a full spectrum of smooth solutions.

Our present results confirm and identify extended abilities of the 
\emph{LEOPARD} code, as a powerful tool for the kinetic wave analysis of 
realistic plasma systems. A great advantage of \emph{LEOPARD} is the 
ability to use arbitrary distribution functions, i.e., not being limited 
to idealized distributions and, therefore, not requiring any pre-derivation 
of the dielectric tensor for a particular distribution model. Premises are 
thus provided to integrate such a code with measurements of velocity 
distributions and describe sufficiently accurate the wave fluctuations 
detected in parallel by the space probes. In addition, our results in 
Section~\ref{sec:results}, can serve as valuable references to test 
similar codes, like the recently introduced \emph{ALPS} 
(Arbitrary Linear Plasma Solver).\cite{Verscharen-et-al-2018}
%

\begin{acknowledgments}
EH acknowledges support from the Ruhr-Universit\"at Bochum, and ML acknowledges support in the framework of the projects G0A2316N (FWO--Vlaanderen) and SCHL~201/35-1 (\emph{DFG}). HF and KS are grateful to the \emph{Deut\-sche For\-schungs\-ge\-mein\-schaft, DFG\/} funding the project SCHE334/10-1. We thank the two anonymous referees for their constructive suggestions and comments on the manuscript.
\end{acknowledgments}

\bibliography{rkd_paper_3}  

\end{document}